\documentclass[sigconf]{acmart}

\renewcommand\footnotetextcopyrightpermission[1]{} 
\usepackage{booktabs} 

\usepackage{algorithmic}
\usepackage{cleveref}
\usepackage[short]{optidef}
\usepackage{physics}
\usepackage{graphicx}
\usepackage{subcaption}
\usepackage{nicefrac}
\usepackage{mathrsfs}
\usepackage{enumitem}

\usepackage[ruled]{algorithm2e} 

\newcommand{\teq}{\! = \!}
\newcommand{\teg}{\! - \!}
\newcommand{\tsum}{\! + \!}

\newcommand{\E}{\mathbb{E}}
\newcommand{\Var}{\mathrm{Var}}

\renewcommand{\Pr}{\mathrm{P}}

\DeclareMathOperator*{\argmax}{\arg \ \max}

\def\clap#1{\hbox to 0pt{\hss#1\hss}} \def\mathclap{\mathpalette\mathclapinternal} \def\mathclapinternal#1#2{%
\clap{$\mathsurround=0pt#1{#2}$}}

\theoremstyle{definition}
\newtheorem{assumption}{Assumption}
\theoremstyle{definition}

\begin{document}
\settopmatter{printfolios=true}
\title{Impact of Social Learning on Privacy-Preserving Data Collection}
  
\author{Abdullah Basar Akbay}
\affiliation{%
	\institution{School of Electrical, Computer\\ and Energy Engineering\\ Arizona State University}
	\city{Tempe}
  	\state{AZ}
    \postcode{85287}}
\email{aakbay@asu.edu}

\author{Weina Wang}
\affiliation{%
	\institution{School of Computer Science\\ Carnegie Mellon University}
	\city{Pittsburgh}
  	\state{PA}
    \postcode{15213}}
\email{weinaw@cs.cmu.edu}

\author{Junshan Zhang}
\affiliation{%
	\institution{School of Electrical, Computer\\ and Energy Engineering\\ Arizona State University}
	\city{Tempe}
  	\state{AZ}
    \postcode{85287}}
\email{junshan.zhang@asu.edu}


\begin{abstract}
We study a model where a data collector obtains  data from users through a payment mechanism, aiming to learn the underlying state from the elicited data. The private signal of each user represents her knowledge about the state; and through social interactions each user can also learn noisy versions of her social friends'  signals, which is called `learned group signals'. Thanks to social learning, users have richer information about the  state beyond their private signals. Based on both her private signal and learned group signals, each user makes strategic decisions to report a privacy-preserved version of her data to the data collector. We develop a Bayesian game theoretic framework to study the impact of social learning on users' data reporting strategies and devise the payment mechanism for the data collector accordingly. Our findings reveal that, in general, the desired data reporting strategy at the Bayesian-Nash equilibrium can be in the form of either a symmetric randomized response (SR) strategy or an informative non-disclosive (ND) strategy. Specifically, a generalized majority voting rule is applied by each user to her noisy group signals to determine which strategy to follow. Further, when a user plays the ND strategy, she reports privacy-preserving data completely based on her group signals, independent of her private signal, which indicates that her privacy cost is zero. We emphasize that the reported data when a user plays the ND strategy is still informative about the underlying state because it is based on her learned group signals. As a result, both the data collector and the users can benefit from social learning which drives down the privacy costs and helps to improve the state estimation at a given payment budget. We further derive   bounds on the minimum total payment required to achieve a given level of state estimation accuracy.

\end{abstract}

%
%



\maketitle

\section{Introduction}
In the era of data analytics, the benefits of personal data collection are pronounced. Gathering personal data, such as customer needs and product reviews, plays an increasingly critical role in a variety of applications, including marketing, scientific research, business and politics. However, recent controversies, such as the Facebook Cambridge Analytica scandal or the reveal of Uber's post-trip customer tracking practices, have given rise to major concerns about the risks of the collection of personal data. In the absence of privacy guarantees, users lose control over their personal data against possible threats once it is submitted. As a result, users can be unwilling to share their personal data, such as political opinions,  movie ratings, product reviews,which can relate to their characters and lifestyles, unless they are sufficiently rewarded and are ensured that the privacy of their shared data is protected  adequately. 

In this work, we study a market model in which users make strategic decisions to sell privacy-preserved versions of their private data to a data collector. Further, our analysis generalizes the existing market models for private data collection \cite{Wang16} by incorporating the ubiquitous social interactions among users  encountered in many settings in our everyday life. Specifically, we ask the question of what is the desired data reporting strategies (from an individual user perspective)  and payment mechanisms (from a data collector perspective), when users can \textit{learn} noisy versions of their friends' data through social interactions. Intuitively, social interactions among the users can help them to become better-informed, which in turn can impact their decision strategies by improving the quality of their data reporting. The focus of this study is to quantify the impact of the social learning on privacy-preserving data collection in this market model.

\subsection{Data Collection, Social Learning and Privacy Leakage}

Consider a real-world setting where an online platform (\textit{data collector}), such as \textit{IMDB}, \textit{Flixster} or \textit{Netflix}, aims to collect, in a cost-effective manner, audience reviews and ratings about a movie. Very likely the rating from an individual will be influenced by her friends, and diverse social relations among the individuals can introduce further complications. This kind of data collection can also be found in many other applications, including product ratings, political campaigns, smartphone applications or hotel and restaurant reviews. A key observation is that social interactions among users  are ubiquitous in many settings in our everyday life. 
Often times users are strategic and  are not bound to truthfully share their personal opinions with the data collector. Furthermore, they can even opt out from data collection and report nothing. Nevertheless, the data collector can utilize a payment mechanism to incentivize participation and reward the users who report informative data. The users are still not compelled to act truthfully and the data collector is not equipped with an instrument to directly authenticate their reported data. We note that  social learning has not been studied in the literature on the design of truthful mechanisms to elicit personal data from strategic users \cite{Chen2013,Wang16,Agarwal2017, Ghosh2014, Nissim2014, Waggoner2015}.

To the best of our knowledge, this paper is the first to investigate the impact of social learning on privacy-preserving data collection. As expected, social learning among users introduces correlation in the reported data and significantly complicate the design. We seek to answer the following key questions: When a user consents to publish her review, what is the best strategy for her to leverage her friends' noisy signals as opposed to her own personal signal in her reported data? Can the users benefit from social learning, and if yes what is the corresponding desired data reporting strategy? Can the data collector design incentive mechanisms to take advantage of social learning? Further, what payment mechanism enables the data collector to minimize the cost in the presence of social learning?

\begin{figure}
\centering
\includegraphics[scale=0.52]{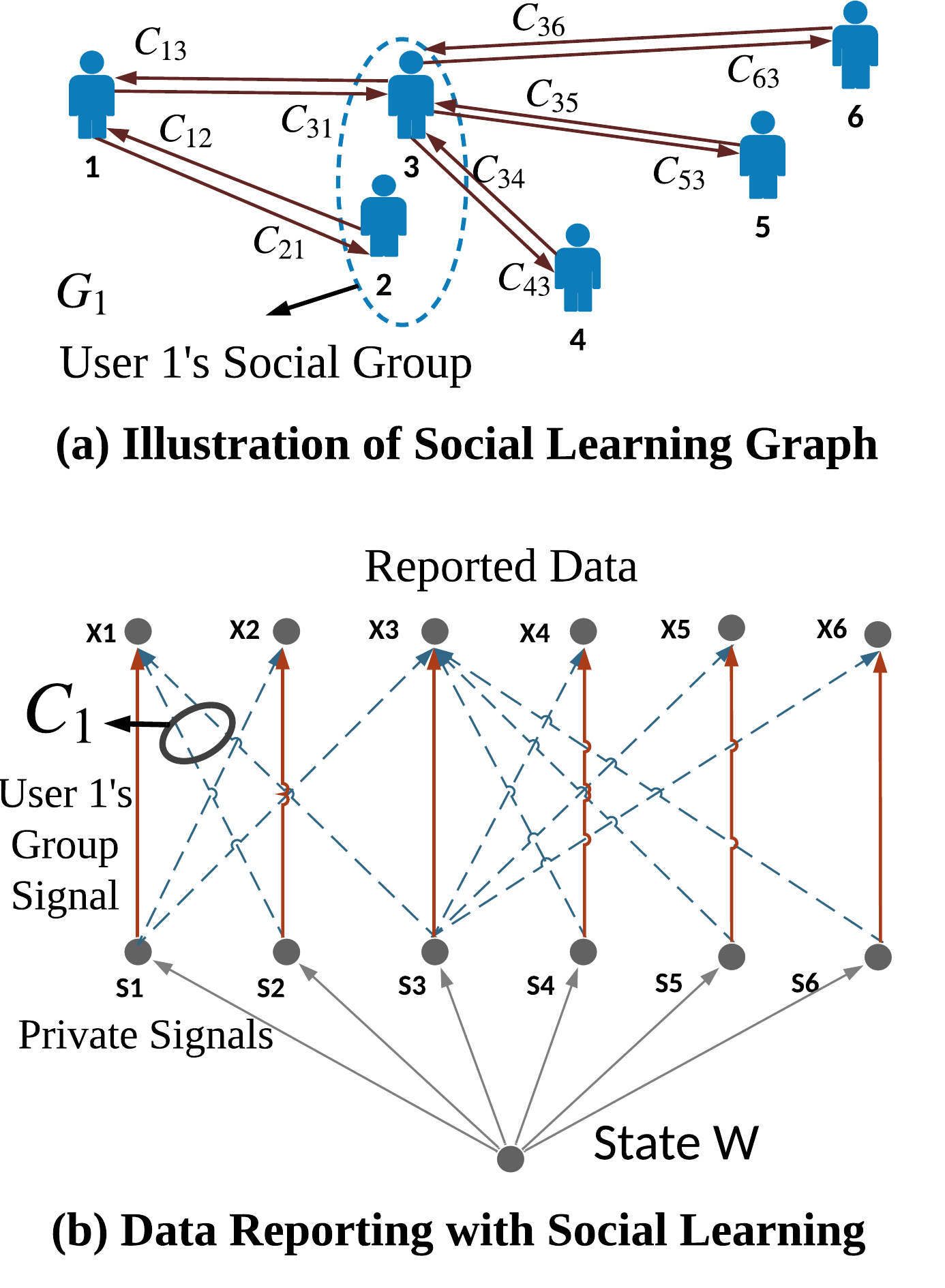}
\vspace{-0.3cm}
\caption{(a) If  two users $i$ and $j$ are friends, each of them gets noisy copies of the other's signals. Vector $C_i$ is called user $i$'s learned group signals and its components are noisy copies of her friends' private signals. (b) The data collector is interested in learning the state $W$, where $W$ is assumed to be a binary random variable. Conditioned on $W$, the private signals $S_1, S_2, \dots, S_N$ are independently and identically distributed, with user $i$'s private signal $S_i \in \{0,1\}$.  Taking $S_i$ and $C_i$ as inputs, user $i$ generates her reported data $X_i$.}
\label{Fig:InfoStructure}
\end{figure}

In this paper, the social learning graph is centered around the  context of learning  the underlying state $W$.  The social learning among privacy-aware users, which can take place in many forms, including in face-to-face meetings and over multiple online social media (e.g., Facebook and Twitter), is captured by a \textit{social learning graph} (or social graph for brevity).  Each vertex of this undirected social graph corresponds to a user and each edge of this social graph points to information exchange between two\footnote{The same  analysis can be carried out for directed graph models. In Section~\ref{Section:NumericalResults}, simulations are provided for both directed and undirected real world networks.}. It is assumed that the user population size is large, and the data collector knows only the degree distribution of the social graph. The data collector attempts to learn the underlying state $W$ based on data collection from the privacy-aware users, by using a payment mechanism to incentivize user participation. Based on her private signal $S_i$ and noisy copies of her friends' signals $C_{ij}$'s (when $i$ and $j$ are friends), each user $i$ reports data $X_i$, which may incur privacy leakage. As a result, each user can either choose to report data or  not to participate. Figure~\ref{Fig:InfoStructure} depicts an illustration of the information flow in this market model. 

\subsection{Challenges}
We develop a Bayesian game theoretic framework to study the impact of social learning on users' data reporting strategies and devise the payment mechanism for the data collector. We assume that the users are rational, risk-neutral and self-interested. In this study, we use ``type" to denote the private signal and learned group signals of each user $i$. Using a variant of the peer-prediction method\footnote{See \cite{FaltingsBook17} for a recent and extensive survey of this field.}, the data collector scores the reported data of user $i$ by comparing it to his estimate of the underlying state $W$, which is computed based on the reported data of all other users. The evaluation of this scoring function requires the computation of the estimation accuracy, which is highly nontrivial because it depends on all the collected data and hence on user types, which are correlated given the underlying state due to social learning. Further, in the existence of social learning, the \textit{quality} of the reports can vary across the users, because different users have different numbers of friends and each user is capable of claiming the control of her privacy level against the data collector. Indeed, the user heterogeneity poses significant challenges to characterize the optimal data reporting strategy and the desired payment mechanism. 

It is worth noting that the data reporting strategies developed under this market model, would result in negligible privacy leakage of friends' signals, as elaborated as follows.
\begin{itemize}
\item As it will be discussed in Section 4, the best response data reporting strategies under the desired payment mechanism are based on the majority voting in which each user locally estimates the underlying state from the sum of the group signal. Given $i$ and $j$ are friends, it can happen only in the worst case scenario for user $i$ that an attacker obtains user $i$'s (noisy) signal by observing $j$'s action: The attacker knows every signal $j$ receives from her friends except $i$ and there is a tie between the number of 1s and 0s in this set. 
\item Any attack attempt which targets to learn a user's private signal from her friends' reported data would require the exact knowledge of social learning graph, which we assume not available. Furthermore, users learn only noisy versions of their friends' signals, where the noise is used to model privacy protection against social learning.
\item In the model under consideration, each user receives $d_i$ signals from her friends and reports either `1' or `0'. Consequently, this implies that the privacy leakage through the reported data, which is a mapping from  the $\{0,1\}^{d_i+1}$ space  to $\{0,1\}$, is minimal in general.

\end{itemize}
Alternatively, one can develop utility functions accounting for the privacy loss of friends' signals.
We will study this in future work.

\subsection{Relevant Work}

Market models where privacy-aware strategic users treat their data as a commodity have recently received much interest \cite{Wang16,Chen2013,Fleischer2012,Ghosh2011,Ghosh2014,Nissim2014,Waggoner2015}. In all these studies, the users are regarded as  individual agents but social learning among them is not accounted for. The market model proposed in \cite{Wang16} can be differentiated in this stream of work where each user directly controls the privacy level of her reported data. Our proposed model can be regarded as a generalization of the model in \cite{Wang16} which assumes that the knowledge of each user is limited to its private signal. User heterogeneity in peer prediction has recently gained attention, but there are very few results on handling its complication \cite{Agarwal2017,Shnayder2016MultiTask}. Further, little attention has been paid to the cases where the reported data is correlated across users given the true state. We shall study both these two issues in the market model in this paper.

As illustrated in Figure~\ref{Fig:InfoStructure}, the users have richer information about the underlying state  beyond their  private signals, thanks to the presence of social learning, and can therefore use this additional information to conceive their reporting strategy which can potentially have significant impact on the data collection. In fact, each user becomes more knowledgeable through social learning and can act as a local data curator. Furthermore, we revisit the notion of \textit{informative strategy} in the framework of data privacy games where the users report their data using randomized response strategies to achieve privacy protection. We show that, in the presence of social learning, this conventional notion of ``informative strategy" would not encompass some desired equilibria where each user reports informative data based on her friends' signals only. Building on this new insight, we introduce informative non-disclosive strategies which allows a user to formulate strategies based on only her learned group signals if there is strong concurrence. This is one of main findings in this study.

We caution that social learning does not create situations akin to herding \cite{Banerjee1992} or information cascades \cite{Bikchandani1992}. In our market model, the users take their actions in parallel, not sequentially, and the reported data in this study, are only revealed to the data collector not to the other participants. Further, in the model, friends do not collude or collaborate when formulating their data reporting strategies based on their private signals and group signals. 

\begin{figure}
	\centering
	\includegraphics[width=0.98\linewidth]{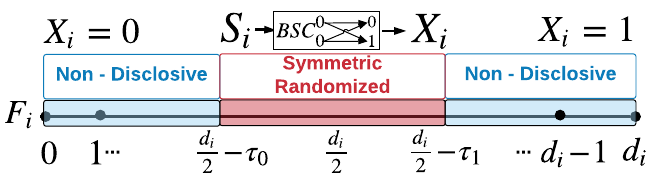}
	\vspace{-0.3cm}
	\caption[Two Symmetric Randomization Regimes]{Strategy profile at Bayesian Nash Equilibria: The data reporting strategy at BNE is in the form of either a non-disclosive strategy or a symmetric randomization strategy. Using $F_i$,  defined as the sum of the social group signals, each user (with degree $d_i$) determines which strategy to follow.}
	\label{Fig:Regimes}
\end{figure}

\subsection{Summary of Main Results}

Our findings reveal that, in general, the data reporting strategies at the Bayesian-Nash equilibria (BNE) can be in the form of either symmetric randomized response (SR) strategy or non-disclosive (ND) strategy (which are formally defined in Section 3), and vary from user to user. Intuitively, when a user plays the ND strategy, her reported data is completely based on her learned group signals, independent of her private signal.  More specifically, under the condition that the social learning graph is sparse\footnote{The average number of friends users have is much smaller than the total number of users.}, we show that the ND strategy is a desired data reporting strategy at Bayesian Nash equilibria,  particularly when her group signals are more reliable, i.e., the signals learned from her friends are less noisy. Intuitively, albeit having a signal different from  the majority of her friends' signals, a user might be better off to pretend in accordance with them. Further, the users' reported data along this line is still informative for the data collector. Our main contributions can be summarized as follows.
\begin{itemize}
\item  We formalize \textit{non-disclosive} (ND) strategies where the reported data by a user   completely depends on her noisy social group signals, independent of her private signal. There are many cases the informative ND strategy is preferred over the symmetric randomized response strategy. Further, both the data collector and the users can benefit from social learning which drives down the privacy costs and helps to improve the state estimation. The underlying rationale is as follows: Thanks to social learning, each individual user has richer information about the underlying state and hence  has more options for data reporting. In particular, some users' privacy cost can be driven to zero when ND strategies are used. This, in turn, benefits the data collector and drives down the overall cost, since his data resources are more informed and can report informative data at lower privacy costs.

\item The data reporting strategy at the BNE is in the form of either an ND strategy or an SR strategy. As illustrated in Figure ~\ref{Fig:Regimes}, a  \textit{majority voting} based data reporting rule is applied by each user to her learned group signals to determine which strategy to follow. Since each user has different social ties, the BNE strategy varies from user to user. In the special case with noiseless group signals, the user employs the ND strategy in which each user reports the majority of her friends' signals, for data reporting, unless there is a tie. In general, the group signals are noisy, and hence the ND strategy requires a 'higher' majority than that in the noiseless case, and otherwise the user would follow the SR strategy. The regime change hinges heavily on the noise levels of her group signals and the privacy requirements.

\item The objective of the data collector is to estimate the underlying state $W$ from the users' reported data. To tackle the technical difficulty that the reported data is correlated across users given the underlying state, we use a Central Limit Theorem for dependence graphs to characterize the statistics of the reported data profile, based on which the data collector can evaluate the estimation error of $W$. The total expected payment is then characterized for a given accuracy target. Our analysis pinpoints to the positive impact of social learning on the privacy-preserving data collection game, in the sense that the data collector can lower the total payment significantly (compared to the case with no learning) and the users incur less privacy costs, thanks to social learning.
\end{itemize}
The rest of the paper is organized as follows. We introduce the models for the privacy-preserving data collection market and the social learning graph in Section~2. We formalize the Bayesian game under this market model in Section~3. We present the main results on the data reporting strategies and the payment mechanism design in Section~4. We discuss the impact of social learning on the payment and accuracy in Section~5. Finally, we summarize and discuss possible extensions and open problems in Section~6.  

\section{System Model} \label{SubInfoStr}

\subsection{Private Signal and Group Signals} \label{Sub:SignalsModel}
\begin{figure}
\centering
\includegraphics[scale=0.40]{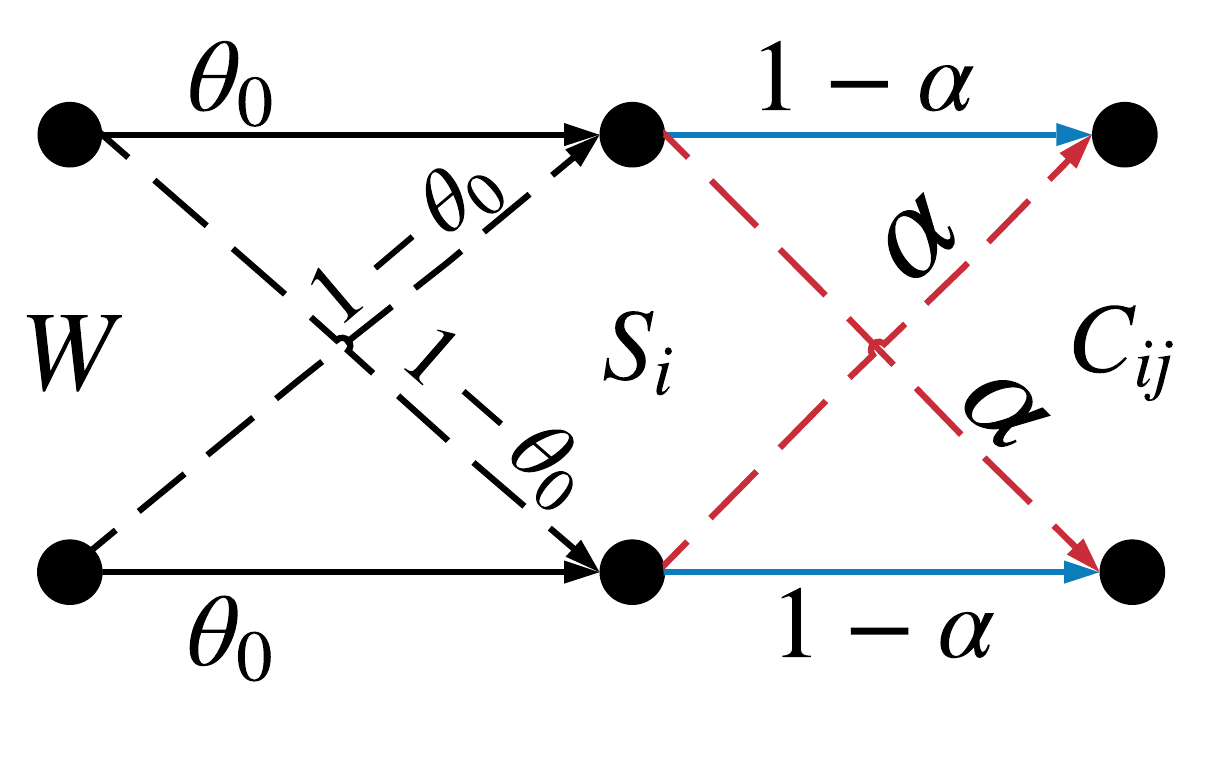}
\vspace{-0.5cm}
\caption{ The parameter $\theta_1$, defined as $\theta_1\teq \theta_0 (1\teg\alpha) + (1\teg\theta_0)\alpha$, measures the quality of the group signal.}
\label{Probs}
\end{figure}

Consider a market model where the data collector is interested in learning the underlying state $W$ from a set $\mathcal{I} = \{1, 2, \dots, N \}$ of $N \geq 2$ users. For ease of exposition, $W$ is assumed to be a binary random variable, for example, representing the product quality as \textit{good} or \textit{bad} \footnote{Binary feedback and review systems are prevalent, e.g. Netflix recently decided to swap out five star rating system for a binary system \cite{atlantic:Netflix}. On many platforms, it is observed that the vast majority of ratings are either the best or the worst option \cite{fradkin2018,Hu2006, Zhou2008}.}. We assume that both $\Pr_W(1)$ and $\Pr_W(0)$ are positive and $\Pr_W (\cdot)$ is common knowledge. As illustrated in Figure~\ref{Fig:InfoStructure}, each user $i$ possesses a binary signal $S_i$, which is her personal data, representing her knowledge about $W$. The private signal profile of the entire population is denoted as $\mathbf{S} = [S_1 \ S_2 \ \cdots \ S_N]$. Given $W$, it is assumed that the binary signals $S_i$'s are independent and identically distributed. Note that the parameter $\theta_0$ with $0.5 < \theta_0 < 1$ determines the \textit{quality of the private signals} for every user:
\begin{subequations} \label{Eq:Theta0}
\begin{align}
\Pr (S_i = 1 | W = 1) = \theta_0, \ \Pr (S_i = 0 | W = 1) = 1- \theta_0, \\
\Pr (S_i = 0 | W = 0) = \theta_0, \ \Pr (S_i = 1 | W = 0) = 1- \theta_0.
\end{align}
\end{subequations}

The \emph{social learning graph} $\mathcal{G} = \{ \mathcal{I}, \mathcal{E} \}$ is used to model t he social coupling among the users. The vertex set is the set of individuals $\mathcal{I}$ and the edge set is given as $\mathcal{E} = \{(i,j) \in \mathcal{I} \times \mathcal{I}: e_{ij} = 1\}$ where $e_{ij} = 1$ if and only if there is a social tie between $i$ and $j$ where $i \neq j$. (Similar studies can be carried out for the direct graph model.) User $i$'s social group $G_i$ is defined as the set of her friends: $G_i = \{j \in \mathcal{I}: e_{i j} = 1 \}$. The number of friends $i$ has is called the degree $D_i$ of that user. We assume that the social graph is a random graph\footnote{We follow the configuration model described in \cite{Newman2003}. The degrees $\{d_i\}_{i= 1}^N$ are independent and identically distributed random integers drawn from $\rho_d$. Pairs of users are chosen at random and edges are formed between them until complete pairing according to the drawn degree sequence. If complete pairing is not possible, one $d_i$ can always be discarded and redrawn from $\rho_d$.} with node degrees following a distribution $\rho_d $ with maximal degree $D_{\mathrm{max}}$ \cite{Newman2003}, and that 
the social graph is sparse satisfying the following conditions \cite{watts99}.
\begin{assumption} \label{Asm:Sparse} 
Maximal degree $D_{\mathrm{max}} = o(N^{\nicefrac{1}{4}})$ and $\E[D^{2+ \triangle}] \! < \! \infty$ for some $\triangle > 0 $.
\end{assumption}
The degree distribution $\rho_d$ is common knowledge for the data collector and the users. However, the data collector does not have any further knowledge about the social graph $\mathcal{G}$. The users know who their friends are, but they do not possess any knowledge about their friends' social groups, i.e., they do not know how many friends of their friends have and who they are. 

Each user $i$ has noisy copies of her friends' personal signals. For user $i$ with social group  $G_i$, let vector $C_i$ denote  her \textit{group signals}: $C_i = [C_{ij_1} C_{ij_2} \dots C_{ij_{d_i}}]$, where $C_{ij}$'s are binary valued. To capture the noise in social learning of group signals, it is assumed that friends' personal signals are ``flipped'' with  \textit{crossover} probability $\alpha$: $P (C_{ij} \teq 1 | S_{j} \teq 0) = P (C_{ij} \teq 0 | S_{j} \teq 1) = \alpha, 0 \leq \alpha < 0.5$. Consequently, the parameter $\theta_1= \theta_0 (1\teg\alpha) + (1\teg\theta_0)\alpha$ points to the \textit{quality of group signals}:
\begin{subequations} \label{Eq:Theta1}
\begin{align}
\Pr (C_{ij} = 1 | W = 1) & = \theta_1, \ \Pr (C_{ij} = 0 | W = 1) = 1- \theta_1, \\
\Pr (C_{ij} = 0 | W = 0) & = \theta_1, \ \Pr (C_{ij} = 1 | W = 0) = 1- \theta_1.
\end{align}
\end{subequations}
Note that these ``flips" are statistically independent with a probability of $\alpha$, i.e. given $j_1$ and $j_2$ are friends with $i$, $\Pr(C_{j_1 i} = s_i| C_{j_2 i} = c_{j_2 i}, S_i = s_i) = \Pr(C_{j_1 i} = c_{j_1 i}| S_i = s_i) = 1-\alpha$.

\subsection{Users' Data Reporting Strategies and Data Collector's Payment Mechanism}\label{SubSec:22}
\begin{figure}[t]
\centering
\includegraphics[scale=0.17]{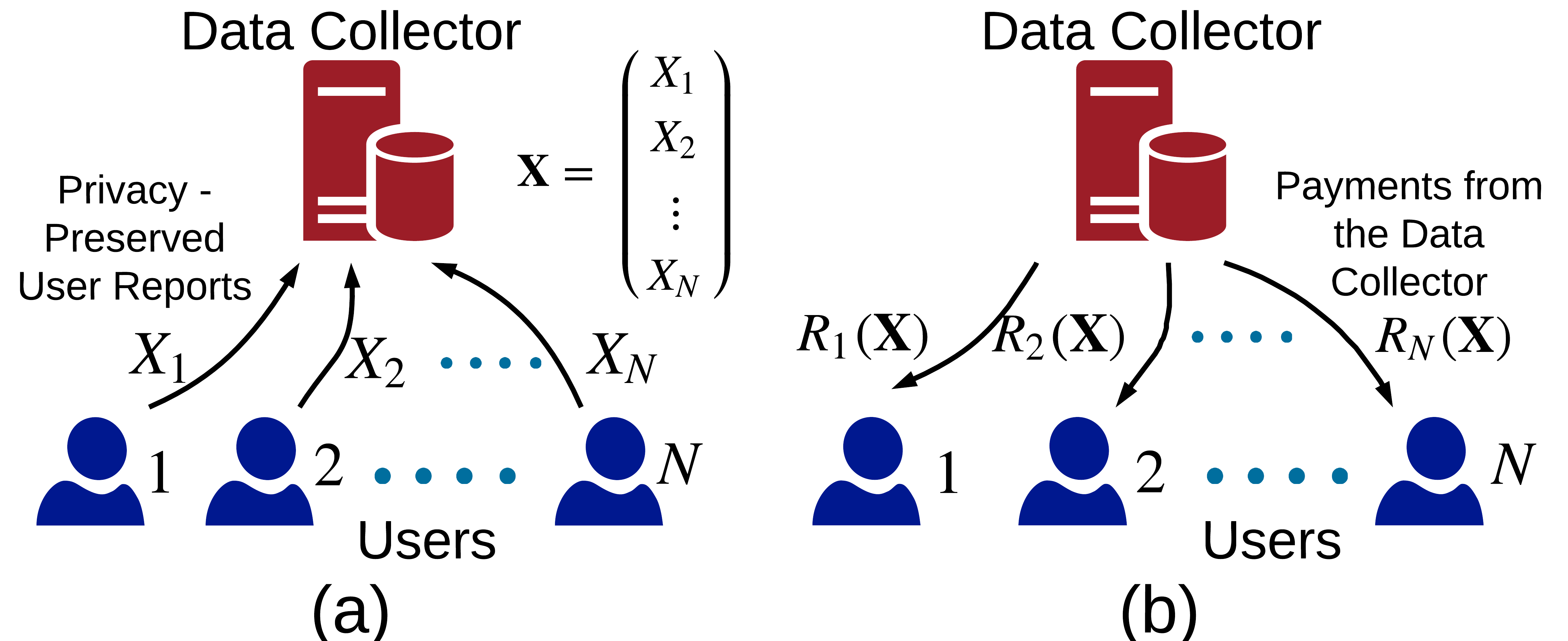}
\caption{A Market Model:  a) Each user plays a Bayesian Game to report her privacy-preserved data to  data collector; b) Data collector determines the amount of reward for each user  based on the entire reported data profile $\mathbf{X} = [X_1 \dots X_N]$.}
\label{Fig:marketFigure}
\end{figure} 
Following the convention in the Bayesian games \cite{Harsanyi2004}, we define \textit{type} $t_i$ as $t_i = [s_i \ c_i]$. Respectively, the type space $\mathcal{T}$ can be defined as $\mathcal{T} = \cup_{k \teq 0}^{D_{\mathrm{max}}} \mathcal{T}_k$ where $\mathcal{T}_k \teq \{0,1 \}^{k\tsum1}$.   The type profile $\mathbf{t}$ is defined as $\mathbf{t} = \{ t_1, t_2, \dots, t_N \}$. Each user $i$ knows her own type vector $t_i$ but lacks the knowledge of other users' type vectors, defined as $\mathbf{t}_{-i}$\footnote{For a given type profile $\mathbf{t}$, define $\mathbf{t}_A = [t_{a_1} \dots t_{a_K}]$ where $A = \{a_1, \dots, a_K \} \subset \mathcal{I}$. The set of all users other than  user $i$ is denoted by "$-i$".}.

The reported data of user $i$ is denoted with $X_i$. It follows that $\mathbf{X} = [ X_1 \ X_2 \ \dots \ X_N ]$ is the reported data profile, where $X_i \in \mathcal{X} = \{0,1,\bot \}$ and $\bot$ represents ``non-participation". Let $\sigma_i(t_i)  \in \triangle (\mathcal{X})$ denote the action user $i$ chooses when her type is $t_i$, where $\triangle(\mathcal{X})$ being the set of all probability distributions over $\mathcal{X}$. The strategy of user $i$,  defined as $\sigma_i = \{\sigma_i(t_i): t_i \in \mathcal{T} \}$, specifies the probabilities $\Pr_{\sigma_i}(X_i \in \mathcal{F}|T_i \teq t_i)$ for all $\mathcal{F} \subseteq \mathcal{X}$. The strategy profile $\sigma$ is defined as $\sigma = \{\sigma_1, \sigma_2, \dots, \sigma_N\}$. A strategy profile $\sigma$ is called \textit{symmetric} if we have:
\begin{equation} \label{Eq:SymProf}
    \sigma_1(t) = \sigma_2(t) = \dots = \sigma_N(t), \quad \forall t \in \mathcal{T}.
\end{equation}
For a given type profile $\mathbf{t}= \{ t_1, t_2, \dots, t_N \}$, $\sigma(\mathbf{t})$ denotes the collection $\sigma(\mathbf{t}) = \{\sigma_1(t_1), \sigma_2(t_2), \dots, \sigma_N(t_N) \}$. 

When $i$ and $j$ are friends, $C_{ij}$, a noisy version of $j$'s personal signal, is a component of $T_i$ (and $C_{ji}$ is a component of $T_j$). Further, if $i$ and $j$ are not friends but they have a common friend $\ell$, then both $i$ and $j$ have noisy copies of $S_\ell$ in their type vectors. To sum up,  $X_i$ and $X_j$ can be correlated given $W$ if user $i$ and $j$ are friends or they have a common friend, because the user strategies depend on the user types. If user $i$ and $j$ are not friends or they do not have a common friend, then $X_i$ and $X_j$ are conditionally independent given the underlying state $W$.

The primary goal of the data collector is to estimate the underlying state $W$ from users' reported data $X_1, X_2, \dots, X_N$. Intuitively, for the data collector's perspective, a desired data reporting strategy produces reported data carrying information about the underlying state $W$.  It is assumed that the data collector cannot impose penalties on the users, in the sense that positive rewards are the only options at his disposal to incentivize informative reporting in the presence of privacy costs as depicted on Fig.~\ref{Fig:marketFigure}. Define the non-negative payment mechanism as $\mathbf{R}: \mathcal{X}^N \to \mathbb{R}^N$,  where $R_i(\mathbf{x})$ specifies the amount of payment for user $i$ given $\mathbf{X} = \mathbf{x}$ and $\mathbf{R}(\mathbf{x}) = [R_1(\mathbf{x}) \ R_2(\mathbf{x}) \ \dots \ R_N(\mathbf{x})]$. Further, the utility of each user is considered as the difference between her expected reward and privacy cost.
\subsection{Data Privacy Model}
Needless to say, the users are subjected to \textit{privacy cost} $g(\sigma_i)$ when they carry out data reporting. Based on the celebrated notion of differential privacy \cite{Dwork2006}, we define the privacy loss inflicted on the users as the level of local differential privacy when using the strategy $\sigma_i$. Roughly speaking, given her group signal $c_i$, user $i$'s privacy cost decreases as her data reporting makes her personal signal $s_i$ more \textit{indistinguishable}.
\begin{definition} \label{priDef} [\textbf{Privacy Level}]
The privacy level of strategy $\sigma_i$, for a given $C_i \teq c_i$, is defined as 
\begin{equation}
\zeta_i(\sigma_i,c_i) \teq \max_{\substack{\mathcal{F} \subseteq \{0, 1, \bot \}, \\ s_i \in \{0,1 \}}}  \ln \Bigg (  \frac{\Pr_{\sigma_i}(X_i \in \mathcal{F} \ | \ C_i \teq c_i, S_i \teq s_i)}{\Pr_{\sigma_i}(X_i \in \mathcal{F} \ | \ C_i \teq c_i, S_i \teq 1 \! - \! s_i)} \Bigg ), \label{priDefEq}
\end{equation}
\end{definition}
\noindent where the convention $0/0 = 1$ is followed. 

To get a more concrete sense, consider an extreme case where the user's reported data is her personal signal, $X_i \teq S_i$ given $C_i \teq c_i$. In this case, the privacy level $\zeta(\sigma_i,c_i) $ is equal to $\infty$, the maximum possible privacy leakage for her. On the contrary, $\zeta(\sigma_i,c_i)$ is equal to 0 when the reported data $X_i$ is independent from the personal signal $S_i$ given $C_i \teq c_i$. This strategy is referred as a \textit{non-disclosive} strategy (including non-informative strategy as a subclass) which incurs the minimum possible privacy leakage for this specific user.

\noindent We next introduce the \textit{privacy cost function} $g(\zeta_i(\sigma_i,c_i))$ and the user utility function. In this work, it is assumed that $g$ is homogeneous across the users and satisfies the following conditions: It is convex, continuously differentiable, increasing, nonnegative and $g(0) = 0$.

\section{Bayesian Game Formulation}
\begin{figure}
\centering
\includegraphics[width=0.95\linewidth]{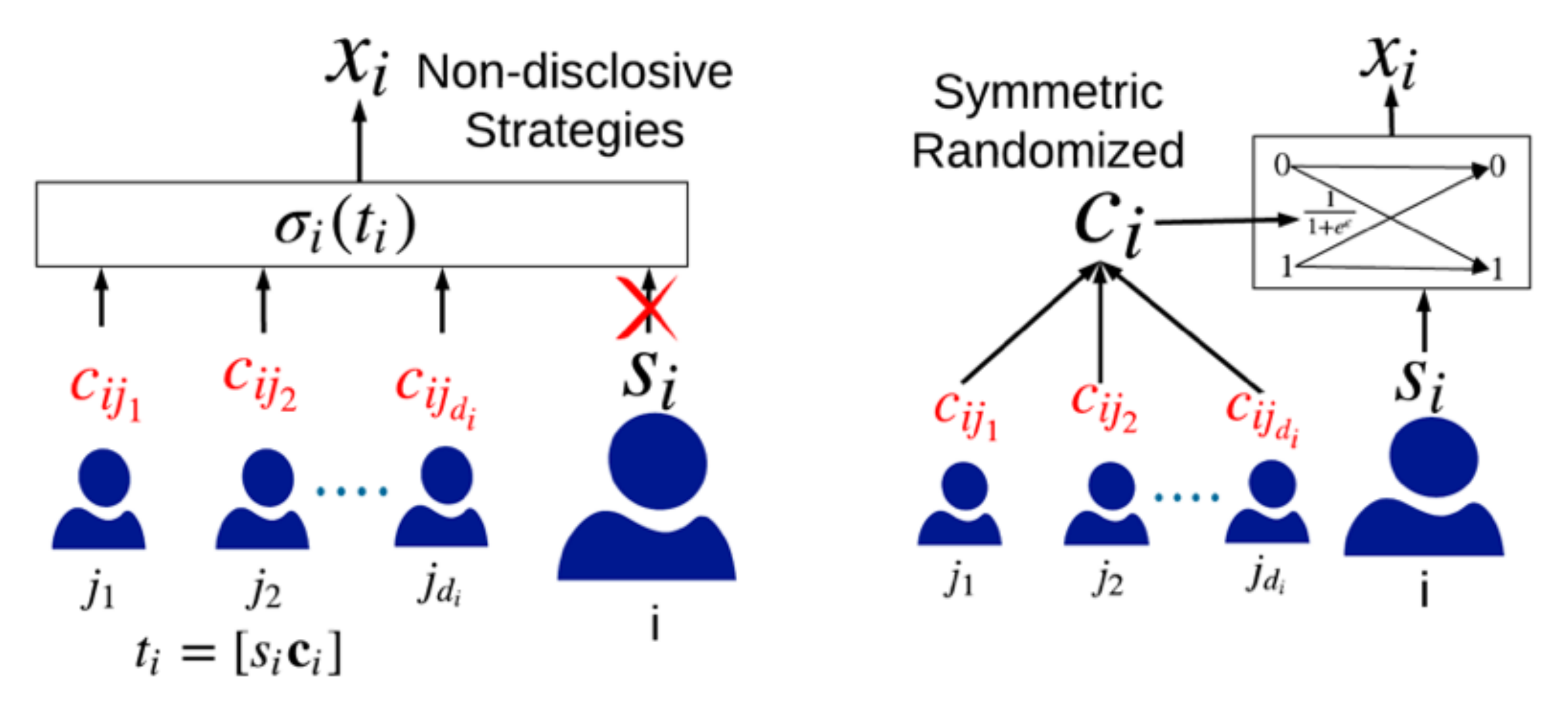}
\vspace{-0.5cm}
\caption[Two Symmetric Randomization Regimes]{Strategy profiles at BNE. (a) Non-Disclosive (ND) Strategies: $X_i$ is independent from the private signal $S_i$. In general, $X_i$ is dependent on the group signal $C_i$. (b) Symmetric Randomized Response (SR) Strategies: $S_i$ and $X_i$ are the input and the output of a noisy binary symmetric channel in which the noise level corresponds to the privacy level of the user. In general, the privacy level depends on the group signals $C_i$.}
\label{Fig:EqStrategies}
\end{figure}
The Bayesian Game under this market model is outlined as follows: The data collector announces a payment mechanism, which actuates a strategic form game where the users are the players aiming to maximize their expected utility, which is the difference between their rewards and their privacy costs. In this game, the common knowledge includes the prior state distribution $\Pr_W$, the signal quality parameter $\theta_0$, the crossover probability $\alpha$, the degree distribution $\rho_d$, the privacy cost function $g$ and the payment mechanism $\mathbf{R}$. In this game with incomplete information, we focus on Bayesian-Nash equilibria where each user has no incentive to unilaterally change her strategy given other users' strategies. Formally, a Bayesian-Nash equilibrium (BNE) is defined in the following\footnote{$(\sigma_i', \mathbf{\sigma}_{-i})$ denotes  the strategy profile $\{ \sigma_1(t_1), \dots, \sigma_{i-1}(t_{i-1}), \sigma_i'(t_i), \sigma_{i+1}(t_{i+1}),$ $\dots, \sigma_N(t_N))\}$ where $i$ plays $\sigma_i'(t_i)$ and the other user play $\sigma_{-i}(\mathbf{t}_{-i})$.}.
\begin{definition}
[\textbf{BNE}]
A strategy profile ${\sigma} = [\sigma_1 \ \sigma_2 \dots \ \sigma_N]$ is a Bayesian-Nash equilibrium (BNE) if, for each user $i \in \mathcal{I}$,
\begin{align}
& \sigma_i \in \argmax_{\sigma_i' \in \triangle (\mathcal{X})} \E_{(\sigma_i',\sigma_{-i})} [R_i(\mathbf{X}) -  g(\zeta(\sigma'_i,C_i))]. \label{Eq:BNEDef}
\end{align}
\end{definition}
We show that users' data reporting strategies at the Bayesian-Nash equilibria are in the form of either \textit{symmetric randomized responses} or \textit{non-disclosive strategies}. Firstly, we formally define the non-disclosive strategies as follows.
\begin{definition} [\textbf{ND Strategy}]
$\sigma_i([\cdot \ c_i])$ is called a non-disclosive (ND) strategy, if $S_i$ and $X_i$ are independent given $C_i$; that is to say, for every $\mathcal{F} \subseteq \{0,1,\bot\}$ and $s_i \in \{0,1\}$ we have 
\begin{equation*}
\Pr_{\sigma_i} (X_i \in \mathcal{F} | S_i \teq s_i, C_i \teq c_i) \teq \Pr_{\sigma_i} (X_i \in \mathcal{F} | C_i \teq c_i).
\end{equation*}
\end{definition}
When an ND strategy is played by user $i$, the reported data $X_i$ is independent from her private signal $S_i$. That is to say, this strategy does not disclose any private information and essentially sets her privacy cost to 0, i.e. $\zeta_i(\sigma_i,c_i) = 0$. Notice that, in general, her reported data still depends on her group signals $C_i$ and is correlated with the underlying state $W$. Next we define the symmetric randomized response, for $\xi(c_i): \{0,1\}^{d_i} \to \mathbb{R}$.
\begin{definition} [\textbf{SR Strategy}] \label{Def:SRStrategies}
$\sigma_i([\cdot \ c_i])$ is called a symmetric randomized response (SR) strategy, if it satisfies the following:
\begin{align*}
& \Pr_{\sigma_i} (X_i \teq 1 | S_i \teq 1, C_i \teq c_i)  \teq \Pr_{\sigma_i} (X_i \teq 0 | S_i \teq 0, C_i \teq c_i) \teq \frac{e^{\xi(c_i)}}{1 \tsum e^{\xi(c_i)}} \\
& \Pr_{\sigma_i} (X_i \teq 0 | S_i \teq 1, C_i \teq c_i)  \teq \Pr_{\sigma_i} (X_i \teq 1 | S_i \teq 0, C_i \teq c_i) \teq \frac{1}{1 \tsum e^{\xi(c_i)}} \\
 & \Pr_{\sigma_i} (X_i \teq \bot | S_i \teq 1, C_i \teq c_i) = \Pr_{\sigma_i} (X_i \teq \bot | S_i \teq 0, C_i \teq c_i) = 0.
\end{align*}
\end{definition}

When an SR strategy is played by user $i$, the privacy level of the strategy, is determined by the injected noise level on the personal signal $S_i$, $\zeta_i(\sigma_i,c_i) = |\xi(c_i)|$. Note that we are interested in the regime $\xi(c_i) > 0$, for the reported data to be useful for the hypothesis testing problem. In general, when an SR strategy is played, both the private signal $S_i$ and the group signal $C_i$ are correlated with the reported data $X_i$. We have the following lemma characterizing the desired strategies of Bayesian-Nash equilibria (the proof can be found in Appendix~\ref{Appendix:ProofNDSR}).
\begin{lemma}
\label{Lemma:SRND}
Under Assumption~\ref{Asm:Sparse}, for any non-negative payment mechanism, a user's data reporting strategy $\sigma_i([\cdot \ c_i])$ in a Bayesian Nash equilibrium is either a symmetric randomized response strategy or a non-disclosive strategy.
\end{lemma}

Note that if a user plays with a strategy where her reported data $X_i$ is independent from both her private data $S_i$ and her group signal $C_i$, then $X_i$ is pure noise. As a result, it is \textit{uninformative} for the data collector, and it is a degenerated form of the ND strategy.
We remark that Lemma~\ref{Lemma:SRND} is a generalization of Lemma~1 in \cite{Wang16}.  When a user $i$ does not have any friends ($D_i \teq 0$), ND strategies reduce to be uninformative. On the other hand, in the presence of social learning ($D_i > 0$), ND strategies can be informative and positively contribute to the data collector's information elicitation.




\section{Data Collector's Payment Mechanism} \label{Section:PaymentMech}
The primary objective of the data collector is to estimate the underlying state $W$ from users' reported data $X_1, X_2, \dots, X_N$ with the minimum error probability. The binary hypothesis testing problem can be stated as 
\begin{equation} \label{Eq:HTProblem}
        \mathcal{H}_0 : W = 0, \quad
        \mathcal{H}_1 : W = 1.
\end{equation}
 Clearly, the estimation of  $W$ from users' reported data $\mathbf{X}$ is viable only if there exists BNE strategies in which $\mathbf{X}$ is informative about $W$. According to Lemma~\ref{Lemma:SRND}, at the BNE, data reporting strategies are either in the form of symmetric randomized responses (SR) or non-disclosive (ND) strategies. 
 In the presence of social learning, in addition to her private signal, each user obtains noisy group signals through social interactions. It is plausible to view each user as a \textit{local decision maker} who processes the data available to her and reports it to the data collector who acts as a \textit{fusion center}. With this insight, we generalize the  notion of informative strategies in the peer-prediction literature to include the ND strategies, and then study the payment mechanism design  for the data reporting strategies where each user chooses between  SR or ND strategies based on their group signals $C_i$.

One main focus of this paper is on designing the payment mechanism $\mathbf{R}$ under which each user can form her data reporting strategies, and the data collector can accurately estimate the underlying state $W$ based on the reported data with minimum error probability. In the presence of social learning, the optimal design hinges heavily upon the user types which are  correlated across users for a given $W$ and involves combinatorial optimization, and hence  is very challenging, if not impossible, to attain for the general case with finite $N$.
To tackle this challenge, we  will tackle the problem first at the local level by considering a hypothetical genie-aided payment mechanism and then study the optimal data reporting strategy  in the asymptotic regime of $N$ (for the sake of tractability). Finally, we present in Sections~\ref{Subsection:PaymentMechDesign} and~\ref{Subsection:NoisyGroup} the desired payment mechanism using peer prediction and the BNE strategies accordingly.
\subsection{Users as Local Data Curators}
We first consider a hypothetical scenario where the data collector has access to the underlying state $W$ and can employ $W$ in a \textit{genie-aided} payment mechanism, $\mathbf{R}^{\mathrm{g}}: \mathcal{X}^N \cross \{0,1 \} \to \mathbb{R}^N$, such that
\begin{equation*}
R^{\mathrm{g}}_i \left((x_i, \mathbf{x}_{-i}),w \right) = R^\mathrm{g}(x_i,w).
\end{equation*}
Observe that in this scenario the payments  users receive do not depend on   reported data of the other users. Furthermore, it is a symmetric payment mechanism, i.e., $R_i(x_i,w) = R_j(x_j,w)$ if $x_i \teq x_j$. In this genie-aided scenario  the mechanism design problem is greatly simplified to a decentralized detection problem in which each user acts as a local decision maker aiming to minimize the probability of error given her type $T_i$. The conditional expected payment of user $i$ at strategy $\sigma_i$ is given by
\begin{equation*}
\E_{\sigma_i}\!\!\left[R^\mathrm{g}_i(X_i, W) | T_i\! \teq t\right]\! \teq \! \sum_{w,x}\! R^\mathrm{g}(x,\!w) \Pr(W\teq w|T_i \teq t) \Pr_{\sigma_i}(X_i \teq x|T_i\teq t).
\end{equation*}
Without loss of generality, assume that $R^\mathrm{g}(0,0)\geq R^\mathrm{g}(1,0)$ and $R^\mathrm{g}(1,1)\geq R^\mathrm{g}(0,1)$ , the payment can be maximized if $X_i$ is selected according to the following decision rule:
\begin{equation*}
\Lambda_i(t) = \frac{\Pr(T_i = t|W = 1)}{\Pr(T_i = t|W = 0)} \mathop{\gtreqless}_{X_i = 0}^{X_i = 1} \frac{\Pr_W(0)}{\Pr_W(1)} \frac{R^\mathrm{g}(0,0)- R^\mathrm{g}(1,0)}{R^\mathrm{g}(1,1)- R^\mathrm{g}(0,1)}
\end{equation*}
where $\Lambda_i(t)$ is the likelihood ratio for a given type $T_i$.  Let $R^\mathrm{g}(1,0)=R^\mathrm{g}(0,1)=0$. Further,  setting the right-hand side (RHS) of the above inequality to $1$ renders the genie-aided payment mechanism as follows:
\begin{equation} \label{Eq:GeniePayMech}
R^\mathrm{g}(x,w) = \begin{cases}
Z^\mathrm{g}/\Pr_W(0) &\mbox{if } x = w = 0,\\
Z^\mathrm{g}/\Pr_W(1) &\mbox{if } x = w = 1, \\
0 &\mbox{otherwise}
\end{cases} 
\end{equation}
where $Z^\mathrm{g}>0$ is a design parameter. Under this payment mechanism, user $i$ employs the maximum-likelihood (ML) decision rule to maximize her conditional expected payment:
\begin{equation*}
  \hat{W}^{\mathrm{ML}}_i(s_i,c_i) = \argmax_{w\in \{0,1\}} \Pr(S_i\teq s_i, C_i \teq c_i|W\teq w),
\end{equation*}
thereby removing the need of using the prior distribution of $P_W(w)$. The ML rule can be rewritten as 
\begin{equation} \label{Eq:UserMLEst}
    \sum_{j \in G_i} c_{ij} \mathop{\gtreqless}_{\hat{W}^{\mathrm{ML}}_i (s_i, c_i) = 0}^{\hat{W}_i^{\mathrm{ML}}(s_i,c_i) = 1} \frac{d_i}{2} - (2s_i -1) \bar{A}
\end{equation}
with $\bar{A}$ given by
\begin{equation} \label{Eq:barA}
\bar{A} = \frac{1}{2} \frac{\log \nicefrac{\theta_0}{(1-\theta_0)}}{\log \nicefrac{\theta_1}{(1-\theta_1)}}.    
\end{equation}
For convenience, denote the sum of the group signals by random variable $F_i$ and its realization by $f_i$:
\begin{equation*}
F_i = \sum_{j \in G_i} C_{ij} \quad \mathrm{and} \quad f_i = \sum_{j \in G_i} c_{ij}.
\end{equation*}
 It follows from \eqref{Eq:UserMLEst} that
\begin{equation}\label{Eq:MLRule}
\hat{W}_i^{\mathrm{ML}}(s_i, f_i) = \begin{cases}
1, & \text{if } f_i > \nicefrac{d_i}{2} + \bar{A}, \\
0, & \text{if } f_i < \nicefrac{d_i}{2} - \bar{A}, \\
s_i, & \text{otherwise. }
\end{cases}
\end{equation}
It can be seen from \eqref{Eq:MLRule} that when $f_i > \nicefrac{D_i}{2}\tsum \bar{A}$ or  $f_i < \nicefrac{D_i}{2}\teg \bar{A}$, the ML rule reduces to reporting the \textit{majority} bit of the group signal and does not involve the private signal $S_i$, thereby incurring  no privacy cost for user $i$.   On the other hand, 
for the case $f_i \in [\nicefrac{d_i}{2}-\bar{A}, \nicefrac{d_i}{2}+\bar{A}]$, the ML rule clearly depends on the private signal $S_i$ and would incur privacy cost.

Based on \eqref{Eq:MLRule}, we formally define the majority voting (MV) based data reporting strategy, denoted by $\sigma^*_i$. It is essentially a modified version of the ML rule. Specifically, when user $i$ needs to incorporate her private signal, $S_i$, into her data reporting strategy; she uses a noisy version of $S_i$.
\begin{definition} \label{Def:MV}
The majority voting (MV) based data reporting, $\sigma_i^*$ has the following form:
\begin{align}
& \bullet \quad \Pr_{\sigma_i^*}  (X_i\teq 1| S_i \teq s_i, C_i \teq c_i) \nonumber \\
   & \hspace{8mm} = \begin{cases} \label{Eq:MV}
            1 & \mbox{if } f_i > d_i/2 \tsum \tau_1, \hfill \mathbf{(ND)} \\
            0 & \mbox{if } f_i < d_i/2 \teg\tau_0, \hfill \mathbf{(ND)} \\ \frac{e^{\xi(f_i)}}{1+e^{\xi(f_i)}}
            & \mbox{if } f_i \in \big[\frac{d_i}{2}\teg \tau_0, \frac{d_i}{2}\tsum \tau_1\big], s_i = 1, \hfill \mathbf{(SR)}\\
            \frac{1}{1+e^{\xi(f_i)}} & \mbox{if } f_i \in \big[\frac{d_i}{2}\teg \tau_0, \frac{d_i}{2}\tsum \tau_1\big], s_i = 0, \hfill \mathbf{(SR)} \\
\end{cases} \\
& \bullet \quad \Pr_{\sigma_i}(X_i\teq 0|  S_i \teq s_i, C_i \teq c_i) = 1 - \Pr_{\sigma_i}(X_i\teq 1| S_i \teq s_i, C_i \teq c_i), \nonumber
\end{align}
where $0 \leq \tau_0,\tau_1 \leq \nicefrac{d_i}{2}$ and $\xi(f_i) \geq 0$. 
\end{definition}
In the MV based data reporting strategies, the sum of the group signals, $f_i$'s, are used. Therefore, with a little abuse of notation, $\zeta(\sigma^*_i,f_i)$ and $\zeta(\sigma^*_i,c_i)$ are used interchangeably in Section 4. When $f_i \in [\nicefrac{d_i}{2}- \tau_0, \nicefrac{d_i}{2} + \tau_1]$, user $i$ employs the SR strategy. In this case, the privacy level of $\sigma^*_i$ for $f_i$ is $\zeta_i(\sigma^*_i,f_i) = \xi(f_i)$. When $f_i \notin  [\nicefrac{d_i}{2}\teg \tau_0, \nicefrac{d_i}{2}\tsum \tau_1]$, user $i$ employs the ND strategy and the privacy level of $\sigma^*_i$ for $f_i$ is $\zeta_i(\sigma^*_i,f_i) = 0$. The privacy level of the SR strategy, $\xi(f_i)$, and the thresholds $\tau_0$ and $\tau_1$, depend on $Z^\mathrm{g}$ from \eqref{Eq:GeniePayMech} and the system model parameters. We will elaborate further on this in Section 4.4.
The majority voting based data reporting strategy profile is denoted by $\sigma^*$ where $\sigma^* = \{ \sigma^*_1, \sigma^*_2, \dots, \sigma^*_N\}$. The next result states that $\sigma^*$ is a BNE in the genie-aided payment mechanism $\mathbf{R}^\mathrm{g}$. Its proof is relegated to Appendix~\ref{Appendix:Proof_Thm_GenieAided}.
\begin{theorem}
\label{Thm:GenieAidedMVbne}
In the payment mechanism $\mathbf{R}^\mathrm{g}$ \eqref{Eq:GeniePayMech}, $\mathbf{\sigma}^*$ is a BNE.
\end{theorem}

Up to this point, we focus on the best response strategies of the users in the genie-aided mechanism, $\mathbf{R}^\mathrm{g}$. In the next subsection, we analyze how the data collector can estimate the underlying state from users' reported data $\mathbf{X}$. In Section~\ref{Subsection:PaymentMechDesign},  building on the genie-aided mechanism $\mathbf{R}^\mathrm{g}$, we devise  a peer-prediction based payment mechanism $\tilde{\mathbf{R}}$ where the data collector obtains the estimate of $W$ from the users' reported data. Finally, in Section~\ref{Subsection:NoisyGroup}, we present the exact details of the BNE $\sigma^*$, and in particular, $\tau_0, \tau_1$ and $\xi(f_i)$, are determined accordingly. 
\subsection{Data Collector as Fusion Center}
\label{Subsection:DataCollectFusion}

Recall that the objective of the data collector is to estimate the underlying state $W$ from the users' reported data $\mathbf{X}$. The conditional distributions of the reported data profile $\mathbf{X}$, given the underlying state $W$, are dictated by the user data reporting strategies.  For a given strategy profile $\sigma$, we restate the binary hypothesis testing problem  \eqref{Eq:HTProblem} as follows:
\begin{equation*}
    \mathcal{H}_0: \mathbf{X} \sim \Pr_\sigma(\mathbf{X} = \mathbf{x}|W = 0), \quad  \mathcal{H}_1: \mathbf{X} \sim \Pr_\sigma(\mathbf{X} = \mathbf{x}|W = 1).
\end{equation*}
The data collector employs the maximum a posteriori (MAP) decision rule, denoted by $\hat{W}_{\sigma}(\mathbf{x})$, in order to minimize the probability of error of the hypothesis testing problem:
\begin{equation} 
    \Lambda_\sigma(\mathbf{x}) \vcentcolon = \frac{\Pr_{\sigma}(\mathbf{X} = \mathbf{x} | W = 1)}{\Pr_{\sigma}(\mathbf{X} = \mathbf{x} | W = 0)} \quad  \mathop{\gtreqless}_{\hat{W}_{\sigma} (\mathbf{x}) = 0}^{\hat{W}_{\sigma}(\mathbf{x}) = 1} \quad \frac{\Pr_W(0)}{\Pr_W(1)} \label{Eq:lhoodEq}
\end{equation}

In general, $X_i$ and $X_j$ are correlated given  $W$, if user $i$ and $j$ are friends or they have common friends owing to the social learning among the users. 
The closed-form evaluation of $\Lambda_\sigma(\mathbf{X})$ is often intractable for dependent observations. In what follows, we present two lemmas to study $\Lambda_\sigma(\mathbf{x})$ in the asymptotic regime of $N$.

Recall that symmetric strategy profiles are defined in \eqref{Eq:SymProf}. The next result shows that $\hat{W}_{\sigma}(\mathbf{x})$ in \eqref{Eq:lhoodEq} depends on  $\sum_{i=1}^N x_i$ when $\sigma$ is a symmetric strategy profile.
\begin{lemma} \label{Lemma:SuffStats}
For every symmetric strategy profile $\sigma$ and $w\in \{0,1\}$, we have
\begin{equation*}
\Pr_\sigma(\mathbf{X} = \mathbf{x}|W = w) = \Pr_\sigma\left(\textstyle \sum_{i=1}^N X_i = \sum_{i=1}^N x_i | W = w \right).
\end{equation*}
\end{lemma}
The proof is relegated to Appendix~\ref{Appendix:ProofSum}. 

Next, we employ a Central Limit Theorem for dependence graphs \cite{janson1988} to characterize the asymptotic statistics of $\sum_{i} X_i$. For any symmetric strategy profile $\sigma$, we define $\mu_w(\sigma)$ as the conditional mean of $X_i$ given $W=w$ with $w\in \{0,1\}$:
\begin{equation} \label{Eq:mu}
    \mu_1 (\sigma) \vcentcolon= \Pr_\sigma(X_i \teq 1|W \teq 1); \quad \ \mu_0 (\sigma) \vcentcolon= \Pr_\sigma(X_i \teq 1|W \teq 0). 
\end{equation}
Recall that $\mathcal{E}_{ij}=1$ if there is a social tie between $i$ and $j$, otherwise $\mathcal{E}_{ij}=0$. Similarly, $B_{ij}=1$ if $i$ and $j$ have a common friend, otherwise $B_{ij}=0$. For convenience, we define $\varsigma_w$ and $\tilde{\varsigma}_w$ for $w \in \{0,1\}$, as follows:
\begin{subequations} \label{Eq:varsigma}
\begin{align}
    \varsigma_{w}(\sigma) & \vcentcolon= \Pr_\sigma(X_i \teq w, X_j \teq w | W \teq w, B_{ij} \teq 0, \mathcal{E}_{ij} \teq 1), \\
    \tilde{\varsigma}_{w}(\sigma) & \vcentcolon= \Pr_\sigma(X_i \teq w, X_j \teq w | W \teq w, B_{ij} \teq 1, \mathcal{E}_{ij} \teq 0), 
\end{align}
\end{subequations}
In the rest of the paper, for purposes of brevity, we drop the dependency of $\mu_w(\sigma), \varsigma_w(\sigma)$ and $\tilde{\varsigma}_w(\sigma)$ on $\sigma$ when it is clear from the context.
We have the following result on the asymptotics of  $\sum_{i=1}^N X_i$
as $N\to \infty$. 
\begin{lemma} \label{CLTLemma}
Under Assumption 1, conditioned on $W\teq w$, for a symmetric data reporting strategy profile $\sigma$,  $\frac{\sum_{i=1}^N X_i - N \mu_w}{\sqrt{N \kappa_w}}$ converges in distribution to a standard normal random variable as $N \to \infty$,   with
\begin{subequations}
\label{Eq:kappa}
\begin{flalign}
    & \kappa_1(\sigma) \vcentcolon = \mu_1 - \mu_{1}^2 + \E[D]\left(\varsigma_1 - \tilde{\varsigma}_1 \right) + \E[D^2]\left(\tilde{\varsigma}_1 - \mu_1^2 \right), \\
    &\kappa_0(\sigma) \vcentcolon =  \mu_0(1\teg \mu_0) \tsum \E[D]\left(\varsigma_0 \teg \tilde{\varsigma}_0 \right) \tsum \E[D^2]\left(\tilde{\varsigma}_0 - (1-\mu_0)^2 \right)\! .
\end{flalign}
\end{subequations}
\end{lemma}
The proof of this lemma is relegated to  Appendix~\ref{Appendix:CLTLemma}. 

Appealing to Lemmas~\ref{Lemma:SuffStats} and \ref{CLTLemma}, for large $N$, the MAP Decision rule $\hat{W}_{\sigma^*}(\mathbf{x})$ can be approximated as follows:
\begin{equation*}
\frac{1}{\kappa_0} \left(\mu_0 \teg \frac{\sum x_i}{N} \right)^2 \hspace{-1mm}- \frac{1}{\kappa_1} \left(\mu_1 \teg \frac{\sum x_i}{N} \right)^2 \mathop{\gtreqless}_{\hat{W}_{\sigma^*} (\mathbf{x}) = 0}^{\hat{W}_{\sigma^*}(\mathbf{x}) = 1} \frac{2}{N} \ln  \sqrt{\frac{\kappa_1}{\kappa_0}} \frac{\Pr_W(0)}{\Pr_W(1)}.
\end{equation*}

\subsection{Payment Mechanism Design}
\label{Subsection:PaymentMechDesign}
Building on  the genie-aided mechanism $\mathbf{R}^\mathrm{g}$, 
next we  turn our attention to  the design of a peer-prediction based payment mechanism $\tilde{\mathbf{R}}$, where the data collector obtains the estimate of $W$ from the users' reported data. Specifically,  in peer prediction  each user $i$'s reported data $X_{i}$ is evaluated by using other users' reported data $\mathbf{X}_{-i}$; and  each user is rewarded with a payment  determined by how her reported data compares with the other users' reported data. In the same spirit, we use majority voting as an effective aggregation method to obtain informative reported data from the users \cite{vonAhn2008,Sheng2008,Liu2016, Wang16}. By rewarding the users whose reported data is in agreement with the the other users' reported data, this payment mechanism incentivizes the users to participate and report informatively using their private and group signals. More specifically, we have the following payment mechanism $\mathbf{\tilde{R}}(\mathbf{X})$:

\begin{enumerate}
\item Each user reports her data, and the data collector counts the number of participants $n$ excluding the users with ``non-participation".
\item For non-participating users, the payment is zero. If $n = 1$, the data collector pays zero to this participant. Otherwise, for each participating user $i$, the data collector computes the majority of the other participants' reported data:
\begin{equation*}
M_{-i} =
\left\{ \begin{array}{ll}
         1 & \text{ if } \sum \limits_{j: x_j \neq \bot, j \neq i} x_j \geq \left \lfloor{\frac{n-1}{2}}\right \rfloor +1 ;\\
        0 & \mbox{ otherwise}.\end{array} \right.
\end{equation*}
\item Compute the payment for user $i$:
\begin{subequations}
\label{Eq:PaymentMech}
\begin{align}
\tilde{R}_i & (1,\mathbf{x}_{-i}) = Z_1  M_{-i}, \label{PaymentMech1} \\
\tilde{R}_i & (0,\mathbf{x}_{-i}) = Z_0 (1-M_{-i}), \label{PaymentMech2}
\end{align}
\end{subequations}
\end{enumerate}
\noindent where $Z_0$ and $Z_1$ are  design parameters to be determined by the data collector. 

In the genie-aided scenario, the payment mechanism $\mathbf{R}^{\mathrm{g}}$ is designed based on the hypothetical scenario where the underlying state $W$ is given. The rationale behind the proposed payment mechanism $\tilde{\mathbf{R}}$ in (15) above  is that the data collector obtains the estimate of $W$ from the noisy user reports and utilizes it in the payment mechanism. Along the same line as in the genie-aided mechanism, in devising the peer prediction based mechanism $\tilde{\mathbf{R}}$, each user first estimates the underlying state $W$ based on her type $t_i$. The next key step lies in the computation of the probability of a user being consistent with  the majority at the BNE strategy profile $\sigma^*$:
\begin{equation} \label{Eq:Beta}
    \beta_w = \Pr_{\sigma^*}(M_{-i} = w | W = w),
\end{equation}
where $w\in \{ 0,1\}$. Clearly, when the number of users is large, the asymptotic statistics of $\sum_{j} X_j$ is the same as the asymptotic statistics of $\sum_{j \in -i} X_j$. Therefore, $\beta_0$ and $\beta_1$ can be computed using Lemma~\ref{CLTLemma}. 

Based on the hypothetical genie-aided payment mechanism $\mathbf{R}^\mathrm{g}$ defined in \eqref{Eq:GeniePayMech}, we obtain the design parameters for the payment mechanism defined in \eqref{Eq:PaymentMech} as  follows: 
\begin{subequations} \label{Eq:Z0Z1}
\begin{align}
Z_0 & = Z \ \frac{\Pr_W(1)\beta_1 + \Pr_W(0)(1-\beta_0)}{(\beta_0 \tsum \beta_1 - 1) \Pr_W(1) \Pr_W(0)}, \\
Z_1 & = Z \ \frac{\Pr_W(1)(1-\beta_1) + \Pr_W(0)\beta_0}{(\beta_0 \tsum \beta_1 - 1) \Pr_W(1) \Pr_W(0)}.
\end{align}
\end{subequations}
where $Z>0$ is a design parameter. For the degenerate case in which the data collector obtains the estimate of $W$ with no error, we have that $\beta_1 = \beta_0 = 1$, indicating that $\tilde{\mathbf{R}}$ reduces to the genie-aided mechanism introduced in \eqref{Eq:GeniePayMech}. Theorem~\ref{MVbne} reveals that there exists a MV based BNE when the data collector employs the payment mechanism $\tilde{\mathbf{R}}$. Its proof is given in Appendix~\ref{ProofThm1}.

\begin{theorem}
\label{MVbne}
In the payment mechanism $\tilde{\mathbf{R}}$ \eqref{Eq:Z0Z1}, $\mathbf{\sigma}^*$ is a BNE.
\end{theorem}

It is worth noting that in the payment mechanism $\tilde{\mathbf{R}}$, the MV based  BNE equilibrium $\sigma^*$ is not the only equilibrium, as expected. At $\sigma^*$, no user can gain by playing uninformative when other users employ the MV rule. However, in $\tilde{\mathbf{R}}$, uninformative equilibria also exist, as it is the case in many peer-prediction and information elicitation mechanisms (\cite{Prelec2004},\cite{Miller04},\cite{jurca2005},\cite{Shnayder2016}). One set of such equilibria is that the users form lying coalitions and collude to report the same uninformative data \footnote{Interested readers can find detailed discussions for different colluding scenarios in the information elicitation mechanism in \cite{Jurca2009}.}. However, we caution that the social learning model does not imply any cooperation and communication among friends.

\subsection{Data Reporting Strategies at BNE }
\label{Subsection:NoisyGroup}

Theorem~\ref{MVbne} establishes the existence of  a MV based BNE under the payment mechanism $\tilde{\mathbf{R}}$. Recall that, $\xi(f_i)$ corresponds to the privacy level of $\sigma_i^*$ when $f_i \in [\nicefrac{d_i}{2}- \tau_0, \nicefrac{d_i}{2} + \tau_1]$. To complete the design of the payment mechanism, it remains to characterize $\tau_0, \tau_1$ and $\xi(f_i)$ as a function of the mechanism design parameter Z.

For convenience, let $\epsilon$ denote the privacy level of $\sigma_i^*$ when $f_i \teq \nicefrac{d_i}{2}$, $\xi(f_i) = \epsilon$. In the following, we show that, $Z$ can be written in terms of $\epsilon$ as follows:
\begin{equation}
Z = g'(\epsilon)  \frac{(e^\epsilon\tsum 1)^2}{2 e^\epsilon (2\theta_0 - 1)}.  
\end{equation}
The following result formalizes this argument and puts forward an algorithm to find the BNE strategies.

\begin{proposition}
\label{Prop:UserAlg}
The BNE strategy profile $\sigma^*$ can be found by using Algorithm~\ref{alg:MV}.
\end{proposition}

For convenience, we define the following functions, which are used in Algorithm~\ref{alg:MV}: 
\begin{subequations}
\label{Eq:Alg1MotherEq}
\begin{equation}
J'(\eta) \teq e^{\eta - \epsilon} \left(\frac{e^\epsilon\tsum 1}{e^{\eta}\tsum 1}\right)^2 \hspace{-0.2cm} \frac{[1\tsum (\nicefrac{\theta_1}{(1-\theta_1)})^{d_i- 2f_i}]g'(\epsilon)/2}{1 \tsum p_W(0)((\nicefrac{\theta_1}{(1-\theta_1)})^{d_i- 2f_i}\teg 1)} \teg g'(\eta), \label{AlgEq2}
\end{equation}
\begin{equation}
\label{AlgEq1}
B(\eta) =  (2\theta_0 \teg 1) 2e^\epsilon \frac{g(\eta)}{g'(\epsilon)} \frac{(e^{\eta}\tsum 1)}{(e^\epsilon \tsum 1)^2},
\end{equation}
\begin{equation}
A_\ell (\eta) = \frac{2\ell\teg 1}{2\ln \left ( \frac{\theta_1}{1 \teg \theta_1} \right )} \ln \left (\frac{e^{\eta}(1\teg \theta_0) \tsum \theta_0 \tsum p_W(\ell) B(\eta)}{e^{\eta} \theta_0 \tsum 1 \teg \theta_0 \teg p_W(1\teg \ell) B(\eta)} \right ), \label{AlgEq3}
\end{equation}
\begin{equation}
    \Upsilon_\ell(\eta) \teq \begin{cases}
        0, \text{ if } A_\ell(\eta) \leq 0\\
        A_\ell(\eta), \text{if } A_\ell(\eta) \! \in \! (0,\bar{A}), \\
        \bar{A}, \text{ if } A_\ell(\eta) \geq \bar{A} \end{cases} \text{for } \ell = 0,1.
\end{equation}
\end{subequations}
Recall that $\bar{A}$ is defined in \eqref{Eq:barA}. Furthermore, we also define $p(s_i,f_i)$ and $q(s_i,f_i)$ as follows:
\begin{subequations} \label{Eq:pqsifi}
\begin{align}
p(s_i,f_i) & = \Pr_{\sigma_i^*} (X_i = 1 | S_i = s_i, F_i = f_i), \label{Eq:psifi}\\
q(s_i,f_i) & = \Pr_{\sigma_i^*} (X_i = 0 | S_i = s_i, F_i =f_i). \label{Eq:qsifi}
\end{align}
\end{subequations}

\begin{algorithm}[h]
\SetAlgoNoLine
\SetEndCharOfAlgoLine{}
\SetKwComment{bsr}{}{}
\KwIn{Type $t_i$, number of friends $d_i$.}
\KwOut{$p(s_i,f_i)$ and $q(s_i,f_i)$}
Solve for $J'(\eta^*) = 0$. Set $\xi(f_i) = \eta^*$.\\
\lIf{$\hspace{0.25cm} 0 \leq f_i \leq d_i/2 \teg \Upsilon_0(\xi(f_i)) \hspace{0.25cm} $}{$\hspace{0.25cm} p(0,f_i) = p(1,f_i) = 0\hspace{0.25cm} $ and $\hspace{0.25cm} q(0,f_i) = q(1,f_i) = 1$ \textbf{(ND Strategy)}}
\lElseIf{$\hspace{0.25cm} f_i \geq d_i/2 \tsum \Upsilon_1(\xi(f_i))   \hspace{0.25cm}$}{$\hspace{0.25cm} p(0,f_i) \teq p(1,f_i) \teq 1 \hspace{0.25cm}$ and $\hspace{0.25cm} q(1,f_i) \teq q(0,f_i) \teq 0$  \textbf{(ND Strategy)}}
\lElse{ $\hspace{0.25cm} p(1,f_i) \teq q(0,f_i) \teq e^{\xi(f_i)}/(1+e^{\xi(f_i)}) \hspace{0.25cm}$ and $\hspace{0.25cm} p_{0,f_i} \teq q_{1,f_i} \teq 1/(1+e^{\xi(f_i)})$  \textbf{(SR Strategy)}) }. 
\caption{MV based BNE strategy $\sigma^*$}
\label{alg:MV}
\end{algorithm}

The proof of proposition~\ref{Prop:UserAlg} can be found in Appendix~\ref{ProofCorAlg}. Simply put, Algorithm~\ref{alg:MV} presents a majority rule for a user to determine $\sigma^*_i$: User $i$ needs to compare the expected utilities of playing the ND and the SR strategies in order to decide upon between them. She first computes the optimal privacy level for the SR strategy, namely $\xi(f_i)$, by solving $J'(\eta^*) \teq 0$ and setting $\xi(f_i) = \eta^*$. Then, she plays the SR strategy with $\xi(f_i)$ if $ \nicefrac{d_i}{2} - \Upsilon_0(\xi(f_i)) \leq f_i \leq \nicefrac{d_i}{2} + \Upsilon_1(\xi(f_i))$;  otherwise, she plays the ND strategy. 

There is a discrepancy between the quality of $S_i$ and $C_{i}$: $\theta_1 < \theta_0$. Therefore, the required majority in $C_i$ for playing ND heavily depends on $\alpha$, the noise level in the group signals. When $\alpha \teq 0$, a simple majority in $C_i$ suffices to determine whether to  play ND or SR. As $\alpha$ increases, the accuracy gap between $S_i$ and $C_i$ widens; therefore, the required majority in $C_i$ for playing ND increases.

\subsection{A Closer Look at Data Reporting Strategies: Two Special Cases}
In general, it gets very complicated to determine $\tau_0$ and $\tau_1$ in terms of $\Upsilon_0(\xi(f_i))$ and $\Upsilon_1(\xi(f_i))$. To get a more concrete sense, in what follows we study two special cases where the majority rule can be further simplified.

\subsubsection{\textbf{The Case with Noiseless Group Signals}}
\label{Subsection:Noiseless}

Our next result reveals that when $\alpha \teq 0$,  at the BNE, a \textit{simple majority} rule within the group signals can be used by each user to determine which strategy to use: If $d_i$ is even, the user plays SR only if $f_i \teq d_i/2$ with privacy level $\epsilon$. If $d_i$ is odd, the user never plays SR.

\begin{corollary} \label{Prop:NoiselessMV}
For the case with noiseless group signals, the BNE strategy profile $\sigma^*$ has the following form:
\begin{align*}
p(s_i,f_i) = \begin{cases}
1 & \mbox{if } f_i > \nicefrac{d_i}{2}, \\
0 & \mbox{if } f_i < \nicefrac{d_i}{2}, \\
\frac{s_i e^\epsilon+1- s_i}{1+e^\epsilon} & \mbox{if } f_i \teq \nicefrac{d_i}{2}, \\
\end{cases}
\quad \text{and}\quad  q(s_i,f_i) \teq 1\teg p(s_i,f_i).
\end{align*}
\end{corollary}

\subsubsection{\textbf{The Case with Equal Priors}}
Our next result reveals that when $p_W(1) \teq p_W(0) \teq 0.5$, we have that $\xi(f_i) = \epsilon$, for $f_i \in \{0,1,\dots, d_i\}$. Therefore, we have $\Upsilon_1(\epsilon) = \Upsilon_0(\epsilon) = \Upsilon(\epsilon)$ as follows:
\begin{equation*}
    \Upsilon(\epsilon) \teq \begin{cases}
        0, \text{ if } A \leq 0\\
        A, \text{ if } A \! \in \! (0,\bar{A}), \\
        \bar{A}, \text{ if } A \geq \bar{A} \end{cases} \hspace{-0.3cm} A \teq \frac{\ln\! \Big( \frac{e^\epsilon[e^\epsilon \theta_0 + 1 - (2\theta_0 - 1)g(\epsilon)/g'(\epsilon)]+ 1-\theta_0}{e^\epsilon[e^\epsilon (1- \theta_0) + 1 + (2\theta_0- 1)g(\epsilon)/g'(\epsilon)]+ \theta_0} \Big )} { 2 \ln \left ( \frac{\theta_1}{1-\theta_1} \right)}.  
\end{equation*}
Consequently, we have $\tau(\epsilon) = \tau_1 = \tau_0$ and $\tau = \Upsilon(\epsilon)$. 

\begin{corollary} \label{Cor:EqPriors}
For the case with equal priors, the optimal SR strategy at the BNE is as follows:
\begin{align*}
p(s_i,f_i) & = \begin{cases}
1 & \mbox{if } f_i > \nicefrac{d_i}{2} + \tau(\epsilon), \\
0 & \mbox{if } f_i < \nicefrac{d_i}{2} - \tau(\epsilon), \\
\frac{s_i e^\epsilon+1- s_i}{1+e^\epsilon} & \mbox{if } f_i \in [\nicefrac{d_i}{2} - \tau(\epsilon),\nicefrac{d_i}{2} + \tau(\epsilon)],  \\
\end{cases}\\
q(s_i,f_i) & = 1 - p(s_i,f_i).
\end{align*}
\end{corollary}
From Corollary \ref{Cor:EqPriors}, it is clear that the MV based data reporting strategies depend heavily on $\alpha$. As $\alpha$ grows, $\tau$ also expands which implies that the user plays the SR strategy more often at the BNE. For conciseness, in the remainder of this section, we suppress the explicit dependence of $\tau$ on $\epsilon$.

Next we study the computation of $\beta_0$ and $\beta_1$ defined in \eqref{Eq:Beta}, under equal priors assumption. We first need to define several terms. Define $\gamma(k;d,p)$ and $\Gamma(k, \ell;d,p)$   corresponding to a Binomial distribution with parameters $m$ (number of trials) and $n$ (probability of success) as follows:
\begin{subequations} \label{Eq:Gammas}
\begin{equation}
\gamma(k;m,n) \teq \begin{cases} \binom{d}{k} n^k (1\teg n)^{m-k} & \text{if } k \in \{0,1,\dots,n \},\\
0 & \text{otherwise};
\end{cases}
\end{equation}
\begin{equation}
    \Gamma(k,\ell ;m,n) \teq \sum_{i = \lceil k\rceil }^{\lfloor \ell\rfloor} \gamma(i; m,n),
\end{equation}
\end{subequations}
where $\lceil k \rceil \vcentcolon =  \min \{m \in \mathbb{Z}: m \geq k \}$ and  $\lfloor \ell \rfloor \vcentcolon = \max \{n \in \mathbb{Z}: n \leq \ell \}$. Next, we define $\nu^{\mathrm{sr}}(d,\tau)$ and $\nu^{\mathrm{nd}}(d,\tau)$:
\begin{subequations} \label{Eq:NuDef}
\begin{align} 
\nu^{\mathrm{sr}}(d,\tau) & \vcentcolon = \Gamma( \nicefrac{d}{2}- \tau,\nicefrac{d}{2}+ \tau;d,\theta_1),\\ 
\nu^{\mathrm{nd}}(d,\tau) & \vcentcolon = \Gamma(\left \lfloor \nicefrac{d}{2} + \tau + 1 \right \rfloor, d;d,\theta_1).
\end{align}
\end{subequations}
Finally, we define $\tilde{\rho}$ as follows:
\begin{equation} \label{Eq:tilderho}
    \tilde{\rho}_{d} \vcentcolon = \Pr(D_i \teq d | D_i > 0) \teq \begin{cases}
    0 , &  \text{ if } d \teq 0; \\
    \rho_{d}/(1 \teg \rho_0), & \text{ else.}
\end{cases}
\end{equation}
Note that, $\tilde{\rho}$ is well defined unless $\rho_0 = 1$ which corresponds to the case there is no social learning among the users. In the rest of the paper, we use the subscript notation $\E_{\tilde{\rho}}$ when we use $\tilde{\rho}$ for the expectation of the user degrees. 
\begin{proposition} \label{Prop:EqualPriorsKappa}
For the case with equal priors, $\kappa_w(\sigma^*)$ and $\mu_w(\sigma^*)$ for $w \in \{0,1\}$ are as follows:
\begin{align}
\mu_1(\sigma^*) & = 1- \mu_0(\sigma^*) = \E[\nu^{\mathrm{nd}}(D, \tau)] + \lambda(\epsilon) \E[\nu^{\mathrm{sr}}(D, \tau)], \label{Eq:muBNE} \\
\kappa_1(\sigma^*) & = \kappa_0(\sigma^*) = \mu_1 - \mu_1^2 + \tilde{\triangle} \E[D^2]+ \triangle(\E[D^2]\teg \E[D]),\label{Eq:kappaBNE}
\end{align}
where $\lambda$, $\triangle$ and $\tilde\triangle$ are defined as
\begin{align*}
& \lambda(\epsilon) \vcentcolon = \frac{\theta_0e^\epsilon\tsum 1\teg \theta_0}{e^\epsilon\tsum 1}, \ \tilde{\triangle} \vcentcolon = \frac{\rho_0}{(1\teg \rho_0)^2} \left(\mu_1^2(2\teg \rho_0) - 2\mu_1\lambda + \rho_0\lambda^2 \right), \\
& \triangle \vcentcolon = \theta_0(1\teg \theta_0)(1\teg 2\alpha) \ \E_{\tilde{\rho}}[(e^\epsilon(1\teg \theta_0)\tsum \theta_0)\ \gamma\left(\left \lfloor\nicefrac{D}{2}\tsum \tau \right \rfloor;D\teg 1,\theta_1\right) \\
&\qquad + (\theta_0 e^\epsilon\tsum 1 \teg \theta_0) \gamma\left(\left \lceil\nicefrac{D}{2}\teg \tau\teg 1\right \rceil;D\teg 1,\theta_1\right)]/(e^\epsilon+1).
\end{align*}
\end{proposition}
The proof of Proposition~\ref{Prop:EqualPriorsKappa} is relegated to Appendix~\ref{Appendix:PropEqual}. 

We can compute $\beta_w = \Pr_{\sigma^*}(M_{-i} = w|W \teq w)$, for $w\in \{0,1\}$. Appealing to Lemma~\ref{CLTLemma}, we have that
\begin{equation} \label{Eq:DeltaMechAcc}
    \beta \vcentcolon = \beta_0 = \beta_1 = \Phi\left(\sqrt{\frac{N - 1}{\kappa_1(\sigma^*)}} \left(\mu_1(\sigma^*) - \frac{1}{2} \right) \right),
\end{equation}
where $\Phi$ denotes the cumulative distribution function of the standard normal distribution. Consequently, for the case with equal priors, the parameters $Z_0$ and $Z_1$ can be found as follows:
\begin{equation*}
    Z = Z_0 = Z_1 =\frac{g'(\epsilon)(e^\epsilon\tsum 1)^2}{e^\epsilon(2\theta_0 \teg 1)(2\beta \teg 1)}.
\end{equation*}
It is clear that $\beta \rightarrow 1$ as  $N$ grows and the proposed payment mechanism $\tilde{\mathbf{R}}$ \eqref{Eq:PaymentMech} boils down to the genie-aided payment mechanism $\mathbf{R}^\mathrm{g}$ \eqref{Eq:GeniePayMech}.

Our next result determines the expected payment of the proposed payment mechanism $\tilde{\mathbf{R}}$.
\begin{theorem} \label{Thm:ExpectedPayment}
For the case with equal priors, the total expected payment at the BNE is the following:
\begin{equation*}
\sum_{i = 1}^N \E_{\sigma^*} [\tilde{R}_i(\mathbf{X})] = Z \left(1-\beta + \frac{\mu_1(\sigma^*)}{2\beta - 1} \right)N.
\end{equation*}
\end{theorem}
Its proof is relegated to Appendix~\ref{Appendix:ExpectedPayment}.

{\color{red} 
}

\section{The Impact of Social Learning}\label{Section:Impact}
In this section, we analyze the impact of social learning on the trade-off between payment and accuracy and that between payment and privacy cost. We also present examples, using social learning graph models based on synthetic data and/or real-world data,    to evaluate the performance of our proposed mechanisms.

\begin{figure*}[t] 
\centering
\begin{subfigure}[b]{0.33\textwidth}
\includegraphics[width=1\textwidth]{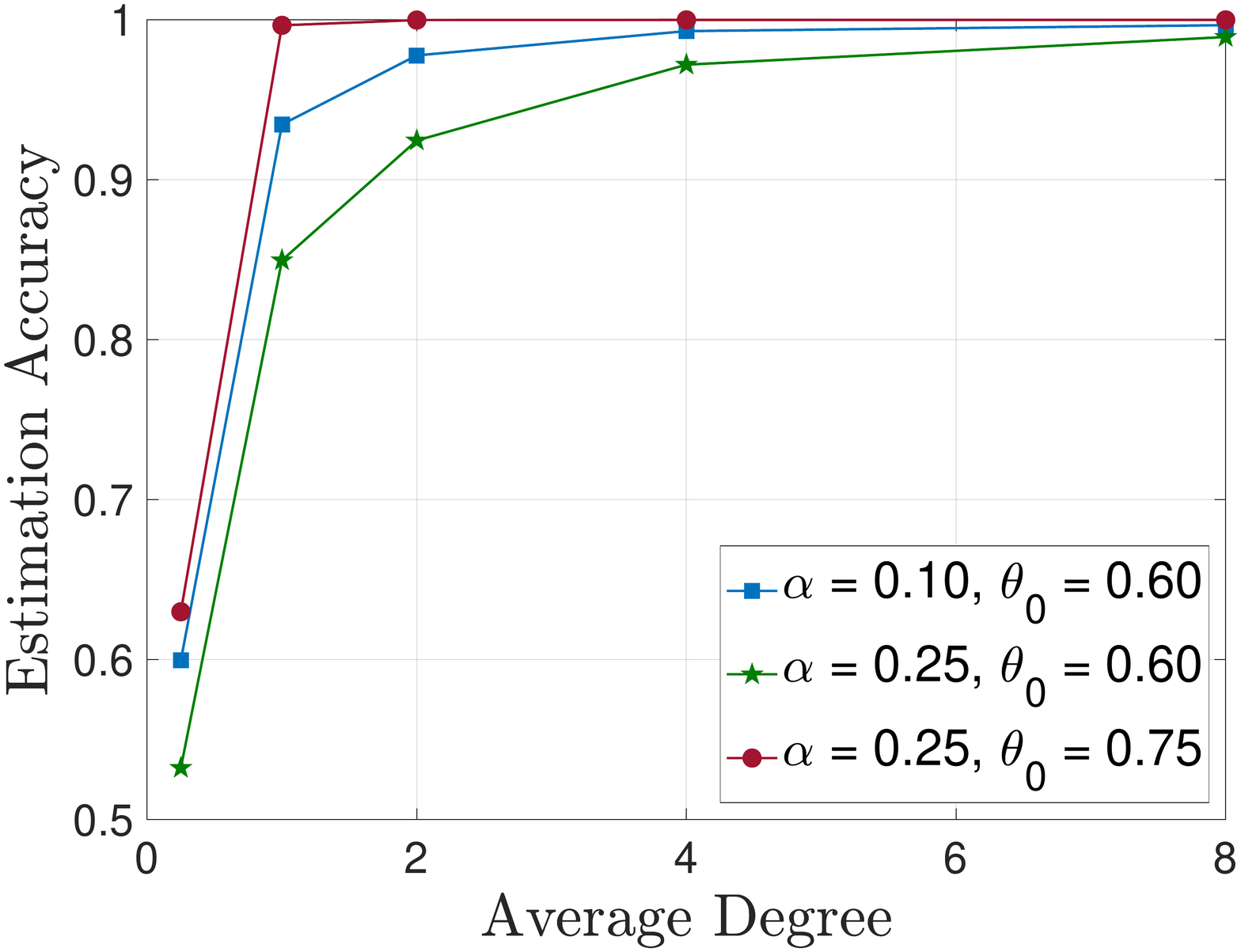} 
\caption{$\Pr_{\sigma^*}(\hat{W}_{\sigma^*}(\mathbf{X}) =W)$ vs. $\E[D_i]$ ($\epsilon\teq 0.1$)}
\end{subfigure}
\begin{subfigure}[b]{0.32\textwidth}
\includegraphics[width=1\textwidth]{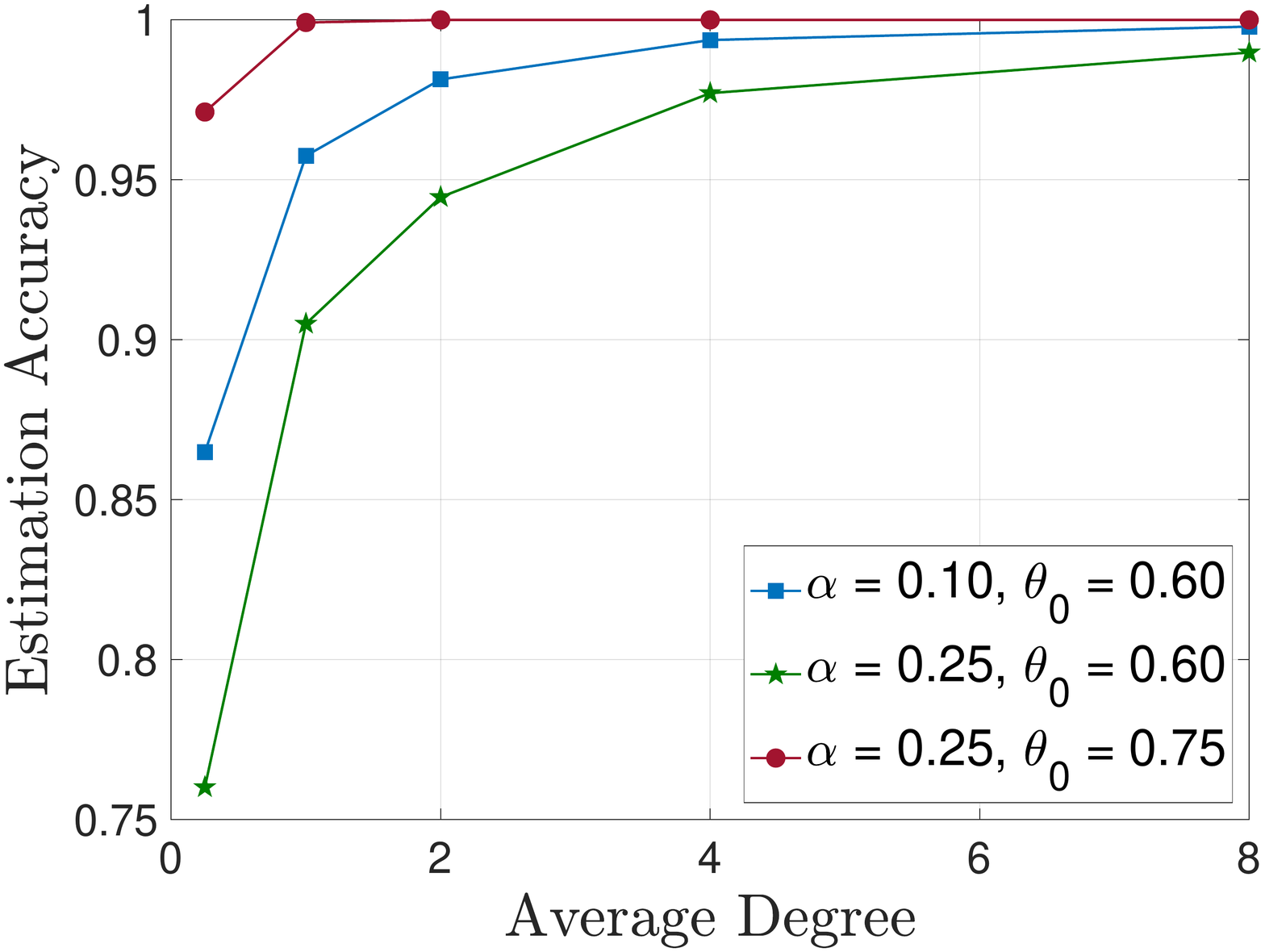} 
\caption{$\Pr_{\sigma^*}(\hat{W}_{\sigma^*}(\mathbf{X}) =W)$ vs. $\E[D_i]$ ($\epsilon\teq 0.5$)}
\end{subfigure}
\begin{subfigure}[b]{0.32\textwidth}
\includegraphics[width=1\textwidth]{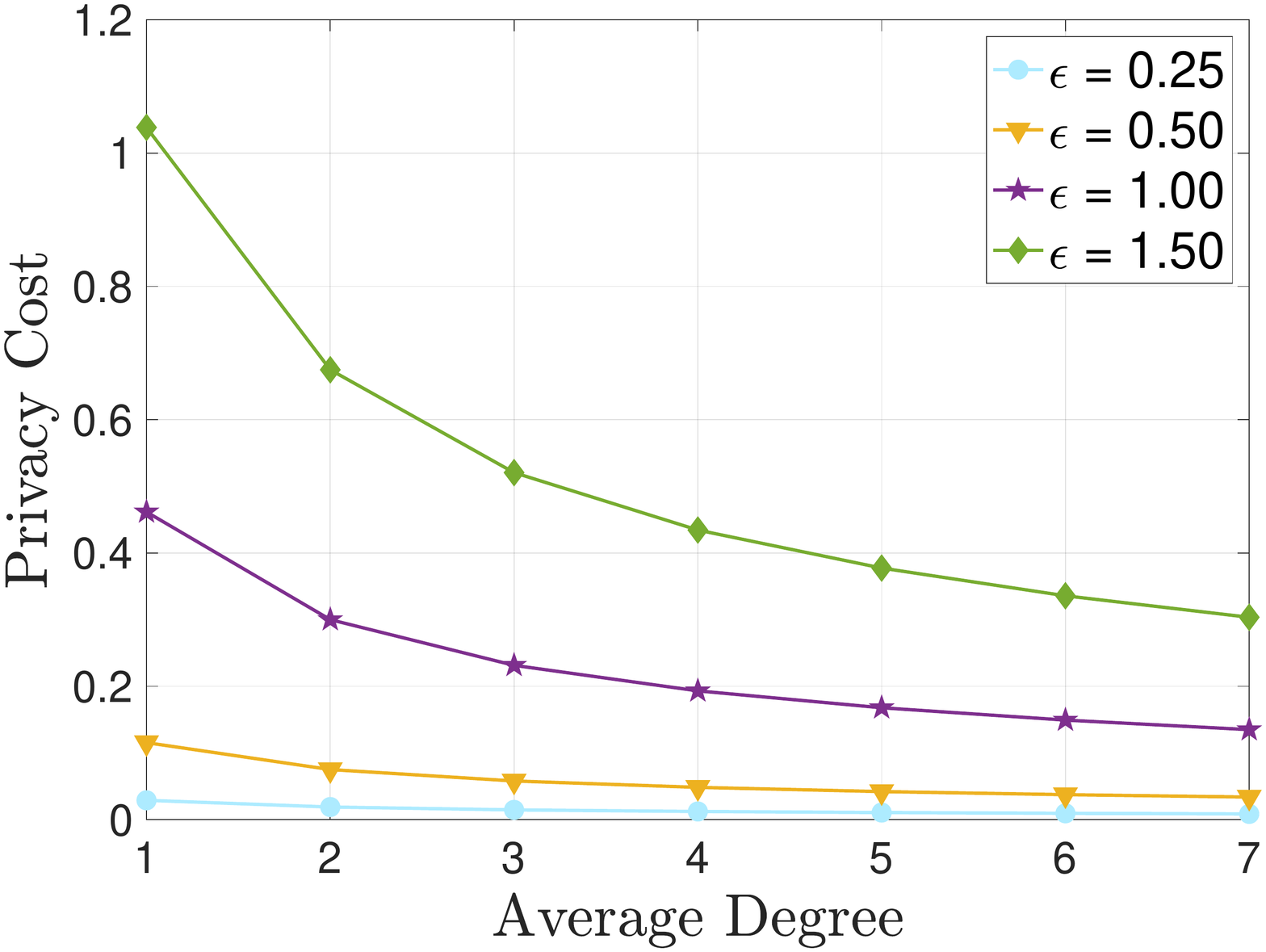}
\caption{$\E[ g(\zeta(\sigma_i^*,C_i))]$ vs $E[D_i]$ ($\alpha\teq 0.25, \theta_0 \teq 0.7$)}
\end{subfigure}
\begin{subfigure}[b]{0.33\textwidth}
\includegraphics[width=1\textwidth]{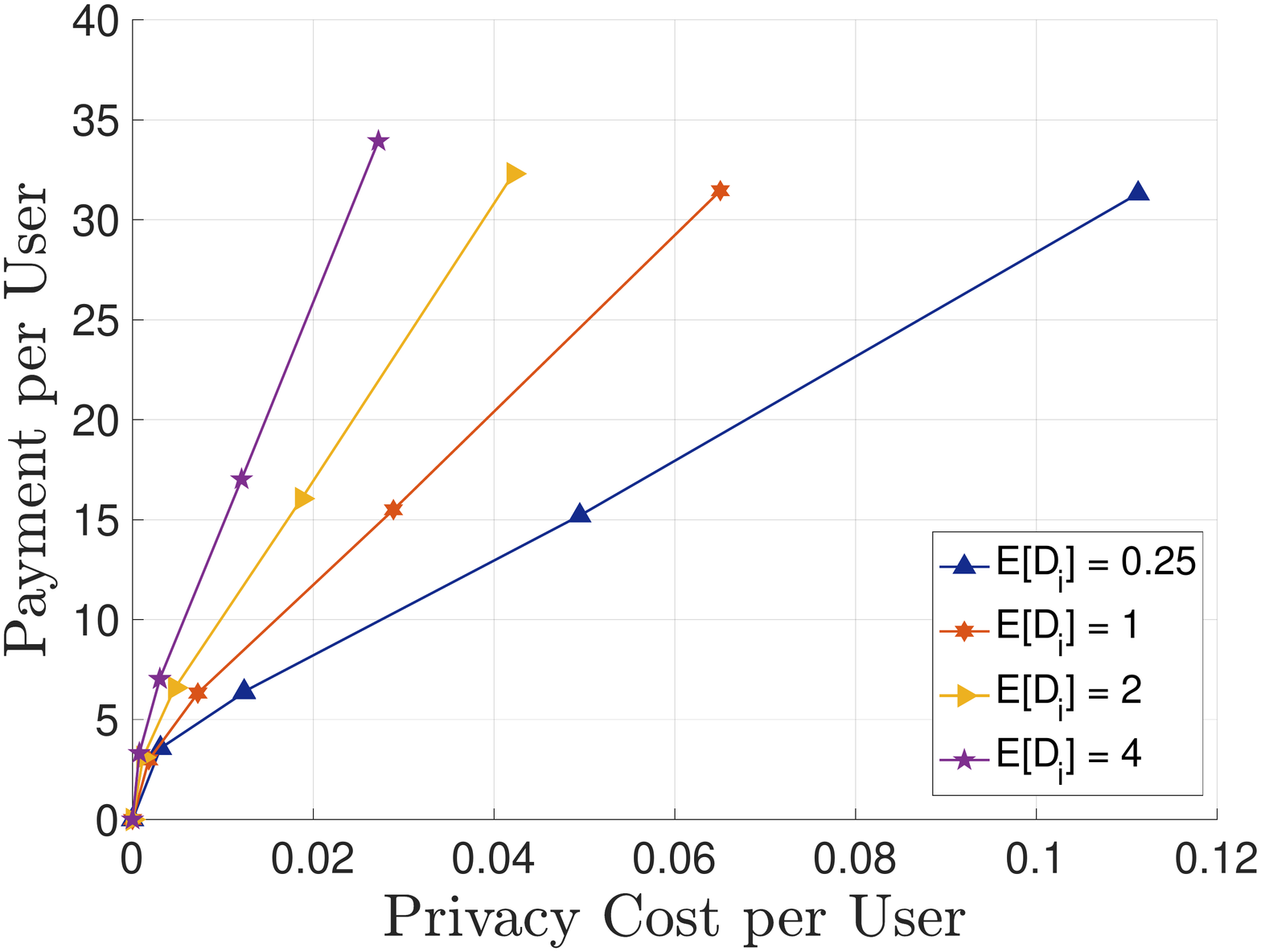}
\caption{\centering $\E_{\sigma^*} [\tilde{R}_i(\mathbf{X})]$ vs. $\E[ g(\zeta(\sigma_i^*,C_i))]$\newline ($\alpha = 0.25,\theta_0 = 0.7$)}
\end{subfigure}
\begin{subfigure}[b]{0.32\textwidth}
\includegraphics[width=1\textwidth]{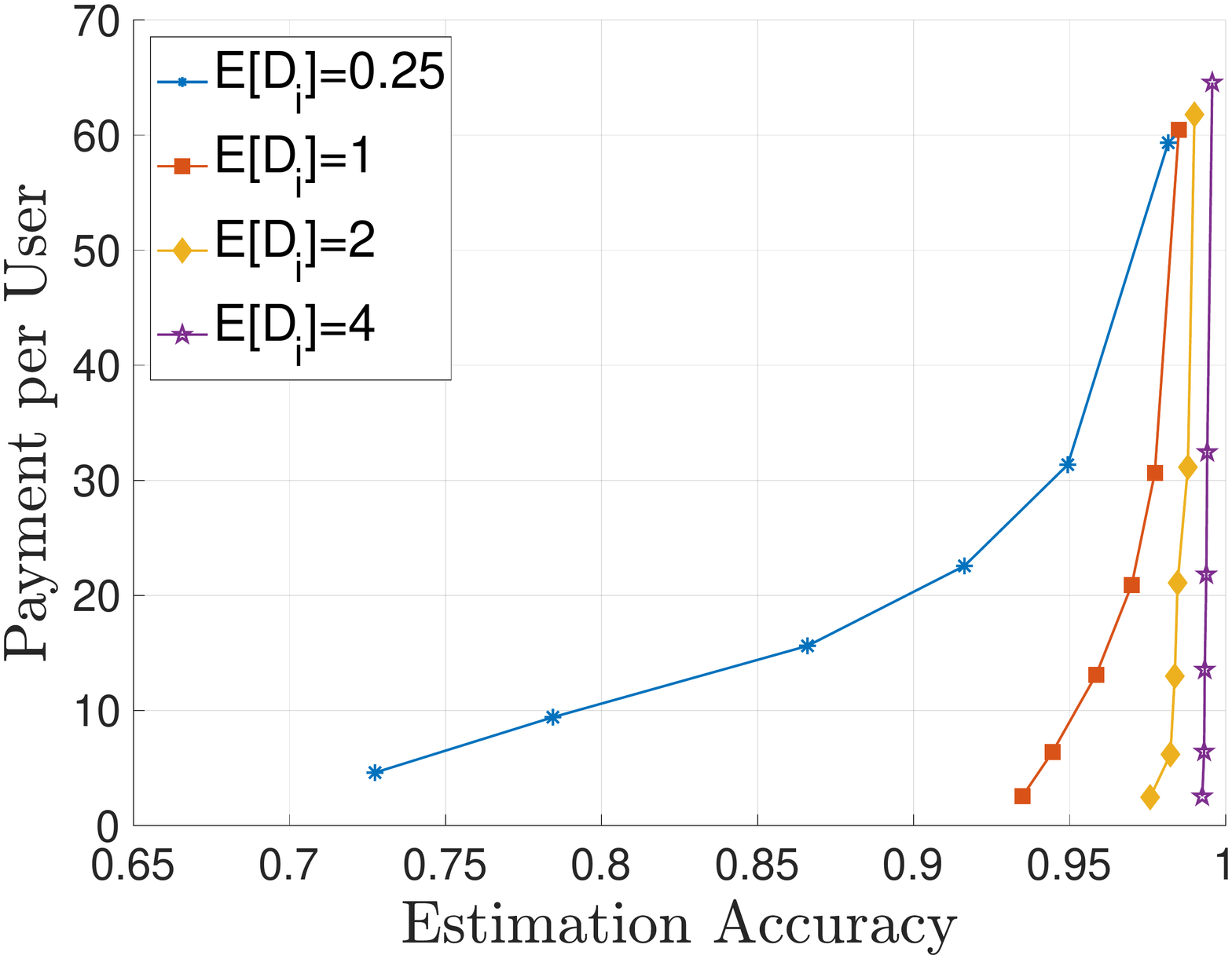}
\caption{\centering$\E_{\sigma^*}[\tilde{R}_i(\mathbf{X})]$ vs. $\Pr_{\sigma^*} (\hat{W}_{\sigma^*} (\mathbf{X}) \teq W)$\newline ($\alpha = 0.10,\! \theta_0= 0.6$)}
\end{subfigure}
\begin{subfigure}[b]{0.32\textwidth}
\includegraphics[width=1\textwidth]{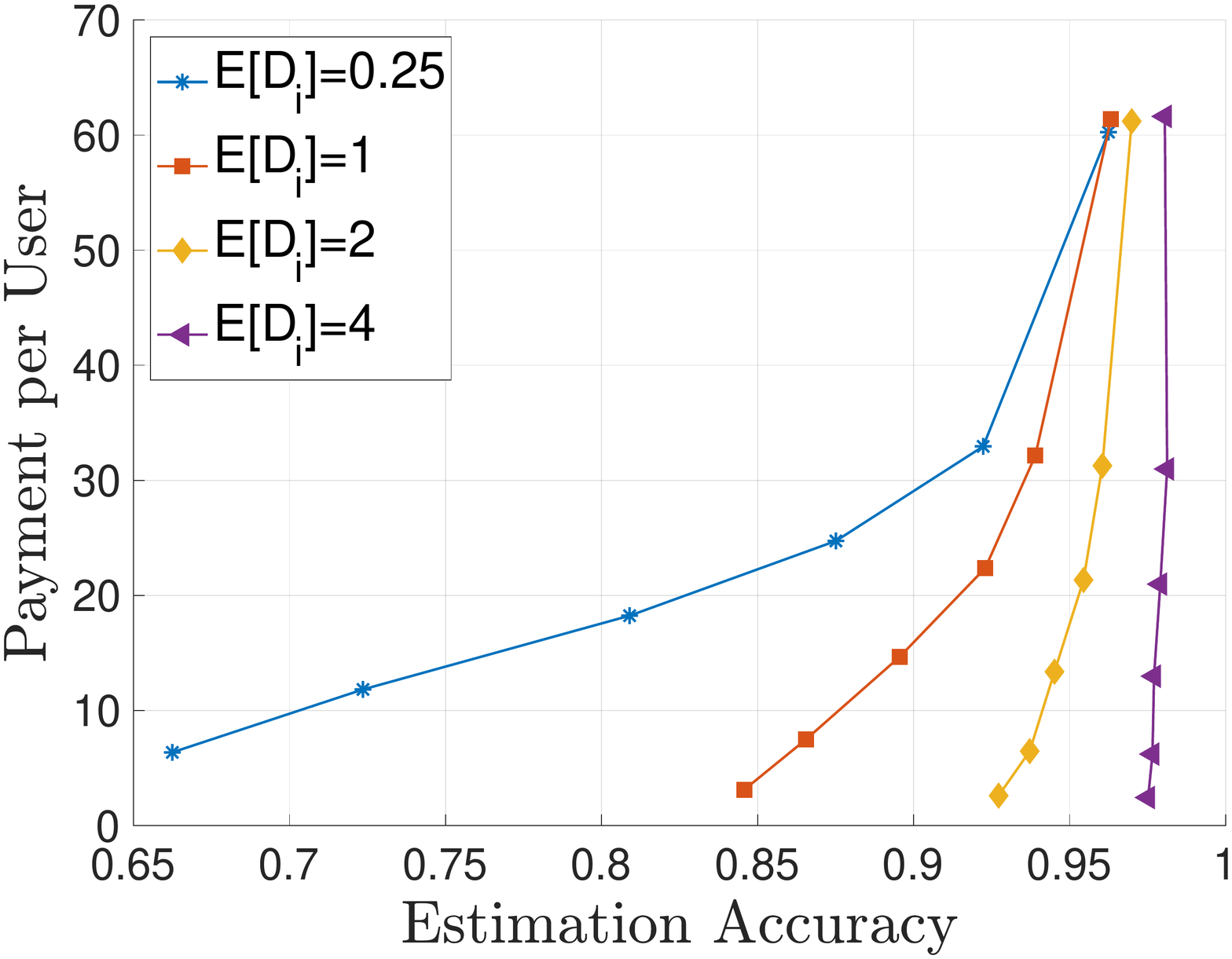}
\vspace{-0.5cm}
\caption{\centering $\E_{\sigma^*}[\tilde{R}_i(\mathbf{X})]$ vs. $\Pr_{\sigma^*} (\hat{W}_{\sigma^*} (\mathbf{X}) \teq W)$\newline ($\alpha \teq 0.25, \theta_0 \teq 0.6$)}
\end{subfigure}
\caption{The impact of social learning: synthetic social learning graphs. (a) (b): State estimation accuracy vs. average degree;  (c) privacy cost per user vs. average degree; (d) payment per user vs. privacy cost per user; (e) (f) payment per user vs. state estimation accuracy.} 
\label{Fig:Synthetic}
\end{figure*}

\subsection{Payment vs. Accuracy}

The data collector aims to minimize the total payment while achieving a given accuracy target in estimating $W$. In particular, the accuracy is measured by the error rate of the MAP detector \eqref{Eq:lhoodEq}. Let $\mathcal{R}(\sigma)$ denote the set of nonnegative payment mechanisms in which $\sigma$ is a BNE. Then, the mechanism design problem for the data collector is given as follows:
\begin{align}
     \min_{\mathbf{R} \in \mathcal{R}(\sigma)} \quad & \sum_{i = 1}^N \E_{\sigma} [R_i(\mathbf{X})], \\
    \mathrm{s.t.} \quad &  \E_\sigma\left[\Pr (\hat{W}_\sigma(\mathbf{X}) \neq W)\right] \leq \mathcal{P}_{e},\nonumber
\end{align}
where the maximum allowable error is represented by $\mathcal{P}_{e}$. It is known that in general it is difficult to characterize the error rate in closed-form at a given BNE strategy profile $\sigma$. Therefore, we measure the accuracy based on an information-theoretic metric which is closely associated to the error rate of the MAP decision rule as follows \cite{Kailath67}:
\begin{equation*}
\E_\sigma \left[ \Pr (\hat{W}_\sigma(\mathbf{X}) \neq W)\right] \leq e^{- \mathcal{B}(\sigma)},
\end{equation*}
where $\mathcal{B}(\sigma)$ denotes the Bhattacharyya distance \cite{Bhattacharyya46}
\begin{equation*}
    \mathcal{B}(\sigma) = - \ln \sum_{\mathbf{x}\in \mathcal{X}^N} \sqrt{\Pr_{\sigma}(\mathbf{X} \teq \mathbf{x}|W \teq 1)\Pr_{\sigma}(\mathbf{X} \teq \mathbf{x}|W \teq 0) }. 
\end{equation*}
Thus, the mechanism design problem can be restated as follows:
\begin{align}
\min_{\mathbf{R} \in \mathcal{R}(\sigma)} \sum_{i = 1}^N \E_{\sigma} [R_i(\mathbf{X})], \quad \mathrm{s.t.} \quad \mathcal{B}(\sigma) \geq - \ln \mathcal{P}_{e}
\end{align}

Let $\mathcal{L}(\mathcal{P}_{e})$ denote the minimum total payment while satisfying the error rate constraint $\mathcal{P}_{e}$. Appealing to Lemma~\ref{Lemma:SuffStats} and~\ref{CLTLemma}, we can simplify the expression for $\mathcal{B}(\sigma)$, for symmetric strategy profiles by approximating  $\Pr_\sigma(\mathbf{X}\teq \mathbf{x}|W \teq w)$ as a Gaussian distribution for large $N$. Thus, $\mathcal{B}(\sigma)$ can be calculated explicitly as follows \cite{Kailath67}:
\begin{equation*}
\mathcal{B}(\sigma) = \frac{N}{4} \frac{(\mu_1(\sigma) - \mu_0(\sigma))^2}{\kappa_1(\sigma) + \kappa_0(\sigma)}.    
\end{equation*}

\subsection{Bounds on Payment}
In this subsection, we investigate an interesting case where all users play  the following non-disclosive strategy,  denoted 
$\sigma^{\mathrm{nd}}$:
\begin{align*}
    &\Pr_{\sigma^{\mathrm{nd}}} (X_i \teq 1|S_i \teq s_i, F_i \teq f_i) = \begin{cases} 1 & \text{ if } f_i > \nicefrac{d_i}{2}, \\
    0 & \text{ if } f_i < \nicefrac{d_i}{2}, \\
    0.5 & \text{ else;}
    \end{cases} \\
    & \Pr_{\sigma^{\mathrm{nd}}} (X_i \teq 0|S_i \teq s_i, F_i\teq f_i) = 1 - \Pr_{\sigma^{\mathrm{nd}}} (X_i \teq 0|S_i \teq s_i, F_i\teq f_i).
\end{align*}
If there is a tie within her group signals, the user tosses a fair coin. It is clear that the above $\sigma^{\mathrm{nd}}$ is a specific form of MV based data reporting strategies with $\tau_1 \teq \tau_0 \teq 0$ and $\xi(f_i) \teq 0$, and further in general $\sigma^{\mathrm{nd}}$ is suboptimal. 
Nevertheless, we show below that if the required estimation accuracy, in terms of $\mathcal{P}_e$, is loose, the total payment can be driven to be arbitrarily small by using the above specific ND strategy, thanks to social learning.

It can be shown that the conditional mean of $X_i$ given $W$ is
\begin{align}\label{Eq:mu1_ND}
\mu_1\big(\sigma^{\mathrm{nd}}\big) & = \E_{\rho}[\Gamma(\lfloor \nicefrac{D}{2}+ 1\rfloor,D;D,\theta_1 + 0.5 \gamma( \nicefrac{D}{2}; D, \theta_1)],\\
\mu_0\big(\sigma^{\mathrm{nd}}\big) & = \E_{\rho}[\Gamma(\lfloor \nicefrac{D}{2}+ 1\rfloor,D;D,1-\theta_1 + 0.5 \gamma( \nicefrac{D}{2}; D, 1-\theta_1)].\nonumber
\end{align}
It is clear that $\mu_0\big(\sigma^{\mathrm{nd}}\big) = 1 -\mu_1\big(\sigma^{\mathrm{nd}}\big)$ and $\mu_1\big(\sigma^{\mathrm{nd}}\big) > \nicefrac{1}{2}$. 
Thus, appealing to Proposition~\ref{Prop:EqualPriorsKappa}, we can find $\kappa_w(\sigma^{\mathrm{nd}})$ for $w\in \{0,1\}$ as follows:
\begin{align}\label{Eq:kappa_ND}
    \kappa_w \big(\sigma^{\mathrm{nd}}\big) \teq \mu_1\big(\sigma^{\mathrm{nd}}\big) \teg \mu_1^2\big(\sigma^{\mathrm{nd}}\big) \tsum\! \tilde{\triangle}^{\mathrm{nd}} \E[D^2] \tsum\! \triangle^{\mathrm{nd}}(\E[D^2]\teg \E[D])
\end{align}
where $\triangle^{\mathrm{nd}}$ and $\tilde{\triangle}^{\mathrm{nd}}$ are found as
\begin{align*}
\tilde{\triangle}^{\mathrm{nd}} & = \big(\mu^2_1\big(\sigma^{\mathrm{nd}}\big) (2-\rho_0)- \mu_1\big(\sigma^{\mathrm{nd}}\big) + 0.25\big)
\rho_0\big / (1-\rho_0)^2,\\
\triangle^{\mathrm{nd}} & \teq (\theta_0 \teg \theta_0^2)(0.5\teg \alpha) \E_{\tilde{\rho}} [\gamma(\lfloor \nicefrac{D}{2} \rfloor;D\teg 1,\! \theta_1)\tsum \gamma(\lceil \nicefrac{D}{2}\teg 1\rceil; D\teg 1,\! \theta_1)].
\end{align*}
After some algebra, we can find $\mathcal{B}(\sigma^{\mathrm{nd}})$ from \eqref{Eq:mu1_ND} and \eqref{Eq:kappa_ND} as follows:
\begin{equation*}
    \mathcal{B}\big(\sigma^{\mathrm{nd}}\big) = \frac{N}{8} \left(\frac{1 \tsum \E\big[D^2\big]\tilde{\triangle}^\mathrm{nd} \tsum \big(\E\big[D^2\big]\teg \E[D]\big)\triangle^\mathrm{nd}}{(2\mu_1(\sigma^{\mathrm{nd}}) - 1)^2} -\frac{1}{4} \right)^{-1}\hspace{-2mm}.
\end{equation*}

Based on the  above, we have the next result  that the data collector can drive the total payment to be arbitrarily small for a given $N$, provided that $\mathcal{P}_e \leq e^{-\mathcal{B}(\sigma^{\mathrm{ND}})}$. 
\begin{proposition} \label{Thm:Pmin}
For the case with equal priors, if $\mathcal{P}_e \geq e^{-\mathcal{B}(\sigma^{\mathrm{ND}})}$, then we have that  $\mathcal{L}(\mathcal{P}_e) = \delta N$ for any $\delta > 0$, indicating that the total payment can be driven to be arbitrarily small. 
\end{proposition}
The proof is relegated to Appendix~\ref{Appendix:EntND}.

\textit{Remarks:} Theorem~\ref{Thm:Pmin} pinpoints to  the positive impact of social learning for all participants of the privacy-preserving data collection game. For the data collector, it implies that he can lower the payment significantly when there are sufficiently many users. From the perspective of the users, each of them incurs zero privacy cost. 

If the error constraint gets tighter, then the data collector can employ the designed payment mechanism, $\tilde{\mathbf{R}}$. Under the equal priors assumption, we can find  $\mathcal{B}(\sigma^*)$ from \eqref{Eq:muBNE} and \eqref{Eq:kappaBNE} as follows:
\begin{equation*}
\mathcal{B}\big(\sigma^* \big) = \frac{N}{8}  \left(\frac{1 \tsum \E\big[D^2\big]\tilde{\triangle} \tsum \big(\E\big[D^2\big]\teg \E[D]\big)\triangle}{(2\mu_1(\sigma^*) - 1)^2} -\frac{1}{4} \right)^{-1}\hspace{-2mm}.
\end{equation*}
Note that $\mathcal{B}(\sigma^*) \geq \mathcal{B}(\sigma^{\mathrm{nd}})$ and the data collector reduces the error rate of the MAP detector $\hat{W}_{\sigma^*}(\mathbf{X})$ by gathering informative reported data from the users who play with the SR strategies. 

Based on Theorem 3, our next result reveals that, when $\mathcal{P}_e < e^{-\mathcal{B}\big(\sigma^{\mathrm{nd}}\big)}$, the expected payment of the payment mechanism $\tilde{\mathbf{R}}$ constitutes an upper bound on $\mathcal{L}(\mathcal{P}_e)$.
\begin{proposition}
For the case with equal priors, when $\mathcal{P}_e < e^{-\mathcal{B}\big(\sigma^{\mathrm{nd}}\big)}$,  we have that
\begin{equation*}
\mathcal{L}(\mathcal{P}_e) \leq Z \left(1-\beta + \frac{\mu_1(\sigma^*)}{2\beta - 1} \right)N.
\end{equation*}
\end{proposition}

\subsection{Numerical Examples} \label{Section:NumericalResults}
In this section, we use examples to 
examine the impact of social learning on the trade-off between payment and accuracy and that between payment and privacy cost, 
using social learning graph models based on synthetic data and/or real-world data.

\begin{figure}[ht] 
\centering
\begin{subfigure}[b]{0.32\textwidth}
\includegraphics[width=1\textwidth]{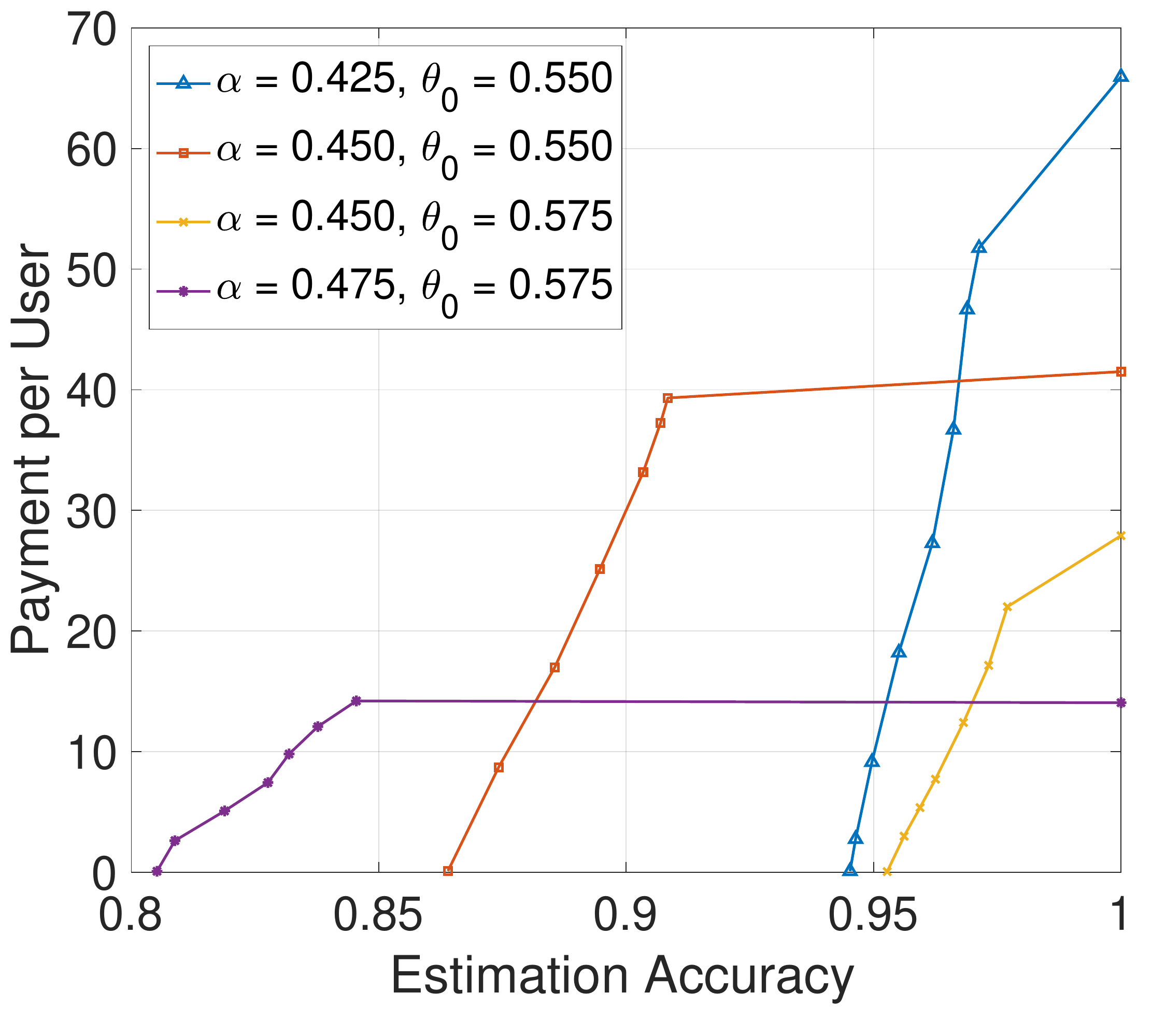}
\vspace{-0.5cm}
\caption{Arxiv GR-QC Collaboration Network}
\end{subfigure}
\begin{subfigure}[b]{0.32\textwidth}
\includegraphics[width=1\textwidth]{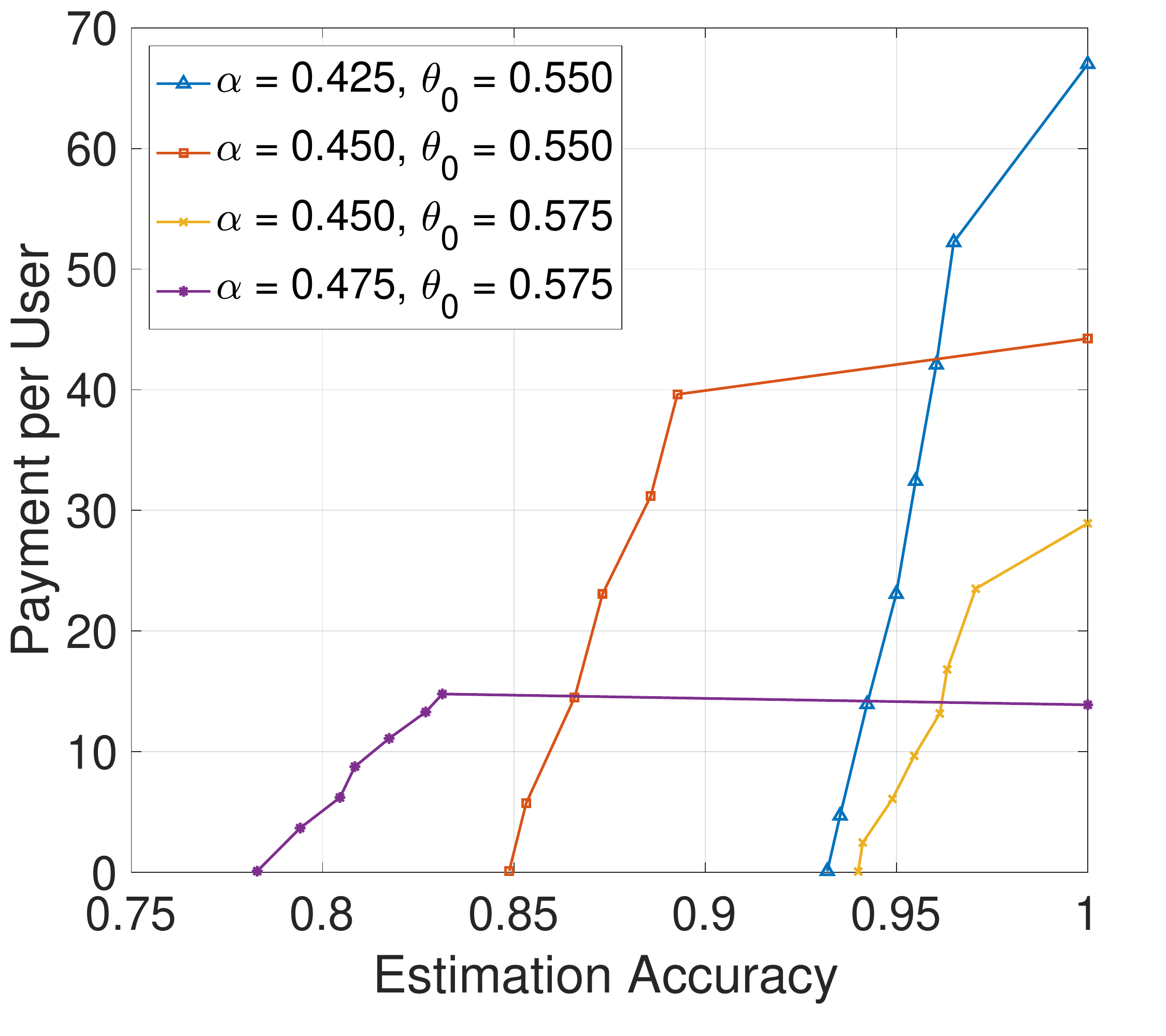}
\vspace{-0.5cm}
\caption{Gnutella Peer-to-Peer Network}
\end{subfigure}
\vspace{-0.2cm}
\caption{$\E[\tilde{R}_i(\mathbf{X})]$ vs. $\Pr_{\sigma^*} (\hat{W}_{\sigma^*} (\mathbf{X}) \teq W)$. The impact of social learning: payment per user vs. accuracy.} \label{Fig:RealWorld}
\end{figure}

\subsubsection{\textbf{Synthetic Social Learning Graphs.}}
To illustrate the impact of different parameters of the social learning graph and the payment mechanism on the performance, we first consider a synthetic model for the social learning graph. In the simulations, we use the Erdos-Renyi random graph model where each social tie is considered to be present with independent probability $\E[D_i]/(N-1)$. For large $N$, the degree distribution can be approximated by the Poisson distribution. We set $N = 250$ and $\Pr_W(0) = \Pr_W(1) = 0.5$ as default values. In the simulations, quadratic cost function is considered: $g(\zeta) = \zeta^2$.

 Fig.~\ref{Fig:Synthetic}a and \ref{Fig:Synthetic}b depict the state estimation accuracy, $\Pr_{\sigma^*} (\hat{W}_{\sigma^*} (\mathbf{X}) \teq W)$, with respect to the average degree of the social learning graph, $\E[D_i]$. To achieve a given accuracy level for the state estimator, the data collector needs to gather informative reported data from users who play the SR strategy, particularly when $\alpha$ is high. In the mechanism $\tilde{\mathbf{R}}$, the payment users receive, when their reported data matches with the majority of the other users' reported data, is an increasing function of $\epsilon$, a parameter determined by the data collector in order to meet the accuracy requirement. The larger $\epsilon$ is, the less noise the reported data would have (albeit the higher privacy cost), and hence the more accurate the state estimator. Fig.~\ref{Fig:Synthetic}(b) depicts $\Pr_{\sigma^*} (\hat{W}_{\sigma^*} (\mathbf{X}) \teq W)$ for $\epsilon=0.5$. Compared to Fig.~\ref{Fig:Synthetic}(a), it is clear that the estimation error is significantly reduced. 

When the social learning among users strengthens, the  privacy cost decreases because they receive informative social group signals $C_i$ more often and hence they play the SR strategy less often.  As illustrated in Fig.~\ref{Fig:Synthetic}(c) and \ref{Fig:Synthetic}(d), when the number of friends of a user increases, it is more likely for this user to play the ND strategy and hence her privacy cost drops. Accordingly, the total payment decreases. Furthermore, at the BNE strategy profile $\sigma^*$, Fig.~\ref{Fig:Synthetic}(e) and \ref{Fig:Synthetic}f depict the average payment each user receives in the payment mechanism $\tilde{\mathbf{R}}$ with respect to the state estimation accuracy, $\Pr_{\sigma^*} (\hat{W}_{\sigma^*} (\mathbf{X}) \teq W)$. It corroborates that the data collector can get an accurate estimate of the underlying state, with a much smaller payment compared to the case with no social learning.

\subsubsection{\textbf{Real World Social Learning Graphs.}}
To evaluate the impact of social learning in practice, we also use two social learning graph models based on real-world data. Firstly, we study Arxiv GR-QC (General Relativity and Quantum Cosmology) collaboration network \cite{snapnets}. The graph contains an edge between authors $i$ and $j$ if they co-authored at least one paper. The graph has 5242 nodes and 14496 edges. Secondly, we use the Gnutella peer-to-peer file sharing network from August 2002 \cite{snapnets}. Nodes represent hosts in the file sharing network and edges represent connections between the Gnutella hosts. It has 6301 nodes and 20777 edges. Fig~\ref{Fig:RealWorld} depicts the state estimation accuracy with respect to the payment per user under the proposed payment mechanism $\tilde{\mathbf{R}}$. These simulation studies also corroborate that the data collector can obtain an accurate estimate of $W$, with small amounts of payments despite the fact that very high noise is injected into  group signals and private signals.

\section{Conclusion and Future Work}

In this paper, we study a market model in which users can learn noisy versions of their social friends' data and make strategic decisions to report privacy-preserved versions of their personal data to a data collector. Thanks to the existence of social learning, the users have richer information about the underlying state beyond their personal signals. We develop a Bayesian game theoretic framework to study the impact of social learning on users' data reporting strategies and devise the payment mechanism for the data collector. Our findings reveal that, in general, the desired data reporting strategy at the Bayesian-Nash equilibria can be in the form of either symmetric randomized response or informative non-disclosive strategy. In particular, 
when a user plays the non-disclosive strategy, she reports her data completely based on her social group signals, independent of her personal signal, which drives her privacy cost to 0. As a result, both the data collector and the users benefit from social learning which lowers the privacy costs and helps to improve the state estimation at a given payment budget.


More specifically, our findings reveal that the desired data reporting strategy at the BNE is in the form of either a non-disclosive strategy or a symmetric randomized strategy. We show that the desired data reporting strategy is a \textit{majority voting} based data reporting rule which is applied by each user to her group signals to determine which strategy to follow. It is worth noting that the payment mechanism is designed to achieve \textit{informative} equilibria, because no user can gain by playing uninformative when other users follow informative data reporting strategies. We caution that the social learning model does not imply any collusion among friends. We use a Central Limit Theorem for dependence graphs to evaluate the estimation error of the underlying state. The total expected payment is characterized subject to a constraint on the estimation error. Our analysis reveals both the data collector and users benefit from social learning: The data collector can get an accurate estimate of the underlying state, with a much smaller payment (compared to the case with no learning), thanks to social learning.

We are currently generalizing this study to account for  the opinion formation dynamics which is based on the fusion of private signals and group signals across heterogeneous users, e.g., diffusion models with influential and stubborn users.  It is also of great interest to investigate the impact of ``fake signals'' (from fake news), and our effort along this line is underway.


\bibliographystyle{ACM-Reference-Format}
\bibliography{citMeet}
\newpage

\appendix
\section{Proof of Lemma 1}
\label{Appendix:ProofNDSR}
 Recall that $T_i = [S_i \ C_i]$ is the type of user $i$ and the type profile is $\mathbf{T} = [ T_1 \ T_2 \ \dots \ T_N ]$. Suppose that user $i$ uses strategy $\sigma'_i$ when other users follow $\tilde{\sigma}_{-i}$.
For convenience, define $\mathcal{U}_i((\sigma'_i,\tilde{\sigma}_{-i}),c_i)$ as follows:
\begin{equation*}
\mathcal{U}_i((\sigma'_i,\tilde{\sigma}_{-i}),c_i) = \E_{(\sigma'_i,\tilde{\sigma}_{-i})}  \left [R_i(\mathbf{X}) - g \left(\zeta(\sigma'_i,C_i) \right) | C_i \teq c_i \right ].  
\end{equation*}
The expected utility $\mathcal{U}_i (\sigma'_i,\tilde{\sigma}_{-i})$ can be written as:
\begin{equation*}
\mathcal{U}_i (\sigma'_i,\tilde{\sigma}_{-i}) = \sum_{d=0}^\infty \rho_{d_i} \hspace{-1mm} \sum_{c_i \in \{0,1\}^{d_i}} \Pr(C_i \teq c_i|D_i \teq d_i) \mathcal{U}_i((\sigma'_i,\tilde{\sigma}_{-i}),c_i).    
\end{equation*}
Then, we have that
\begin{align*}
& \mathcal{U}_i((\sigma'_i,\tilde{\sigma}_{-i}),c_i) = \\
& \sum_{x_i,s_i,\mathbf{t}_{-i}} \Pr (S_i \teq s_i | C_i \teq c_i) \Pr_{\sigma'_i} (X_i \teq x_i | T_i \teq t_i)\Pr(\mathbf{T}_{-i}\teq \mathbf{t}_{-i}| T_i \teq t_i) \\
& \quad \quad \times \Big( \E_{\tilde{\sigma}_{-i}} \big [  R_i(x_i,\mathbf{X}_{-i}) \big | T_i \teq t_i, \mathbf{T}_{-i} \teq \mathbf{t}_{-i} \big ]  \teg g(\zeta(\sigma'_i,c_i)) \Big ). \label{Eq:Lemma1CondExp}
\end{align*}
For convenience, define $p(s_i,c_i)$ and $q(s_i,c_i)$ as follows:
\begin{subequations} \label{Eq:pqsici}
\begin{align}
    p(s_i,c_i) & = \Pr_{\sigma'_i} (X_i \teq 1|S_i \teq s_i, C_i \teq c_i), \\
    q(s_i,c_i) & = \Pr_{\sigma'_i} (X_i \teq 0|S_i \teq s_i, C_i \teq c_i).
\end{align}
\end{subequations}
Based on Definition \ref{priDef}, we have that
\begin{align*}
& \zeta ( \sigma_i', c_i ) \teq \max \! \bigg \{  \left | \ln \left ( \frac{p(1,c_i)}{p(0,c_i)} \right )  \right |, \left | \ln \left ( \frac{1 \teg p(1,c_i)}{1 \teg p(0,c_i)} \right ) \right | , \left | \ln \left ( \frac{1\teg q(1,c_i)}{q(0,c_i)} \right ) \right |  , \nonumber \\
& \left | \ln \left ( \frac{q(1,c_i)}{q(0,c_i)} \right ) \right | \!, \! \left | \ln \left ( \frac{p(1,c_i)\tsum q(1,c_i)}{q(0,c_i) \tsum p(0,c_i)} \right ) \right |,  \left | \ln \left ( \frac{1\teg p(1,c_i)\!+\!q(1,c_i)}{1 \teg q(0,c_i)\tsum p(0,c_i)} \right ) \right | \bigg \}.
\end{align*}
For notation consistency, we use $g(p(1,c_i),p(0,c_i), q(1,c_i), q(0,c_i))$ to denote the privacy cost. It follows that
\begin{align*}
 \mathcal{U}_i((\sigma'_i,&\tilde{\sigma}_{-i}),c_i) \teq p(1,c_i) K(1,c_i) + p(0,c_i) K(0,c_i) + q(1,c_i)L(1,c_i) \\& + q(0,c_i)L(0,c_i) - g(p(1,c_i),p(0,c_i), q(1,c_i), q(0,c_i)),
\end{align*}
where $K(s_i,c_i)$ and $L(s_i,c_i)$ are given by
\begin{align*}
K(s_i,c_i) & =  \sum_{\mathbf{t}_{-i}} \Pr(S_i \teq s_i | C_i \teq c_i)\ \Pr(\mathbf{T}_{-i} \teq \mathbf{t}_{-i} |  T_i \teq t_i) \\
& \hspace{1.2cm} \times \E_{\tilde{\sigma}_{-i}} \big ( R_i (1, \mathbf{X}_{-i}) | T_i \!=\! t_i, \mathbf{T}_{-i} \!=\! \mathbf{t}_{-i} \big), \\
L(s_i,c_i) & =  \sum_{\mathbf{t}_{-i}} \Pr(S_i \teq s_i | C_i \teq c_i) \Pr(\mathbf{T}_{-i} \teq \mathbf{t}_{-i} |  T_i \teq t_i) \\
& \hspace{1.2cm} \times \E_{\tilde{\sigma}_{-i}} \big ( R_i (0, \mathbf{X}_{-i}) | T_i \!=\! t_i, \mathbf{T}_{-i} \!=\! \mathbf{t}_{-i} \big). 
\end{align*}
For any non-negative payment mechanism, $K(s_i,c_i),L(s_i,c_i) \geq 0$. Furthermore, for any $c_i \in \{0,1\}^{d_i}$, we have $K(1,c_i)$ and $K(0,c_i)$ are either both equal to zero or both positive. Similarly, for a given $c_i \in \{0,1\}^{d_i}$, $L(1,c_i)$ and $L(0,c_i)$ are either both equal to zero or both positive. When $K(s_i,c_i) = 0$ and $L(s_i,c_i) > 0$, the best response of user $i$ is to play ND and always report $X_i \teq 0$. When $L(s_i,c_i)\teq 0$ and $K(s_i,c_i)> 0$, then the best response of user i is to play ND and always report $X_i = 1$. When $L(s_i,c_i) \teq K(s_i,c_i)\teq 0$, the expected reward of the user is 0 and she can follow any ND strategy.

For positive $K(s_i,c_i), L(s_i,c_i)$ functions, determining the best response strategy $\sigma_i'$ reduces to the following optimization problem:
\begin{maxi}[2]
{\substack{p(1,c_i), p(0,c_i), \\ q(1,c_i),q(0,c_i)}} {\mathcal{U}_i((\sigma'_i,\tilde{\sigma}_{-i}),c_i)\hspace{3cm}}{}{} \nonumber
\addConstraint{0 \leq}{p(1,c_i)  \leq 1,  \quad 0  \leq q(1,c_i)  \leq 1}
\addConstraint{0  \leq}{p(1,c_i) + q(1,c_i)  \leq 1} 
\addConstraint{0  \leq}{p(0,c_i)  \leq 1, \quad 0 < q(0,c_i) < 1}
\addConstraint{0  \leq}{p(0,c_i) + q(0,c_i) \leq 1} 
\addConstraint{0 <}{p(1,c_i) + q(1,c_i) + p(0,c_i) + q(0,c_i).} \end{maxi}


This optimization problem has the same form of problem $(P)$ in \cite{Wang16}. Based on \cite{Wang16}, the solution of this optimization problem, $p^*(s_i,c_i),q^*(s_i,c_i)$, must satisfy the following conditions:
\begin{enumerate}
    \item $p^*(1,c_i) + q^*(1,c_i) = p^*(0,c_i) + q^*(0,c_i)  = 1$
    \item If $p^*(1,c_i) + p^*(0,c_i)  \neq 1$, then $p^*(1,c_i) = p^*(0,c_i)$.
\end{enumerate}
It follows that if the optimal action is not playing a symmetric randomization strategy ($p^*(1,c_i) = q^*(0,c_i)$), then it is playing a non-disclosive strategy ($p^*(1,c_i) = p^*(0,c_i)$).
\section{Proof of Theorem 1}
\label{Appendix:Proof_Thm_GenieAided}
Consider the payment mechanism defined in \eqref{Eq:GeniePayMech}. Observe that the expected utility of user $i$ at strategy $\sigma'_i$ is independent from the strategies of the rest of the users: 
\begin{equation*}
\mathcal{U}_i  (\sigma'_i,\sigma^*_{-i}) = E_{\sigma'_i}[R^{\mathrm{g}}_i(X_i, W) - g(\zeta(\sigma'_i,C_i))].
\end{equation*}
Thus, $\mathcal{U}_i  (\sigma'_i,\sigma^*_{-i}) = \mathcal{U}_i(\sigma'_i)$. Define, $\mathcal{U}_i(\sigma'_i,c_i)$ as follows:
\begin{equation*}
\mathcal{U}_i(\sigma'_i,c_i) = E_{\sigma'_i}[R^{\mathrm{g}}_i(X_i, W) - g(\zeta(\sigma'_i,C_i))|C_i \teq c_i].  
\end{equation*}
Note that, we have
\begin{equation*}
\mathcal{U}_i(\sigma'_i) = \sum_{d=0}^\infty \rho_{d_i} \sum_{c_i \in \{0,1\}^{d_i}} \Pr(C_i \teq c_i|D_i \teq d_i) \mathcal{U}_i(\sigma'_i,c_i).    
\end{equation*}
Next, we write $\mathcal{U}_i(\sigma'_i,c_i)$ as follows:
\begin{align*}
    \mathcal{U}_i(\sigma'_i,c_i) =&  \sum_{s_i =0}^1 \Pr(S_i = s_i|C_i = c_i) \sum_{x_i = 0}^1 \Pr_{\sigma'_i}(X_i \teq x_i|T_i \teq t_i)  \\
    & \times \sum_{w = 0}^1 \Pr_W(w|T_i \teq t_i) [R^{\mathrm{g}}_i(x_i, w) \teg g(\zeta(\sigma'_i,c_i))],
\end{align*}
where we used the facts that $D_i$ and $W$ are independent, $S_i$ and $C_i$ are conditionally independent given $D_i$ and $W$, and that $X_i$ and $W$ are independent given $T_i$ under strategy $\sigma'_i$. 

Recall that $p(s_i,c_i)$ and $q(s_i,c_i)$ are defined in \eqref{Eq:pqsici}. Following the rationale in the proof of Lemma~\ref{Lemma:SRND}, we have that
\begin{align*}
     \mathcal{U}_i(\sigma'_i,c_i) = & \sum_{s_i=0}^1 p(s_i,c_i) K(s_i,c_i) + q(s_i,c_i)L(s_i,c_i) \\
    & \quad - g\big(p(1,c_i),p(0,c_i), q(1,c_i), q(0,c_i),
\end{align*}
where $K(s_i,c_i)$ and $L(s_i,c_i)$ are defined as
\begin{align*}
K(s_i,c_i) & = \Pr(S_i \teq s_i | C_i \teq c_i) \sum_{w=0}^1 \Pr_W(w|  T_i \teq t_i)\ R_i^{\mathrm{g}}(1,w), \\
L(s_i,c_i) & = \Pr(S_i \teq s_i | C_i \teq c_i) \sum_{w=0}^1 \Pr_W(w|  T_i \teq t_i)\ R_i^{\mathrm{g}}(0,w).
\end{align*}
From \eqref{Eq:GeniePayMech}, we have that
\begin{align*}
K(s_i,c_i) = Z^{\mathrm{g}} \Pr(S_i \teq s_i | C_i \teq c_i) \Pr_W(1|  T_i \teq t_i)/\Pr_W(1),\\
L(s_i,c_i) = Z^{\mathrm{g}} \Pr(S_i \teq s_i | C_i \teq c_i) \Pr_W(0|  T_i \teq t_i)/\Pr_W(0).
\end{align*}
Observe that $K(s_i,c_i)$ and $L(s_i,c_i)$ are positive for every $s_i$ and $c_i$. Therefore, by the proof of Lemma~\ref{Lemma:SRND}, non-participation decision cannot be the best response and hence $p(s_i, c_i) + q(s_i, c_i) = 1$. The privacy level depends solely on $p(1,c_i)$ and $p(0,c_i)$ and the privacy level corresponding to strategy $\sigma_i'(t_i)$ can be written as:
\begin{align*}
\zeta(\sigma_i',c_i) = \max \bigg \{  \left | \ln \left ( \frac{p(1,c_i)}{p(0,c_i)} \right )  \right |, \left | \ln \left ( \frac{1 \teg p(1,c_i)}{1 \teg p(0,c_i)} \right ) \right | \bigg \}.
\end{align*}
For brevity, we denote the cost function as $g(p(1,c_i),p(0,c_i))$.  Then,
\begin{align*}
     \mathcal{U}_i(\sigma'_i,c_i) = & \  p(1,c_i) \overline{K}(1,c_i) + p(0,c_i)\overline{K}(0,c_i) \\
     & + \overline{K}(c_i) - g\big(p(1,c_i),p(0,c_i)\big)
\end{align*}
where
\begin{align*}
\overline{K}(s_i,c_i) \teq K(s_i,c_i) \teg L(s_i,c_i),\ \overline{K}(c_i) \teq L(1,c_i) + L(0,c_i).
\end{align*}
Recall that $f_i$ denotes the sum signal $f_i \teq \sum_{j \in G_i} c_{ij}$. For convenience, define $Y_0$ and $Y_1$ as
\begin{equation*}
    Y_1 \vcentcolon = \theta_1^{f_i} (1 - \theta_1)^{d_i - f_i}, \ Y_0 \vcentcolon = \theta_1^{d_i - f_i}(1 - \theta_1)^{f_i}.
\end{equation*}
After some algebra, we can write $\overline{K}(s_i,c_i)$ and $\overline{K}(c_i)$ as follows:
\begin{subequations} \label{Eq:AppendixBKBarEqs}
\begin{align}
\overline{K}(1,c_i) = & \ Z^{\mathrm{g}} \frac{\theta_0 Y_1 - (1 - \theta_0) Y_0}{\Pr_W(1) Y_1 + \Pr_W(0) Y_0},\\
\overline{K}(0,c_i) = & \ Z^{\mathrm{g}} \frac{(1-\theta_0) Y_1 - \theta_0 Y_0}{\Pr_W(1) Y_1 + \Pr_W(0) Y_0},\\
\overline{K}(c_i) = & \ Z^{\mathrm{g}} \frac{Y_0}{\Pr_W(1) Y_1 + \Pr_W(0) Y_0}.
\end{align}
\end{subequations}
Note that, $\overline{K}(s_i,c_i)$ and $\overline{K}(c_i)$ can also be written as $\overline{K}(s_i,f_i)$ and $\overline{K}(f_i)$. Therefore, in the rest of the proof, we use this notation. By Lemma~\ref{Lemma:SRND}, it suffices to consider SR and ND strategies:

\textbf{(I)} \textit{Symmetric Randomized Response}: The strategy is in the form of $p(1,f_i) + p(0,f_i) = 1$. Note that, an SR strategy can also be written as
\begin{equation*}
    p(1,f_i) = \frac{e^{\xi(f_i)}}{1+e^{\xi(f_i)}}, \ p(0, f_i) = \frac{1}{1+e^{\xi(f_i)}}.
\end{equation*}
Then the expected utility can be written as $\mathcal{U}_i  (\sigma'_i,f_i) = J[\xi(f_i)]$ where
\begin{equation*}
J[\xi(c_i)] \vcentcolon = \overline{K}(1,f_i) \frac{e^{\xi(f_i)}}{1\tsum e^{\xi(f_i)}} + \overline{K}(0,f_i) \frac{1}{1\tsum e^{\xi(f_i)}} + \overline{K}(f_i) \teg g (|\xi(c_i)|).
\end{equation*}
The optimal privacy level as a function of $f_i$ can be found by solving 
\begin{equation*}
    \eta^* = \argmax_{\eta\geq 0} J(\eta),
\end{equation*}
and setting $\xi(f_i) = \eta^*$. Observe that $\overline{K}(1,f_i) > \overline{K}(0,f_i)$ as $\theta_0 > 0.5$; therefore, $\eta^*$ is always non-negative. For any $\eta \geq 0$,
\begin{equation*}
J'(\eta) = \left (\overline{K}(1,f_i)- \overline{K}(0,f_i) \right) \frac{e^\eta}{\left (e^{\eta} + 1 \right)^2} - g'(\eta)
\end{equation*}
\begin{equation*}
J''(\eta) = - \left (\overline{K}(1,f_i)- \overline{K}(0,f_i) \right) \frac{e^\eta (e^\eta- 1)}{\left (e^{\eta} + 1 \right)^3} - g''(\eta)
\end{equation*} 
By the convexity of $g$,  $J''(\eta) \leq 0 $ for any $\eta\geq 0$. Therefore,  $\eta^*$ can be found by solving  the first order condition $J'(\varphi)\teq 0$.

\textbf{(II)} \textit{Non-disclosive Strategy}: $p(1,f_i)= p(0,f_i)$. Simplify the notation since $p(f_i) \teq p(1,f_i)\teq p(0,f_i)$. Define $\tilde{J}[p(f_i)]$ as
\begin{equation*}
\tilde{J}[p(f_i)] = p(\overline{K}(1,f_i) + \overline{K}(0,f_i)) + \overline{K}(f_i).
\end{equation*}
Then, we can find the optimal $p(f_i)$ as 
\begin{equation*}
\varphi^*(f_i) = \frac{\mathrm{sgn}(\overline{K}(1,f_i) + \overline{K}(0,f_i)) + 1}{2},
\end{equation*}
and setting $p(f_i) = \varphi^*$. 

The sum of the group signal $f_i$ determines whether user $i$ plays with the SR strategy or the ND strategy.
Define $h(f_i)$ as follows:
\begin{equation*}
h(f_i) =  J[\eta^*] - \tilde{J}[\varphi^*].   
\end{equation*}
User $i$ employs the SR strategy if $J[\eta^*]>\tilde{J}[\varphi^*]$ and employs the ND strategy if $J[\eta^*] < \tilde{J}[\varphi^*]$:
\begin{equation}\label{Eq:hci}
p(s_i,f_i) = \begin{cases} 
\varphi^* &\mbox{if } h(f_i) < 0 \text{ (ND) } \\ 
e^{\eta^*}/(1\tsum e^{\eta^*}) &\mbox{if } h(f_i) > 0 \text{ (SR) } \end{cases}
\end{equation}
If $h(f_i) = 0$, then the user can randomly decide whether she plays with SR or ND.

First, we consider the case $f_i \teq \nicefrac{d_i}{2}$. For the ND strategy, we have
\begin{equation*}
p(\nicefrac{d_i}{2}) = \varphi^* = 0.5 \text{ and } \tilde{J}[0.5] = \overline{K}(d_i/2).  
\end{equation*}
For the SR strategy, if there exists some $\epsilon > 0$ such that
\begin{equation*}
Z^{\mathrm{g}} = \frac{g'(\epsilon)}{2} \frac{1}{2\theta_0 \teg 1} \frac{(e^\epsilon+1)^2}{e^\epsilon},
\end{equation*}
then 
\begin{equation*}
\xi(\nicefrac{d_i}{2}) = \eta^* = \epsilon \text{ and } J(\epsilon) = \frac{g'(\epsilon)(e^{2\epsilon}-1)}{2e^\epsilon} + \overline{K}(\nicefrac{d_i}{2}).
\end{equation*}
This implies $h(\nicefrac{d_i}{2})>0$ and the user plays SR. If there does not exist such an $\epsilon>0$, then $\xi(\nicefrac{d_i}{2}) = \eta^* = 0$ and $J(0) = \overline{K}(d_i/2)$. In this case, $h(\nicefrac{d_i}{2}) = 0$; hence the user can toss a fair coin to decide whether she plays with SR or ND.

When $f_i \neq d_i/2$, we have that
\begin{equation*}
h(f_i) \teq \begin{cases}
\!\overline{K}(1,f_i) \frac{e^{\xi(f_i)}}{1 + e^{\xi(f_i)}} \tsum
\overline{K}(0,f_i) \frac{1}{1 + e^{\xi(f_i)}} \teg g(\xi(f_i)) \ \mathrm{if } f_i \! < \! \nicefrac{d_i}{2}, \vspace{2mm} \\ 
- \overline{K}(1,f_i) \frac{1}{1 + e^{\xi(f_i)}} \teg \overline{K}(0,f_i)  \frac{e^{\xi(f_i)}}{1+ e^{\xi(f_i)}} \teg g(\xi(f_i)) \ \mathrm{if } f_i \! > \! \nicefrac{d_i}{2}.
\end{cases}
\end{equation*}
Recall that $\bar{A}$ is defined in \eqref{Eq:barA}. If $f_i \geq d_i/2 + \overline{A}$, then $\overline{K}(0,f_i),\overline{K}(1,f_i)\geq 0$; and hence the ND strategy with $p(f_i) \teq 1$ is the best response. Similarly, if $f_i \leq d_i/2 - \overline{A}$, then $\overline{K}(0,f_i),\overline{K}(1,f_i)\leq 0$; and hence the ND strategy with $p(f_i) \teq 0$ is the best response. Note that, both $\overline{K}(1,f_i)$ and $\overline{K}(0,f_i)$ are monotonic functions and non-decreasing in $f_i$. Then, there exist $\tau_0, \tau_1 < d_i/2$ such that
\begin{equation}
\label{Eq:T1PrfHFormSgn}
\mathrm{sgn}(h(f_i)) = \begin{cases}
-1 &\mbox{if } f_i < d_i/2 \teg \tau_0 \\
-1 &\mbox{if } f_i > d_i/2 \tsum \tau_1 \\
1 &\mbox{else. } 
\end{cases}
\end{equation}

\section{Proof of Lemma 2}
\label{Appendix:ProofSum}
Recall that $\mathcal{E}$ is the adjacency matrix of the social graph $\mathcal{G}$. Let $\vec{\mathcal{E}_i}$ denote the $i^{th}$ row of the adjacency matrix $\mathcal{E}$: $\vec{\mathcal{E}_i} = [\mathcal{E}_{i1} \dots \mathcal{E}_{iN}]$. Recall that $\mathbf{S}$ is the signal profile $\mathbf{S} = [S_1 \ S_2 \dots S_N]$ and $\mathbf{X}$ is the reported data profile $\mathbf{X} = [X_1 \ X_2 \dots X_N]$. Conditioned on $W$, the joint distribution of $\mathbf{X}$ can be written as
\begin{align*}
    \Pr_{\sigma} (\mathbf{X}  \teq \mathbf{x}|W \teq w) = \sum_{\mathbf{s} \in \{0,1\}^N} \Pr(\mathbf{S} \teq \mathbf{s}|W \teq w) \ \Pr_{\sigma} (\mathbf{X} \teq \mathbf{x}|\mathbf{S} \teq \mathbf{s}).
\end{align*}
Recall that we consider the configuration model described in \cite{Newman2003} for the social graph $\mathcal{G}$. The degrees $\mathbf{D} = [D_1 \ D_2 \dots D_N]$ are independent and identically distributed random integers drawn from $\rho_d$. Pairs of users are chosen at random and edges are formed between them until complete pairing according to the drawn degree sequence. If complete pairing is not possible, one $d_i$ can always be discarded and redrawn from $\rho_d$. This procedure generates every possible adjacency matrix $\mathcal{E}$ with the given degree sequence $\mathbf{d} = [d_1 \ d_2 \dots d_N]$ with equal probability. Given $W$, $S_i$'s are independent and identically distributed with the quality parameter $\theta_0$. Thus, it follows that $\mathcal{E}$ and $\mathbf{S}$ are independent.

Define $\mathscr{S}_N$ as the set of all symmetric and binary matrices in which the entries of the main diagonals are all zeros. The conditional distribution of $\mathbf{X}$ given $\mathbf{S}$ can be written as
\begin{align*}
\Pr_{\sigma}  (\mathbf{X} \teq \mathbf{x}| \mathbf{S} \teq \mathbf{s}) & = \sum_{\mathbf{e} \in \mathscr{S}_N} \Pr_{\sigma}(\mathbf{X} \teq \mathbf{x}| \mathbf{S} \teq \mathbf{s}, \mathcal{E} \teq \mathbf{e}) \Pr(\mathcal{E} \teq \mathbf{e}| \mathbf{S} \teq \mathbf{s})\\
& = \sum_{\mathbf{e} \in \mathscr{S}_N} \Pr_{\sigma}(\mathbf{X} \teq \mathbf{x}| \mathbf{S} \teq \mathbf{s}, \mathcal{E} \teq \mathbf{e}) \Pr(\mathcal{E} \teq \mathbf{e}).
\end{align*}
Recall that, given user type $T_i = t_i$, the statistics of $X_i$ is determined by the user strategy $\sigma_i(t_i)$. Therefore, given the type profile $\mathbf{T} = [T_1 \ T_2 \dots T_N]$, $X_1, \dots, X_N$ are conditionally independent. Thus,
\begin{flalign*}
    & \Pr_{\sigma}(\mathbf{X} \teq \mathbf{x}|\mathbf{S} \teq \mathbf{s}, \mathcal{E} \teq \mathbf{e})  &&\\
    & = \sum_{\mathbf{t}\in \mathcal{T}^N} \Pr(\mathbf{T} \teq \mathbf{t}|\mathbf{S} \teq \mathbf{s}, \mathcal{E} \teq \mathbf{e}) \Pr_{\sigma} (\mathbf{X} \teq \mathbf{x}|\mathbf{S} \teq \mathbf{s}, \mathcal{E} \teq \mathbf{e}, \mathbf{T} \teq \mathbf{t}),  \\
    & = \sum_{\mathbf{t} \in \mathcal{T}^N} \prod_{i \teq 1}^N \Pr(T_i \teq t_i|\mathbf{S} \teq \mathbf{s}, \mathcal{E} \teq \mathbf{e}) \ \Pr_\sigma(X_i \teq x_i | \mathbf{S} \teq \mathbf{s}, \mathcal{E} \teq \mathbf{e}, \mathbf{T} \teq \mathbf{t}),
\end{flalign*}
\begin{align*}
    & = \sum_{\mathbf{t} \in \mathcal{T}^N} \prod_{i} \Pr(T_i \teq t_i|\mathbf{S} \teq \mathbf{s}, \vec{\mathcal{E}}_i \teq \vec{e}_i) \ \Pr_\sigma(X_i \teq x_i | \mathbf{S} \teq \mathbf{s}, \vec{\mathcal{E}}_i \teq \vec{e}_i, T_i \teq t_i),\\
    & = \sum_{\mathbf{t} \in \mathcal{T}^N} \prod_i \Pr_\sigma(X_i= x_i, T_i= t_i | \mathbf{S}= \mathbf{s}, \vec{\mathcal{E}}_i= \vec{e}_i),
\end{align*}
\begin{flalign*}
    & = \sum_{t_1 \in \mathcal{T}} \dots \sum_{t_N \in \mathcal{T}} \prod_i \Pr_\sigma(X_i= x_i, T_i = t_i | \mathbf{S} = \mathbf{s}, \vec{\mathcal{E}}_i = \vec{e}_i), &&\\
    & = \sum_{t_1 \in \mathcal{T}} \Pr_\sigma(X_1 \teq x_1, T_1 \teq t_1 | \mathbf{S} \teq \mathbf{s}, \vec{\mathcal{E}}_1 \teq \vec{e}_1) \dots &&\\
    & \hspace{1cm}\sum_{t_N \in \mathcal{T}} \Pr_\sigma(X_N \teq x_N, T_N \teq t_N | \mathbf{S} \teq \mathbf{s}, \vec{\mathcal{E}}_N \teq \vec{e}_N), &&\\
    & = \prod_i \Pr_\sigma(X_i \teq x_i|\mathbf{S} \teq \mathbf{s}, \vec{\mathcal{E}}_i \teq \vec{e}_i).
\end{flalign*}
Consequently, we have
\begin{align*}
    \Pr_\sigma & (\mathbf{X} \teq \mathbf{x}| \mathbf{S} \teq \mathbf{s}) = \sum_{\mathbf{e}} \Pr(\mathcal{E} \teq \mathbf{e}) \prod_i \Pr_{\sigma}(X_i \teq x_i|\mathbf{S} \teq \mathbf{s}, \vec{\mathcal{E}}_i \teq \vec{e}_i), \\
    & = \sum_{\mathbf{e}} \Pr(\mathcal{E} \teq \mathbf{e}) \prod_i \Pr_{\sigma}(X_i \teq x_i|\mathbf{S} \teq \mathbf{s}) \frac{\Pr_{\sigma_i}(\vec{\mathcal{E}}_i \teq \vec{\mathbf{e}}_i| X_i \teq x_i, \mathbf{S} \teq \mathbf{s})}{\Pr(\vec{\mathcal{E}_i} \teq \vec{e}_i)}.
\end{align*}
Next, we define $\mathcal{Y}_1(\mathbf{s}, \mathbf{x})$ and $\mathcal{Y}_2(\mathbf{s}, \mathbf{x})$ as
\begin{align*}
    \mathcal{Y}_1(\mathbf{s}, \mathbf{x}) \vcentcolon & = \sum_{\mathbf{e}} \Pr(\mathcal{E} \teq \mathbf{e}) \prod_i \frac{\Pr_{\sigma_i}(\vec{\mathcal{E}}_i \teq \vec{\mathbf{e}}_i| X_i \teq x_i, \mathbf{S} \teq \mathbf{s})}{\Pr(\vec{\mathcal{E}_i} \teq \vec{\mathbf{e}}_i)},\\
    \mathcal{Y}_2(\mathbf{s}, \mathbf{x}) \vcentcolon & = \prod_i \Pr_{\sigma_i}(X_i \teq x_i|\mathbf{S} \teq \mathbf{s}).
\end{align*}
It follows that
\begin{equation*}
  \Pr_{\sigma^*} (\mathbf{X} \teq \mathbf{x}|W \teq w) = \sum_{\mathbf{s}} \Pr(\mathbf{S} \teq \mathbf{s}| W \teq w) \mathcal{Y}_1(\mathbf{s}, \mathbf{x}) \mathcal{Y}_2(\mathbf{s}, \mathbf{x}).
\end{equation*}
Let $\mathcal{I}_N$ be the collection of all permutations on the set indices $\mathcal{I} = \{1,2,\dots,N\}$. Then, for $\pi \in \mathcal{I}_N$, $\mathbf{x}_{\pi} = [x_1 \dots x_N]_\pi$ denotes the permuted sequence $[x_{\pi(1)} \dots x_{\pi(N)}]$. By the symmetry of $\sigma$, for any $\pi \in \mathcal{I}_N$, we have that
\begin{equation*}
    \mathcal{Y}_1(\mathbf{s}, \mathbf{x}) = \mathcal{Y}_1(\mathbf{s}_\pi, \mathbf{x}_\pi), \quad  \mathcal{Y}_2(\mathbf{s}, \mathbf{x}) = \mathcal{Y}_2(\mathbf{s}_\pi, \mathbf{x}_\pi).
 \end{equation*}
Consequently, for any $\pi \in \mathcal{I}_N$, we have the following:
\begin{align*}
\Pr_{\sigma} & (\mathbf{X} \teq \mathbf{x}|W \teq w) = \sum_{\mathbf{s} \in \{0,1\}^\mathrm{N}} \Pr(\mathbf{S} \teq \mathbf{s}| W \teq w) \mathcal{Y}_1(\mathbf{s}, \mathbf{x}) \ \mathcal{Y}_2(\mathbf{s}, \mathbf{x}), \\
& = \sum_{\mathbf{s} \in \{0,1\}^\mathrm{N}} \Pr(\mathbf{S} \teq \mathbf{s}| W \teq w)\ \mathcal{Y}_1(\mathbf{s}_\pi, \mathbf{x}_\pi) \mathcal{Y}_2(\mathbf{s}_\pi, \mathbf{x}_\pi), \\
& = \sum_{\mathbf{s} \in \{0,1\}^\mathrm{N}} \Pr(\mathbf{S} \teq \mathbf{s}_\pi| W \teq w)\ \mathcal{Y}_1(\mathbf{s}_\pi, \mathbf{x}_\pi)\ \mathcal{Y}_2(\mathbf{s}_\pi, \mathbf{x}_\pi), \\
& = \sum_{\mathbf{s} \in \{0,1\}^\mathrm{N}} \Pr(\mathbf{S} \teq \mathbf{s}| W \teq w)\ \mathcal{Y}_1(\mathbf{s}, \mathbf{x}_\pi)\ \mathcal{Y}_2(\mathbf{s}, \mathbf{x}_\pi), \\
& \Pr_{\sigma} (\mathbf{X} = \mathbf{x}_\pi|W = w).
\end{align*}
Thus,
\begin{equation*}
\Pr_{\sigma} (\mathbf{X} \teq \mathbf{x}|W \teq w) = \Pr_{\sigma} \left(\sum_{i} X_i \teq \sum_{i} x_i \bigg|W \teq w \right).
\end{equation*}

\section{Proof of Lemma 3}
\label{Appendix:CLTLemma}
In what follows, we use a Central Limit Theorem (CLT) for dependence graphs \cite{janson1988} to characterize $\Pr_\sigma (\nicefrac{1}{\sqrt{N}}\sum_{i\teq 1}^N X_i < y|W = w)$. In this proof, for purposes of brevity, we drop the dependency of $\Pr_\sigma (X_i)$ on $\sigma$. Similarly, $\mu_w$, $\varsigma_w$ and $\overline{\varsigma}_w$ are used to denote $\mu_w(\sigma)$, $\varsigma_w(\sigma)$ and $\overline{\varsigma}_w(\sigma)$. We firstly present the proof for the case with $W=1$. For convenience, define $Y_N$ and $\tilde{Y}_N$ as
\begin{equation*}
Y_N \vcentcolon = \frac{1}{\sqrt{N}} \sum_{i = 1}^N X_i, \quad \tilde{Y}_N \vcentcolon = \frac{Y_N - \E[Y_N]}{\sqrt{\Var[Y_N]}}
\end{equation*}
Let $G_{ij} \vcentcolon= G_i \cap G_j$ be the set of users who are friends with both $i$ and $j$. Recall that $e_{ij}=1$ if there is a social tie between $i$ and $j$. To apply the CLT, we first evaluate $\Var [Y_N | W = 1]$:
\begin{align}
    & \Var [Y_N |W \teq 1 ] \teq \frac{1}{N} \mkern-4mu \left ( \sum_{i,j} \Pr( X_i \! \teq X_j \teq \! 1 | W \! \teq 1) \teg \left ( \sum_{i} \! \Pr(X_i \! \teq 1 | W \! \teq 1) \right )^{\!2} \right) \nonumber \\
    & \teq \! \frac{1}{N} \! \Bigg [ \hspace{-.75mm} \sum_{\substack{i,j \teq 1,\\ j\neq i}}^N \!  \sum_{k = 0}^{N - 2} \! \sum_{e = 0}^1 \! \Pr(|G_{ij}| \! \teq k,\! e_{ij} \! \teq e) \Pr (X_i \! \teq X_j \! \teq 1 | W \! \teq 1,\! |G_{ij}| \! \teq k,\! e_{ij} \! \teq e)  \nonumber \\
    & \quad +  \sum_{i\teq 1}^N  \Pr(X_i \teq 1 | W \teq 1) - \left ( \sum_{i\teq 1}^N  \Pr(X_i \teq 1 | W \teq 1) \right )^{\! 2} \Bigg]. \nonumber 
\end{align}
Next, we turn our attention to the evaluation of $\Pr(|G_{ij}| \teq k, e_{ij} \teq e)$. If users $i$ and $j$ are friends, and they also have a mutual friend $\ell$; then, $\{i,j,\ell\}$ is a 3-cycle in the social graph $\mathcal{G}$. Furthermore, consider the case where $i$ and $j$ are not friends but they have 2 common friends $\ell$ and $\hat{\ell}$. Then, $\{i,
j,\ell,\hat{\ell}\}$ is a 4-cycle in the social graph $\mathcal{G}$. Let $\mathscr{C}_3 (\mathcal{G})$ and $\mathscr{C}_4 (\mathcal{G})$ be the set of 3-cycles and 4-cycles in $\mathcal{G}$, respectively. It follows that
\begin{align*}
\sum_{\substack{i,j = 1,\\ j\neq i}}^N \sum_{k = 1}^{N-2} \Pr(|G_{ij}| \teq k, e_{ij} \teq 1) & \leq \sum_{\substack{i,j,\ell = 1,\\ \ell \neq j\neq i}}^N \Pr(\{ i,j,\ell\} \in \mathscr{C}_3 (\mathcal{G})), \\
\sum_{\substack{i,j = 1,\\ j\neq i}}^N \sum_{k = 2}^{N-2} \Pr(|G_{ij}| \teq k, e_{ij} \teq 0) & \leq \sum_{\substack{i,j,\ell, \hat{\ell} = 1,\\ \ell \neq \hat{\ell} \neq j\neq i}}^N \Pr(\{ i,j,\ell, \hat{\ell} \} \in \mathscr{C}_4 (\mathcal{G})).
\end{align*}
Then
\begin{align*}
\Pr(\{ i,j,\ell\} \in \mathscr{C}_3 (\mathcal{G})) & = \sum_{m_3} \Pr(|\mathscr{C}_3(\mathcal{G})| = m_3) \frac{m_3}{\binom{N}{3}} = \frac{\E[|\mathscr{C}_3(\mathcal{G})|]}{\binom{N}{3}}, \\
\Pr(\{ i,j,\ell, \hat{\ell}\} \in \mathscr{C}_4 (\mathcal{G})) & = \sum_{m_4} \Pr(|\mathscr{C}_4(\mathcal{G})| = m_4) \frac{m_4}{\binom{N}{4}} = \frac{\E[|\mathscr{C}_4(\mathcal{G})|]}{\binom{N}{4}}.
\end{align*}
Consequently,
\begin{align*}
\sum_{\substack{i,j = 1,\\ j\neq i}}^N \sum_{k = 1}^{N-2} \Pr(|G_{ij}| \teq k, e_{ij} \teq 1) & \leq 6 \E [ |\mathscr{C}_3| ], \\
\sum_{\substack{i,j = 1,\\ j\neq i}}^N \sum_{k = 2}^{N-2} \Pr(|G_{ij}| \teq k, e_{ij} \teq 0) & \leq 24 \E [|\mathscr{C}_4|].
\end{align*}
A key next step is to quantify $\E[|\mathscr{C}_3 (\mathcal{G})|]$ and $\E[|\mathscr{C}_4(\mathcal{G})|]$. Let $\mathbf{D} \teq [D_1 \dots D_N]$ denote the degree profile. Appealing to Theorem 2 in \cite{Bollobas1980}, we have that $|\mathscr{C}_3(\mathcal{G})|$ and $|\mathscr{C}_4(\mathcal{G})|$ are asymptotically independent Poisson random variables with conditioned expectations:
\begin{align*}
    & \E \Big [|\mathscr{C}_3(\mathcal{G})| \Big|\mathbf{D} \teq \mathbf{d} \Big] \teq \frac{1}{6} \left( \frac{\sum_{i\teq 1}^N d_i^2}{\sum_{i\teq 1}^N d_i} \teg 1 \right)^3 \\
    & \E \Big [|\mathscr{C}_4(\mathcal{G})| \Big|\mathbf{D} \teq \mathbf{d} \Big] \teq \frac{1}{8} \left ( \frac{\sum_{i\teq 1}^N d_i^2}{\sum_{i\teq 1}^N d_i} \teg 1 \right )^4.
\end{align*}
It follows that
\begin{subequations}
\label{Eq:UIProof}
\begin{align}
    \lim_{N \to \infty} \E[|\mathscr{C}_3(\mathcal{G})|] & = \frac{1}{6} \lim_{N \to \infty} \E \left [ \left ( \frac{\sum_{i\teq 1}^N D_i^2}{\sum_{i\teq 1}^N D_i} - 1 \right )^3 \right],  \\ 
    \lim_{N \to \infty} \E[|\mathscr{C}_3(\mathcal{G})|] & =  \frac{1}{8} \lim_{N \to \infty} \E \left [ \left ( \frac{\sum_{i\teq 1}^N D_i^2}{\sum_{i\teq 1}^N D_i} - 1 \right )^4 \right].
\end{align}
\end{subequations}






For convenience, define $Z_n$ and $\tilde{Z}_n$ for $n \in \{1,2, \dots, N \}$ as follows:
\begin{equation*}
    Z_n = \left( \frac{\sum_{i=1}^n D_i^2}{\sum_{i=1}^n D_i} - 1 \right)^4,  \ \ \tilde{Z}_n = \left( \frac{\sum_{i=1}^n D_i^2}{\sum_{i=1}^n D_i} - 1 \right)^3.
\end{equation*}
Next, we show that $\{Z_n: n\geq 1 \}$ is uniformly integrable (UI) as $n$ grows.  If $D_i > 0$, then
\begin{align*}
Z_n \leq & \left( \frac{\sum_{i= 1}^n D_i^2}{\sum_{i= 1}^n D_i} \right)^4 \leq \left( \frac{\sum_{i= 1}^n D_i^2}{n} \right)^4 = \frac{1}{n^4} \Bigg( \sum_{\substack{i \neq j \neq k \neq \ell}} D_i^2 D_j^2 D_k^2 D_\ell^2 \\
& + \sum_{i=1}^n D_i^8 + \sum_{\substack{i\neq j}} (4D_i^6 D_j^2 + 3 D_i^4 D_j^4) + \sum_{\substack{i \neq j \neq k}}  6 D_i^4 D_j^2 D_k^2 \Bigg).
\end{align*}
Then, from Assumption~\ref{Asm:Sparse}, for some $0 < \delta < \nicefrac{\triangle}{2}$ it directly follows that
\begin{align*}
& \E\left[ \left(\sum \! D_i^4 D_j^4 \Big /n^4 \right)^{\!1\tsum \delta} \right] \! < \! 1 \! < \! \infty, \E\left[ \left(\sum \! D_i^8 \Big/ n^4 \right)^{\!1\tsum \delta} \right] < n^{-2(1\tsum \delta)} \!<\! \infty,\\
&\E\left[\left(\sum D_i^2 D_j^2 D_k^2 D_\ell^2 \Big/n^4 \right)^{1\tsum \delta} \right] \leq \left(\E\left[D^{2+2\delta}\right]n^{-1-\delta} \right)^4 < \infty, \\
& \E\left[ \left(\sum D_i^6 D_j^2 \Big /n^4 \right)^{1\tsum \delta} \right] < \E\left[D^{2+2\delta}\right]n^{-\frac{7}{4}(1+\delta)} < \infty, \\
& \E\left[ \left(\sum D_i^4 D_j^2 D_k^2 \Big /n^4 \right)^{1\tsum \delta} \right] < \left( \E\left[D^{2\tsum 2\delta} \right] \right)^2 < \infty.
\end{align*}
Thus, $Z_n$ is UI \cite{Billingsley2012}. Clearly, if $Z_n$ is UI for $D_i > 0$, then $Z_n$ is also UI for $D_i \geq 0$. Since $|\tilde{Z}_n| \leq Z_n$ and $Z_n$ is UI, it follows that $\tilde{Z}_n$ is also UI. Therefore, the limits and expectations can be interchanged in \eqref{Eq:UIProof}. Then
\begin{align*}
\lim_{N \to \infty} & \E [|\mathscr{C}_3(\mathcal{G})|] \teq \frac{1}{6} \E \left [ \left ( \frac{\lim \limits_{N \to \infty} \frac{1}{N} \sum_{i=1}^N D_i^2}{ \lim \limits_{N \to \infty} \frac{1}{N} \sum_{i = 1}^N D_i} \teg 1 \right )^3 \right] \teq \frac{1}{6} \left ( \frac{\E [D^2]}{\E [D]} \teg 1 \right )^{\!3} \! \!, \\
\lim_{N \to \infty} & \E [|\mathscr{C}_4(\mathcal{G})|] \teq \frac{1}{8} \E \left [ \left ( \frac{\lim \limits_{N \to \infty} \frac{1}{N} \sum_{i=1}^N D_i^2}{ \lim \limits_{N \to \infty} \frac{1}{N} \sum_{i = 1}^N D_i} \teg 1 \right )^{\!4} \right] \teq \frac{1}{8} \left ( \frac{\E [D^2]}{\E [D]} \teg 1 \right )^{\!4} \! \!.
\end{align*}
Consequently, we have
\begin{align*}
\lim_{N \to \infty} \! \frac{1}{N} \mkern-5mu \sum_{\substack{i,j = 1,\\ j\neq i}}^N \! \sum_{k = 1}^{N-2} \Pr(|G_{ij}| \teq k, e_{ij} \teq 1) & \leq \lim_{N \to \infty} \frac{1}{N} \left ( \frac{\E [D^2]}{\E [D]} \teg 1 \right )^{\!3} \teq 0, \\
\lim_{N \to \infty} \! \frac{1}{N} \mkern-5mu \sum_{\substack{i,j = 1,\\ j\neq i}}^N \! \sum_{k = 2}^{N-2} \Pr(|G_{ij}| \teq k, e_{ij} \teq 0) & \leq \lim_{N \to \infty} \frac{3}{N} \left ( \frac{\E [D^2]}{\E [D]} \teg 1 \right )^{\!4} \teq 0.
\end{align*}
Next, we evaluate the expected number of user pairs $(i,j)$ such that $i$ and $j$ are friends with no common friend, and vice versa. Appealing to \cite{Newman2001}, we have that
\begin{align*}
\lim\limits_{N \to \infty} & \ \frac{1}{N} \sum_{i=1}^N  \sum_{j=1, j\neq i}^N \Pr(|G_{ij}| = 0, e_{ij} = 1) \ \cdot \\
& \Pr (X_i = X_j = 1 | W = 1, |G_{ij}| = 0, e_{ij} = 1) = \varsigma_1 \E[D], \\
\lim\limits_{N \to \infty} & \ \frac{1}{N}  \sum_{i=1}^N \sum_{j=1, j\neq i}^N \Pr(|G_{ij}| = 1, e_{ij} = 0) \ \cdot \\
\hspace{-5mm} &\Pr (X_i = X_j = 1 | W = 1, |G_{ij}| = 1, e_{ij} = 0) = \tilde{\varsigma_1} (\E[D^2] - \E[D]).
\end{align*}
For the case with $|G_{ij}| = 0$ and $e_{ij} = 0$, we have that
\begin{align*}
    & \Pr(|G_{ij}| = 0, e_{ij} = 0) = 1 - \Pr(|G_{ij}| \geq 1, e_{ij} = 0) - \Pr(e_{ij} = 1) , \\
    & \Pr(X_i = X_j = 1| W = 1, |G_{ij}| = 0, e_{ij}= 0) = \mu_1^2.
\end{align*}
Then,
\begin{align*}
    & \lim_{N \to \infty} \frac{1}{N} \Bigg ( \! \sum_{i \neq j} \! \Pr(|G_{ij}| \! \teq 0,\! e_{ij} \! \teq 0) \Pr (X_i \! \teq X_j \! \teq 1 | W \! \teq 1,\! |G_{ij}| \! \teq 0,\! e_{ij} \! \teq 0) \\
    & \hspace{1.5cm}- \left( \sum_{i} \Pr(X_i \teq 1|W \teq 1) \right)^{\! 2} \Bigg) = -\mu_1^2 \left( 1 + \E[D^2]\right).
\end{align*}
Consequently, for $\Var[Y_N |W = 1 ]$ we have that
\begin{align*}
    \lim_{N \to \infty} \Var[Y_N|W\teq 1] \teq \mu_1 \teg \mu_1^2 \tsum \E[D] \left(\varsigma_1 \teg \tilde{\varsigma}_1 \right) \tsum \E[D^2] \left (\tilde{\varsigma}_1 \teg \mu_1^2\right) \teq \kappa_1.
\end{align*}
Similarly, the variance of $Y_N$ given $W \teq 0$ is computed as
\begin{align*}
    & \lim_{N \to \infty}  \Var[Y_N|W\teq 0] = \lim_{N \to \infty} \Var \left[\frac{1}{N} \sum_{i\teq 1}^N (1\teg X_i) \ |W\teq 0 \right], \\
    & \quad = (1\teg \mu_0) \teg (1\teg \mu_0)^2 \tsum  \E[D] \left(\varsigma_0^0 \teg \varsigma_0^1 \right) \tsum  \E[D^2] \left (\varsigma_0^1 \teg (1\teg \mu_0)^2\right).
\end{align*}
Next, appealing to  Theorem 2 of \cite{janson1988}, we have that  for $\gamma < 2/3$,
\begin{equation*}
    \lim_{N \to \infty} \left( \frac{N}{D_{\mathrm{max}}} \right)^\gamma \frac{(D_{\mathrm{max}})^2/N}{\Var[Y_N|W\teq 1]} \to 0.
\end{equation*}
Thus, it follows that conditioned on $W = w$, 
\begin{equation*}
    \lim_{N \to \infty} \frac{Y_N - \E[Y_N|W = w]}{\sqrt{\Var[Y_N| W = w]}} =  \lim_{N \to \infty} \frac{\sum_{i=1}^N X_i - N \mu_w}{\sqrt{N \kappa_w}} \xrightarrow{d} \mathscr{N}(0,1).
\end{equation*}

\section{Proof of Theorem 2}
\label{ProofThm1}
Consider the payment mechanism defined in \eqref{PaymentMech1} and \eqref{PaymentMech2}. When other users use $\mathbf{\sigma}^*_{-i}$, the expected utility of user $i$ at strategy $\sigma'_i$ is given as
\begin{align*}
& \mathcal{U}_i  (\sigma'_i,\sigma^*_{-i})  = \E_{(\sigma'_i,\sigma^*_{-i})}  \Big [\tilde{R}_i(\mathbf{X}) - g\left(\zeta(\sigma'_i,C_i)\right) \Big ] = \sum_{d_i} \rho_{d_i} \cdot \\
& \sum_{c_i} \Pr (C_i\teq c_i|D_i \teq d_i) \sum_{s_i} \Pr (S_i \teq s_i|C_i\teq c_i) \sum_{x_i} \Pr_{\sigma'_i}(X_i \teq x_i|T_i \teq t_i) \ \cdot \\
&\sum_{w} \! \Pr_W\!(w|  T_i\teq t_i) \E_{(\sigma'_i,\sigma^*_{-i})} \! \! \left [\! \tilde{R}_i(\mathbf{X}) \teg g(\zeta(\sigma'_i,C_i)) \Big | T_i \teq t_i, \! X_i \teq x_i, \! W \teq w \! \right ] 
\end{align*}

Next, observe that 
\begin{align*}
& \E_{(\sigma'_i,\sigma^*_{-i})} \Big [\tilde{R}_i(\mathbf{X}) - g(\zeta(\sigma'_i,C_i)) \Big | T_i \teq t_i, X_i \teq x_i, W \teq w \Big ] \nonumber \\
& = \E_{\sigma_{-i}^*}\big [\tilde{R}_{i}(x_i, \mathbf{X}_{-i})|T_i \teq t_i, W \teq w \big ] - g\left(\zeta(\sigma'_i,c_i)\right) 
\end{align*}
where the equality follows from the conditional independence between $X_i$ and $\mathbf{X}_{-i}$ given $T_i$ and $W$. When the payment mechanism in \eqref{PaymentMech1} and \eqref{PaymentMech2} is applied, we have that
\begin{subequations}
\label{Eq:THMEQ2}
\begin{align}
& E_{\sigma_{-i}^*} \big [\tilde{R}_{i}(1, \mathbf{X}_{-i})| T_i \! \teq \! t_i,W \! \teq \! w\big ] \teq Z_1 \Pr_{\sigma_{-i}^*}(M_{-i} \! \teq \! 1| T_i \! \teq \! t_i,W \! \teq \! w) \label{THMEQ21}\\
& \E_{\sigma_{-i}^*} \big [\tilde{R}_{i}(0, \mathbf{X}_{-i})| T_i \! \teq \! t_i,W \! \teq \! w\big ] \teq Z_0 \Pr_{\sigma_{-i}^*}(M_{-i}\! \teq \! 0| T_i \! \teq \! t_i,W \! \teq \! w) \label{THMEQ22}
\end{align}
\end{subequations}
Recall that $G_i$ is the set of user $i$'s friends and $\overline{G}_i$ is the set of her second neighbors, i.e. friends of her friends. In the case of $D_i = 0$, given $X_i$, $T_i = [S_i]$ is conditionally independent from $\mathbf{X}_{-i}$. However, for $D_i > 0$, in general, $T_i$ is correlated with the reported data of users from $G_i$ and $\bar{G}_i$.  It can be shown that
\begin{align*}
& \Pr_{\sigma^*_{-i}}(M_{-i} = 1 | T_i = t_i, W = w) \\
& \quad = \Pr_{\sigma^*_{-i}} \left (\sum_{j \in G_i \cup \overline{G}_i} \hspace{-3mm} X_j \ + \sum_{\ell\in \mathcal{I} \setminus (G_i \cup \bar{G}_i)} \hspace{-4mm} X_\ell \ > N\triangle \ \big | \ T_i \teq t_i, W \teq w \right).
\end{align*}
Since, $X_i$'s are Bernoulli random variables we have
\begin{align*}
& \Pr_{\sigma^*_{-i}} \left (\hspace{6mm}\sum_{\mathclap{\ell \in \mathcal{I} \setminus (G_i \cup \overline{G}_i) }} X_\ell > N \triangle \bigg | W \teq w \right ) \ < \ \Pr_{\sigma^*_{-i}} (M_{-i} \teq 1 | T_i \teq t_i, W \teq w) \\
& \quad < \ \Pr_{\sigma^*_{-i}} \left ( \sum_{\ell \in \mathcal{I} \setminus (G_i \cup \overline{G}_i) } \hspace{-2mm} X_\ell > N \triangle \teg |G_i \cup \overline{G}_i| \ \bigg | W \teq w \right ).
\end{align*}

Under Assumption~\ref{Asm:Sparse} the maximal degree $\overline{D}$ is bounded and hence $|G_i \cup \overline{G}_i|$ is also bounded. Then, using a Sandwich argument, it can be shown that when the population size $N$ is large,  $\Pr (M_{-i} \teq 1|T_i\teq t_i, W \teq w) $ can be well approximated by $ \Pr (M_{-i} \teq 1| W \teq w)$ and that $ \Pr (M_{-i} \teq 0|T_i\teq t_i, W \teq w)$ by $\Pr (M_{-i} \teq 0| W \teq w)$. 
Recall that $\beta_0 = \Pr (M_{-i} \teq 0| W \teq 0)$ and $\beta_1 = \Pr (M_{-i} \teq 1| W \teq 1)$. Then the expected rewards in \eqref{Eq:THMEQ2} can be written as
\begin{subequations}
\begin{align}
& \E_{\sigma_{-i}^*} \big [\tilde{R}_{i}(1, \mathbf{X}_{-i})| W \! \teq 1\big ] \teq Z_1  \beta_1 , \E_{\sigma_{-i}^*} \big [\tilde{R}_{i}(0, \mathbf{X}_{-i})| W \! \teq 1\big ] \teq Z_0  (1\teg \beta_1), \nonumber \\
&\E_{\sigma_{-i}^*} \big [\tilde{R}_{i}(0, \mathbf{X}_{-i})| W \! \teq 0\big ] \teq Z_0  \beta_0, \E_{\sigma_{-i}^*} \big [\tilde{R}_{i}(1, \mathbf{X}_{-i})| W \! \teq 0\big ] \teq Z_1  (1\teg \beta_0). \nonumber
\end{align}
\end{subequations}
Along the same lin as in the proof of Lemma~\ref{Lemma:SRND}, we can write the expected reward of user $i$ at the strategy $\sigma'_i$ in terms of $p(s_i,c_i)$ and $q(s_i, c_i)$ as follows:
\begin{align*}
    \E_{(\sigma'_i,\sigma^*_{-i})}  \left [\tilde{R}_i(\mathbf{X}) | D_i = d_i \right] & =\sum_{c_i} \Pr(C_i = c_i|D_i = d_i) \ \cdot \\
    & \hspace{-5mm} \sum_{s_i} p(s_i,c_i) \ K(s_i, c_i) + q(s_i,c_i) \ L(s_i,c_i)
\end{align*}
\begin{align*}
\mathrm{where} \hspace{1mm} & K(s_i,c_i) = \Pr(S_i = s_i | C_i = c_i) \ \sum_{w} \E_{\sigma_{-i}^*} \big [\tilde{R}_{i}(1, \mathbf{X}_{-i}) | W = w \big ] \\ 
& \hspace{3.5cm}  \times \Pr(W = w |  S_i = s_i, C_i = c_i) , \\
& L(s_i,c_i) = \Pr(S_i = s_i | C_i = c_i) \ \sum_{w} \E_{\sigma_{-i}^*} \big [\tilde{R}_{i}(0, \mathbf{X}_{-i}) | W = w \big ] \\ 
& \hspace{3.5cm}  \times \Pr(W = w |  S_i = s_i, C_i = c_i). \\
\end{align*}
It can be shown that $K(s_i,c_i)$ and $L(s_i,c_i)$ are positive for every $s_i$ and $c_i$. Therefore, by the proof of Lemma~\ref{Lemma:SRND}, non-participation decision cannot be the best response and hence $p(s_i, c_i) + q(s_i, c_i) = 1$. The privacy level depends solely on $p(1,c_i)$ and $p(0,c_i)$ and the privacy level corresponding to strategy $\sigma_i'(t_i)$ can be written as:
\begin{align*}
\zeta(\sigma_i',c_i) = \max \bigg \{  \left | \ln \left ( \frac{p(1,c_i)}{p(0,c_i)} \right )  \right |, \left | \ln \left ( \frac{1 \teg p(1,c_i)}{1 \teg p(0,c_i)} \right ) \right | \bigg \}.
\end{align*}
For brevity, we denote the cost function as $g(p(1,c_i),p(0,c_i))$. The expected utility can be written as 
\begin{align*}
    & \E_{(\sigma'_i,\sigma^*_{-i})}  \left [\tilde{R}_i(\mathbf{X}) - g\left(\zeta(\sigma'_i,C_i)\right) | D_i \teq d_i \right] \teq \sum_{c_i} \Pr(C_i = c_i | D_i = d_i) \cdot \\ 
    & \hspace{5mm} \left ( p(1,c_i) \overline{K}(1,c_i) + p(0,c_i)\overline{K}(0,c_i) + \overline{K}(c_i) - g (p(1,c_i),p(0,c_i)) \right )
\end{align*}
where $\overline{K}(s_i,c_i)$ and $\overline{K}(c_i)$ are defined as
\begin{equation}\label{Eq:KBar}
\overline{K}(s_i,c_i) \vcentcolon = K(s_i,c_i) - L(s_i,c_i),\quad  \overline{K}(c_i) \vcentcolon = L(1,c_i) + L(0,c_i). 
\end{equation}
Recall that $f_i$ denotes the sum signal $f_i \teq \sum_{j \in G_i} c_{ij}$. For convenience define 
\begin{equation*}
    Y_1 \vcentcolon = \theta_1^{f_i} (1 \teg \theta_1)^{d_i \! - f_i}, \ Y_0 \vcentcolon = \theta_1^{d_i \! - f_i}(1\teg \theta_1)^{f_i}.
\end{equation*}
It follows that $\overline{K}(1,c_i)$ and $\overline{K}(0,c_i)$ can be written as
\begin{subequations}
\label{Eq:T1PrfKBar} 
\begin{equation}
\overline{K}(1,c_i) = Z \frac{\theta_0 Y_1 - (1 - \theta_0) Y_0}{\Pr_W(1) Y_1 + \Pr_W(0) Y_0},
\end{equation}
\begin{equation}
\overline{K}(0,c_i) = Z \frac{(1-\theta_0) Y_1 - \theta_0 Y_0}{\Pr_W(1) Y_1 + \Pr_W(0) Y_0},
\end{equation}
\end{subequations}
Note that, $\overline{K}(s_i,c_i)$ and $\overline{K}(c_i)$ can also be written as $\overline{K}(s_i,f_i)$ and $\overline{K}(f_i)$. Furthermore, \eqref{Eq:T1PrfKBar} has the form of \eqref{Eq:AppendixBKBarEqs} from Appendix~\ref{Appendix:Proof_Thm_GenieAided}. Therefore, the rest of the proof follows directly from the proof of Theorem 1.

\section{Proof of Proposition 1}
\label{ProofCorAlg}
In  Algorithm~\ref{alg:MV}, the optimal SR privacy level $\eta^*$ can be found by solving the first order condition $J'(\eta^*)\teq 0$. It follows from \eqref{Eq:T1PrfKBar} that, if $f_i \leq d_i/2 - \bar{A}$, the user plays the ND strategy with $p(1,f_i) \teq p(0,f_i) \teq 0$ and $\tau_0 \leq \bar{A}$. Similarly, if $f_i \geq d_i/2 + \bar{A}$, user plays the ND strategy with $p(1,c_i) \teq p(0,c_i) \teq 1$ and $\tau_1 \leq \bar{A}$. Recall that $h(f_i)$ is defined in Appendix B. For $f_i < d_i/2$, we have that
\begin{align*}
    h(f_i) \teq Z \frac{\theta_0 e^{\eta^*} \tsum 1 \teg \theta_0 \teg (\theta_0 \tsum (1\teg\theta_0)e^{\eta^*})(\theta_1/(1\teg \theta_1))^{d_i\teg 2 f_i}}{(p_W(1) \tsum (\theta_1/(1\teg \theta_1))^{d_i\teg 2 f_i})} \teg g(\eta^*).
\end{align*}

It follows that for $f_i < d_i/2$,  the condition $h(f_i) > 0$ for playing the SR strategy reduces to
\begin{equation*}
    f_i \!>\! \frac{d_i}{2} \teg \frac{\log \! \left ( \! \frac{e^{\eta^*}\theta_0 + 1 \teg \theta_0 - \Pr_W(1) (2\theta_0 \teg 1) 2 e^{\eta^*} g(\eta^*)(e^{\eta^*} \tsum 1)/ (g'(\epsilon)e^\epsilon \tsum 1)^2)}{e^{\eta^*}(1\teg \theta_0) + \theta_0 + \Pr_W(0) (2\theta_0 \teg 1) 2 e^{\eta^*} g(\eta^*)(e^{\eta^*} \tsum 1) / (g'(\epsilon)(e^\epsilon \tsum 1)^2)} \! \right )}{2 \log( \theta_1/(1-\theta_1))}, 
\end{equation*}
which yields $A_0(\eta^*)$. 
Similarly, for $f_i > d_i/2$, the condition $h(c_i) > 0$ for playing the SR strategy reduces to
\begin{equation*}
    f_i \!<\! \frac{d_i}{2} \tsum \frac{\log \! \left ( \! \frac{e^{\eta^*}(1 \teg \theta_0) \tsum \theta_0 \tsum \Pr_W(1) (2\theta_0 \teg 1) 2 e^{\eta^*} g(\eta^*)(e^{\eta^*} \tsum 1)/ (g'(\epsilon)e^\epsilon \tsum 1)^2)}{e^{\eta^*}\theta_0 + 1 \teg \theta_0 \teg \Pr_W(0) (2\theta_0 \teg 1) 2 e^{\eta^*} g(\eta^*)(e^{\eta^*} \tsum 1) / (g'(\epsilon)(e^\epsilon \tsum 1)^2)} \! \right )}{2 \log( \theta_1/(1-\theta_1))}, 
\end{equation*}
which yields $A_1(\eta^*)$.

\section{Proof of Proposition 2} \label{Appendix:PropEqual}
\textbf{(I)} In what follows,  we first evaluate $\mu_1(\sigma^*)$. Note that
\begin{align*}
& \mu_1(\sigma^*)= \Pr_{\sigma^*}(X_i \teq 1|W \teq 1) = \Pr_{\sigma^*}(F_{i}>\nicefrac{D_i}{2}\tsum \tau|W \teq 1) \\
& \quad + \lambda(\epsilon) \Pr_{\sigma^*}(F_i\in[\nicefrac{D_i}{2}\teg \tau, \nicefrac{D_i}{2} \tsum \tau]|W \teq 1),\\
& = \E_\rho \left[ \Gamma\left(\left \lfloor \nicefrac{D}{2}\tsum \tau\tsum1\right \rfloor , D;D,\theta_1 \right) \tsum \lambda(\epsilon) \Gamma \left( \nicefrac{D}{2}\teg \tau,\nicefrac{D}{2}\tsum \tau;D,\theta_1 \right)\right],
\end{align*}
where $\lambda(\epsilon)$ is defined as
\begin{equation*}
\lambda(\epsilon) \vcentcolon = \Pr_{\sigma^*} (X_i \teq 1|W \teq 1, D_i \teq 0) = \frac{\theta_0 e^\epsilon\tsum 1\teg \theta_0}{e^\epsilon\tsum 1}.
\end{equation*}

\textbf{(II)} For convenience, $\mathcal{J}_{k \ell}$ and $\tilde{\mathcal{J}}_{k \ell}$ are defined for $k, \ell \in \{0,1\}$ as  follows:
\begin{align*}
\mathcal{J}_{k \ell} & \vcentcolon = \Pr_{\sigma^*}(X_i = 1|W = 1, B_{ij}= 0, \mathcal{E}_{ij}= 1, S_i= k, C_{ij}= \ell),\\
\tilde{\mathcal{J}}_{k \ell} & \vcentcolon = \Pr_{\sigma^*}(X_i= 1|W= 1, B_{ij}= 0, \mathcal{E}_{ij}= 1, S_i= k, S_j= \ell).
\end{align*}
Next, we also define $F_{i,-j}$ as follows:
\begin{equation*}
    F_{i,-j} = \sum_{\ell \in G_i \setminus \{ j\}} C_{i\ell}.
\end{equation*}
Recall that 
\begin{equation*}
    \tilde{\rho}_{d} \vcentcolon = \Pr(D_i \teq d | D_i > 0) \teq \begin{cases}
    0 , &  \text{ if } d \teq 0; \\
    \nicefrac{\rho_{d}}{(1 \teg \rho_0)}, & \text{ else.}
\end{cases}
\end{equation*}
Rewrite $\mathcal{J}_{k\ell}$ as
\begin{align*}
\mathcal{J}_{k\ell} & = \Pr_{\sigma^*}(F_{i,-j} > \tau \teg \ell \tsum \nicefrac{D}{2}|W \teq 1, D>0)\\
& + \frac{k e^\epsilon\tsum 1\! -\! k}{e^\epsilon\tsum1}\ \Pr_{\sigma^*}(F_{i,-j} \in [\nicefrac{D}{2} \teg \tau \teg \ell, \nicefrac{D}{2}\tsum \tau \teg \ell]|W \teq 1, D>0)\\
& = \E_{\tilde{\rho}} \bigg[ \Gamma \left(\left \lfloor\frac{D}{2}+ \tau + 1 - \ell\right \rfloor, D - 1; D - 1, \theta_1 \right) \\
& \quad + \frac{k e^\epsilon + 1 - k}{e^\epsilon\tsum1}\ \Gamma \left(\frac{D}{2} - \tau - \ell, \frac{D}{2} + \tau - \ell; D - 1, \theta_1 \right) \bigg]
\end{align*}
and $\tilde{\mathcal{J}}_{k\ell}$ can be written in terms of $\mathcal{J}_{k\ell}$ as
\begin{equation*}
    \tilde{\mathcal{J}}_{k\ell} = (1-\alpha)\mathcal{J}_{k\ell} + \alpha \mathcal{J}_{k(1-\ell)}.
\end{equation*}

\textbf(III) Next, we define $\tilde{\mu}_1(\sigma^*)$ as the the mean of $X_i$ at strategy profile $\sigma$ when it is given that $W \teq 1$ and $i$ has at least one friend:
\begin{equation*}
\tilde{\mu}_1 (\sigma^*)  \vcentcolon = \Pr_{\sigma^*} [X_i=1|W= 1, D_i>0].  
\end{equation*}
From the proof of Lemma~\ref{CLTLemma}, we know that the probability of users $i$ and $j$ being friends and having an common friend together is negligible. Thus, $\tilde{\mu}_1(\sigma^*)$ can be written in terms of $\tilde{\mathcal{J}}_{k\ell}$:
\begin{align*}
\tilde{\mu}_1 (\sigma^*)&  \teq \sum_{s_i,k} \Pr(S_i \teq s_i, S_{j} \teq k|W \teq 1) \Pr_{\sigma^*}(X_i \teq 1|W \teq 1, S_i \teq s_i, S_{j} \teq k), \\
& = \theta_0^2 \tilde{\mathcal{J}}_{11} + (1\teg \theta_0)^2 \tilde{\mathcal{J}}_{00} + \theta_0 (1\teg \theta_0)(\tilde{\mathcal{J}}_{01}+\tilde{\mathcal{J}}_{10}).
\end{align*}
After some algebra, $\tilde{\mu}_1(\sigma^*)$ can also be written in terms of $\mathcal{J}_{k\ell}$ as
\begin{align*}
    \tilde{\mu}_1 = \theta_0((1\teg \alpha) \mathcal{J}_{11} \tsum \alpha \mathcal{J}_{10}) + (1\teg \theta_0)((1\teg \alpha) \mathcal{J}_{00} \tsum \alpha \mathcal{J}_{01}) + \mathcal{V}_1
\end{align*}
where $\mathcal{V}_1$ is defined as
\begin{equation*}
\mathcal{V}_1 \vcentcolon = \theta_0(1\teg \theta_0)(1\teg 2\alpha) (\mathcal{J}_{10}+\mathcal{J}_{01}-\mathcal{J}_{00}-\mathcal{J}_{11}).
\end{equation*}
Observe that
\begin{align*}
\mathcal{V}_1 =&  \theta_0(1\teg \theta_0)(1\teg 2\alpha)(e^\epsilon\teg 1)/(e^\epsilon\tsum 1) \\
& \times  \E_{\tilde{\rho}} \left[\gamma \left(\left \lfloor \nicefrac{D}{2} \tsum \tau \right \rfloor;D\teg 1, \theta_1 \right) \teg \gamma \left(\left \lceil \nicefrac{D}{2} \teg \tau \teg 1 \right \rceil;D\teg 1, \theta_1 \right) \right].
\end{align*}
For small $\alpha$, it follows from Corollary~\ref{Cor:EqPriors} that $\tau$ is small and we have
\begin{equation*}
    \E_{\tilde{\rho}} \left[\gamma \left(\left \lfloor \nicefrac{D}{2} \tsum \tau \right \rfloor;D\teg 1, \theta_1 \right) \teg \gamma \left(\left \lceil \nicefrac{D}{2} \teg \tau \teg 1 \right \rceil;D\teg 1, \theta_1 \right) \right] \cong 0.
\end{equation*}
Thus, $\tilde{\mu}_1 >> \mathcal{V}_1$. On the other hand, for large $\alpha$, we have very small $(1\teg 2\alpha)$ and hence $\tilde{\mu}_1 >> \mathcal{V}_1$. It follows that
\begin{equation*}
    \tilde{\mu}_1(\sigma^*) \cong \theta_0((1\teg \alpha) \mathcal{J}_{11} \tsum \alpha \mathcal{J}_{10}) + (1\teg \theta_0)((1\teg \alpha) \mathcal{J}_{00} \tsum \alpha \mathcal{J}_{01}).
\end{equation*}

\textbf{(IV)} Next, we evaluate $\varsigma_1(\sigma^*)$:
\begin{align*}
& \varsigma_1(\sigma^*) = \Pr_{\sigma^*}(X_i = X_j = 1| W = 1, B_{ij} = 0, \mathcal{E}_{ij} = 1), \\
& = \sum_{s_i, s_j}\Pr(S_i \teq s_i, S_j \teq s_j|W \teq 1) \sum_{k,\ell} \Pr(C_{ij} \teq k, C_{ji} \teq \ell|S_i \teq s_i, S_j \teq s_j)\\
& \times\!  \Pr_{\sigma^*}(X_i \teq 1|W \teq 1,\! S_i \teq s_i, C_{ij} \teq k)\Pr_{\sigma^*}(X_j \teq 1|W \teq 1,\! S_j \teq s_j, C_{ji} \teq \ell),
\end{align*}
\begin{align*}
&=\theta_0^2 ((1\teg \alpha)^2 \mathcal{J}_{11}^2 \tsum \alpha^2  \mathcal{J}_{10}^2\tsum 2\alpha(1\teg \alpha)\mathcal{J}_{10}\mathcal{J}_{11})\\
&+(1\teg \theta_0)^2((1\teg \alpha)^2 \mathcal{J}_{00}^2 \tsum \alpha^2  \mathcal{J}_{01}^2\tsum 2\alpha(1\teg \alpha)\mathcal{J}_{01}\mathcal{J}_{00})\\
&\tsum 2\theta_0(1\teg \theta_0)((1\!\teg\! \alpha)^2 \mathcal{J}_{10}\mathcal{J}_{01} \tsum \alpha^2  \mathcal{J}_{11}\mathcal{J}_{00} \tsum \alpha(1\!\teg\!\alpha)(\mathcal{J}_{10}\mathcal{J}_{00}\tsum \mathcal{J}_{11}\mathcal{J}_{01})).
\end{align*}
After some algebra, $\varsigma_1(\sigma^*)$ can be written as
\begin{align*}
    \varsigma_1(\sigma^*) = & \left(\theta_0((1\teg \alpha) \mathcal{J}_{11} \tsum \alpha \mathcal{J}_{10}) + (1\teg \theta_0)((1\teg \alpha) \mathcal{J}_{00} \tsum \alpha \mathcal{J}_{01})\right)^2 \tsum \mathcal{V}_0
\end{align*}
where $\mathcal{V}_0$ is defined as
\begin{equation*}
\mathcal{V}_0 = 2 \theta_0(1-\theta_0)(1-2\alpha) (\mathcal{J}_{01}\mathcal{J}_{10} - \mathcal{J}_{11}\mathcal{J}_{00}).    
\end{equation*}
Observe that, $\mathcal{V}_0$ can be expressed as
\begin{align*}
    \mathcal{V}_0 &=  2 \theta_0(1\teg \theta_0)(1\teg 2\alpha) \frac{e^\epsilon\teg 1}{e^\epsilon\tsum 1}  \Big(\E_{\tilde{\rho}}\left[\Gamma\left(\left \lfloor\nicefrac{D}{2}\tsum\tau\tsum1\right \rfloor,D\teg 1;D\teg 1,\theta_1\right)\right]\\
    &\times \E_{\tilde{\rho}}\left[\gamma\left(\left \lfloor\nicefrac{D}{2}\tsum \tau \right \rfloor;D\teg 1,\theta_1\right) \teg \gamma\left(\left \lceil\nicefrac{D}{2}\teg \tau\teg 1\right \rceil;D\teg 1,\theta_1\right) \right]\\
    & + \E_{\tilde{\rho}}\left[\Gamma\left(\nicefrac{D}{2}\teg \tau,\nicefrac{D}{2}\tsum \tau;D\teg 1,\theta_1\right)\right]\E_{\tilde{\rho}}\left[\gamma\left(\left \lfloor\nicefrac{D}{2}\tsum \tau\right \rfloor;D\teg 1,\theta_1\right)\right] \Big).
\end{align*}
For small $\alpha$, it follows from Corollary~\ref{Cor:EqPriors} that $\tau$ is small and
\begin{align*}
& \E_{\tilde{\rho}}\left[\Gamma\left(\nicefrac{D}{2}\teg \tau,\nicefrac{D}{2}\tsum \tau;D\teg 1,\theta_1\right)\right] \cong 0,\\
& \E_{\tilde{\rho}}[\gamma\left(\left \lfloor\nicefrac{D}{2}\tsum \tau \right \rfloor;D\teg 1,\theta_1\right) \teg \gamma\left(\left \lceil\nicefrac{D}{2}\teg \tau\teg 1\right \rceil;D\teg 1,\theta_1\right)] \cong 0.
\end{align*}
Thus, $\varsigma_1 >> \mathcal{V}_0$. On the other hand, for large $\alpha$, we have very small $(1\teg 2\alpha)$ and hence $\varsigma_1 >> \mathcal{V}_0$. It follows that
\begin{equation} \label{Eq:AppFVarsigma1}
    \varsigma_1 \cong \left(\theta_0((1\teg \alpha) \mathcal{J}_{11} \tsum \alpha \mathcal{J}_{10}) + (1\teg \theta_0)((1\teg \alpha) \mathcal{J}_{00} \tsum \alpha \mathcal{J}_{01})\right)^2.
\end{equation}

\textbf{(V)} Next, we refer to the observation from the Proof of Lemma~\ref{CLTLemma} that the probability of users $i$ and $j$ having more than one common friend together is negligible. Thus, $\tilde{\varsigma}_1(\sigma^*)$ can be evaluated as:
\begin{align*}
 \tilde{\varsigma}_1  & = \Pr_{\sigma^*}(X_i \teq X_j \teq 1| W \teq 1, B_{ij} \teq 1, \mathcal{E}_{ij} \teq 0) \teq \sum_{s_\ell} \Pr(S_\ell \teq s_\ell|W \teq 1)\\
&\times \Pr_{\sigma^*}(X_i \teq X_j \teq 1|W \teq 1, \mathcal{E}_{i\ell} \teq \mathcal{E}_{j\ell} \teq 1, B_{i\ell} \teq B_{j\ell} \teq 0, S_\ell \teq s_\ell)
\end{align*}
\begin{align*}
= & \sum_{s_\ell} \Pr(S_\ell \teq s_\ell|W \teq 1) \Pr^2_{\sigma^*}(X_i \teq 1| W \teq 1, \mathcal{E}_{i\ell} \teq 1, B_{i\ell} \teq 0, S_\ell \teq s_\ell)\\
= & \theta_0 (\theta_0 \tilde{\mathcal{J}}_{11} + (1\teg \theta_0)\tilde{\mathcal{J}}_{01})^2 + (1\teg \theta_0) (\theta_0 \tilde{\mathcal{J}}_{10} \tsum (1\teg \theta_0)\tilde{\mathcal{J}}_{00})^2.
\end{align*}
After some algebra, $\tilde{\varsigma}_1(\sigma^*)$ can also be written as:
\begin{align}\label{Eq:AppFTildeVarSigma}
\tilde{\varsigma}_1 \teq \tilde{\mu}_1^2 \tsum \theta_0(1\teg \theta_0)(1\teg 2\alpha)(\theta_0(\mathcal{J}_{11} \teg \mathcal{J}_{10}) \tsum (1\teg \theta_0)(\mathcal{J}_{01} \teg \mathcal{J}_{00})).
\end{align}

\textbf{(VI)} Recall the definition of $\kappa_1(\sigma^*)$:
\begin{align*}
\kappa_1 & = \mu_1 - \mu_{1}^2 + \E[D]\left(\varsigma_1 - \tilde{\varsigma}_1 \right) + \E[D^2]\big(\tilde{\varsigma}_1 - \mu_1^2 \big),\\
& = \mu_1 \teg \mu_1^2 + \E[D^2](\tilde{\mu}_1^2 \teg \mu_1^2) 
\tsum \E[D]\left(\varsigma_1 \teg \tilde{\varsigma}_1 \right) \tsum \E[D^2](\tilde{\varsigma}_1 \teg \tilde{\mu}_1^2).
\end{align*}
From \eqref{Eq:AppFVarsigma1} and \eqref{Eq:AppFTildeVarSigma}, it follows that
\begin{align*}
\E[D]\left(\varsigma_1(\sigma^*) \teg \tilde{\varsigma}_1(\sigma^*) \right) \tsum \E[D^2](\tilde{\varsigma}_1(\sigma^*) \teg \tilde{\mu}_1^2(\sigma^*)) = \triangle (\E[D^2]\teg \E[D])
\end{align*}
where $\triangle$ is defined as
\begin{align*}
\triangle \vcentcolon = \theta_0(1\teg \theta_0)(1\teg 2\alpha) (\theta_0(\mathcal{J}_{11} \teg\! \mathcal{J}_{10}) \tsum (1\teg\! \theta_0)(\mathcal{J}_{01} \teg\! \mathcal{J}_{00})).
\end{align*}
After some algebra, $\triangle$ can be found as
\begin{align}
&\triangle \vcentcolon = \theta_0(1\teg \theta_0)(1\teg 2\alpha) \ \E_{\tilde{\rho}}[(e^\epsilon(1\teg \theta_0)\tsum \theta_0)\ \gamma\left(\left \lfloor\nicefrac{D}{2}\tsum \tau \right \rfloor;D\teg 1,\theta_1\right) \nonumber \\
&\qquad + (\theta_0 e^\epsilon\tsum 1 \teg \theta_0) \gamma\left(\left \lceil\nicefrac{D}{2}\teg \tau\teg 1\right \rceil;D\teg 1,\theta_1\right)]/(e^\epsilon+1). \nonumber
\end{align}
For convenience, we define
\begin{align*}
\tilde{\triangle} = \tilde{\mu}_1^2-\mu_1^2.
\end{align*}
Note that, we have
\begin{equation*}
\tilde{\mu}_1 = (\mu_1 -\rho_0 \lambda)/(1-\rho_0).
\end{equation*}
Thus, 
\begin{equation*}
\tilde{\triangle} \vcentcolon = \frac{\rho_0}{(1\teg \rho_0)^2} \left(\mu_1^2(2\teg \rho_0) - 2\mu_1\lambda + \rho_0\lambda^2 \right).
\end{equation*}
Consequently, $\kappa_1(\sigma^*)$ can be written as:
\begin{equation*}
    \kappa_1 = \mu_1 \teg \mu_1^2 + \tilde{\triangle}\E[D^2] + \triangle(\E[D^2]\teg \E[D]).
\end{equation*}

\textbf{(VII)} Next, we evaluate $\mu_0(\sigma^*)$:
\begin{align*}
\mu_0(\sigma^*) & = \Pr_{\sigma^*}(X_i \teq 1|W \teq 0) = \Pr_{\sigma^*}(F_{i}>\nicefrac{D_i}{2}\tsum \tau|W \teq 0)\\
& \quad  + (1\teg \lambda(\epsilon))\Pr_{\sigma^*}(F_i\in[\nicefrac{D_i}{2}\teg \tau, \nicefrac{D_i}{2} \tsum \tau]|W \teq 0),\\
& = \E_{\rho}[\Gamma \left( \left \lfloor \nicefrac{D}{2}\tsum \tau\tsum1\right \rfloor,D;D,1\teg\theta_1 \right)] \\
& \quad + (1- \lambda(\epsilon)) \E_{\rho}[\Gamma \left( \nicefrac{D}{2}\teg \tau,\nicefrac{D}{2}\tsum \tau;D,1\teg\theta_1 \right)].
\end{align*}
Observe that
\begin{align*}
& \E_{\rho}[\Gamma \left( \nicefrac{D}{2}\teg \tau,\nicefrac{D}{2}\tsum \tau;D,1\teg\theta_1 \right)]=\E_{\rho}[\Gamma \left( \nicefrac{D}{2}\teg \tau,\nicefrac{D}{2}\tsum \tau;D,\theta_1 \right)],\\
& \E_{\rho}[\Gamma \left( \left \lfloor \nicefrac{D}{2}\tsum \tau\tsum1\right \rfloor,D;D,1\teg\theta_1 \right)] \\
& = 1\teg \E_{\rho}[\Gamma \left( \nicefrac{D}{2}\teg \tau,\nicefrac{D}{2}\tsum \tau;D,\theta_1 \right)]\teg \E_{\rho}[\Gamma \left( \left \lfloor \nicefrac{D}{2}\tsum \tau\tsum1\right \rfloor,D;D,\theta_1 \right)].
\end{align*}
Thus, it follows that
\begin{align*}
& \mu_0\teq 1 \teg \E_{\rho}[\Gamma (\lfloor \nicefrac{D}{2}\tsum \tau\tsum1 \rfloor,D;D,\theta_1)] \teg \lambda \E_{\rho}[\Gamma( \nicefrac{D}{2}\teg \tau,\nicefrac{D}{2}\tsum \tau;D,\theta_1)],\\
& = 1 - \mu_1.
\end{align*}

\textbf{(VIII)} Recall the definition of $\kappa_0(\sigma^*)$:
\begin{equation*}
\kappa_0 = \mu_0(1\teg \mu_0) \tsum \E[D]\left(\varsigma_0 \teg \tilde{\varsigma}_0 \right) \tsum \E[D^2]\left(\tilde{\varsigma}_0 - (1-\mu_0)^2 \right).
\end{equation*}
Note that, we have
\begin{align*}
& \Pr_{\sigma^*}(X_i = 0|W = 0, S_i = s, C_{ij} = k) \\
& \quad = \Pr_{\sigma^*}(X_i = 1|W = 1, S_i = 1- s, C_{ij} = 1- k)
\end{align*}
and
\begin{align*}
& \Pr_{\sigma^*}(X_i = 0| W = 0, \mathcal{E}_{i\ell} = 1, B_{i\ell} = 0, S_\ell = s) \\
& \quad = \Pr_{\sigma^*}(X_i = 1| W = 1, \mathcal{E}_{i\ell} = 1, B_{i\ell} = 0, S_\ell = 1-s).
\end{align*}
Thus,
\begin{equation*}
\varsigma_1(\sigma^*) = \varsigma_0(\sigma^*), \quad \tilde{\varsigma}_1(\sigma^*) = \tilde{\varsigma}_0(\sigma^*).
\end{equation*}
Consequently,
\begin{equation*}
\kappa_0(\sigma^*) = \mu_1(1\teg \mu_1) \tsum \E[D](\varsigma_1\teg \tilde{\varsigma}_1)\tsum \E[D^2](\tilde{\varsigma}_1 \teg \mu_1^2) = \kappa_1(\sigma^*).
\end{equation*}
\section{Proof of Theorem 3} \label{Appendix:ExpectedPayment}
We consider the payment mechanism $\tilde{\mathbf{R}}$ that is proposed in Theorem~\ref{MVbne}. The expected payment to user $i$ at $\sigma^*$ can be written as
\begin{equation*}
    \E_{\sigma^*}[\tilde{R}_i(\mathbf{X})|D_i \teq d] \teq \sum_{f=0}^{d} \Pr(F_i \teq f|D_i \teq d) \E_{\sigma^*}[\tilde{R}(\mathbf{X})|F_i \teq f, D_i \teq d].
\end{equation*}
For convenience, define $\overline{K}_i(s,f,d)$ and $\overline{K}_i(f,d)$ as:
\begin{align*}
\overline{K}_i&(s,f,d) \teq \Pr(S_i \teq s|F_i \teq f, D_i \teq d) \sum_{w = 0}^1 \Pr_W(w|S_i \teq s, F_i \teq f, D_i\teq d)\\
& \times \left(\E_{\sigma_{-i}^*} \big [\tilde{R}_{i}(1, \mathbf{X}_{-i}) | W = w \big ] -\E_{\sigma_{-i}^*} \big [\tilde{R}_{i}(0, \mathbf{X}_{-i}) | W = w \big ] \right),\\
\overline{K}_i&(f,d) \teq \sum_{w=0}^1 \Pr_W(w|F_i \teq f, D_i \teq d) \ \E_{\sigma_{-i}^*} \big [\tilde{R}_{i}(0, \mathbf{X}_{-i}) | W = w \big ].
\end{align*}
It follows from the Proof of Theorem~\ref{MVbne} that we have
\begin{align*}
    \E_{\sigma^*}& [\tilde{R}_i(\mathbf{X})|F_i \teq f,D_i \teq d] \\
    & = \begin{cases}
    & \overline{K}_i(f,d) \hspace{3.4cm} \mathrm{ if } f_i < d_i/2 - \tau, \\
    & \overline{K}_i(f,d) \tsum \overline{K}_i(1,f,d) \tsum \overline{K}_i(0,f,d) \quad \mathrm{ if } f_i > d_i/2 + \tau, \\
    & \overline{K}_i(f,d) + \left (\overline{K}_i(1,f,d)e^\epsilon + \overline{K}_i(0,f,d)\right)\frac{1}{e^\epsilon+1} \\
    & \hspace{2.7cm} \mathrm{ if } \quad d_i/2-\tau \leq f_i \leq d_i/2+\tau.
    \end{cases}
\end{align*}
Thus, 
\begin{align*}
\E_{\sigma^*}[\tilde{R}(\mathbf{X})&|D_i \teq d]  = \hspace{-2.5mm} \sum_{f= \lfloor \nicefrac{d}{2}+ \tau + 1\rfloor}^d \hspace{-5mm} \Pr(F_i \teq f | D_i \teq d ) \! \left (\overline{K}_i(1,f,d) \tsum \overline{K}_i(0,f,d) \right)\\
& + \sum_{f = \lceil \nicefrac{d}{2} - \tau \rceil}^{\lfloor\nicefrac{d}{2} + \tau \rfloor} \Pr(F_i \teq f | D_i \teq d )\ \frac{\overline{K}_i(1,f,d)e^\epsilon \tsum \overline{K}_i(0,f,d)}{1 + e^\epsilon}\\
& + \quad \sum_{f=0}^d \Pr(F_i \teq f | D_i \teq d ) \overline{K}_i(f,d).
\end{align*}
Recall that $\overline{Z}(\epsilon)$ is defined in the Proof of Theorem~\ref{MVbne} as:
\begin{equation*}
    \overline{Z}(\epsilon) = \frac{g'(\epsilon)(e^\epsilon + 1)^2}{2e^\epsilon}.
\end{equation*}
We have that
\begin{align*}
\overline{K}_i(1,f,d) + \overline{K}_i&(0,f,d) = 2\overline{Z}(\epsilon)(2\theta_0 - 1)^{-1}\\
& \times \left(\Pr_W(1|F_i \teq f, D_i \teq d)-\Pr_W(0|F_i \teq f, D_i \teq d) \right).
\end{align*}
By the definition of $\nu^{\mathrm{sr}}(d,\tau)$ and $\nu^{\mathrm{nd}}(d,\tau)$ \eqref{Eq:NuDef}, we have that
\begin{align*}
 \nu^{\mathrm{sr}}(d,\tau) &= \Pr(F_i \in [\nicefrac{d}{2} - \tau, \nicefrac{d}{2} + \tau]|W = 1, D_i = d), \\
&= \Pr(F_i \in [\nicefrac{d}{2} - \tau, \nicefrac{d}{2} + \tau]|W=0, D_i= d);\\
\nu^{\mathrm{nd}}(d,\tau) &= \Pr(F_i > \nicefrac{d}{2} + \tau]|W = 1, D_i = d),\\
& = \Pr(F_i < \nicefrac{d}{2}- \tau]|W= 0, D_i= d).
\end{align*}
Thus, we have that
\begin{align*}
& \hspace{-2.5mm} \sum_{f = \lfloor \nicefrac{d}{2}+ \tau + 1\rfloor}^d \hspace{-3mm} \Pr(F_i \teq f | D_i \teq d ) \left (\overline{K}_i(1,f,d) \tsum \overline{K}_i(0,f,d) \right)  = \frac{\overline{Z}(\epsilon)}{2\theta_0\teg 1} \\
&  \times \hspace{-2.5mm} \sum_{f = \lfloor \nicefrac{d}{2}+ \tau + 1\rfloor}^d \hspace{-5mm} \Pr(F_i \teq f|W\teq 1, D_i \teq d) \teg \Pr(F_i \teq f|W\teq 0,D_i \teq d) \\
& =  g'(\epsilon)(e^\epsilon\tsum1)^2/(2e^\epsilon(2\theta_0\teg 1)) (2\nu^{\mathrm{nd}}(d,\tau) + \nu^{\mathrm{sr}}(d,\tau) - 1 ).
\end{align*}
Following the same rationale, we have
\begin{align*}
 \sum_{f={\nicefrac{d}{2}-\tau}}^{\nicefrac{d}{2}+\tau} & \frac{1}{1+ e^\epsilon} \Pr(F_i = f | D_i = d ) \left(\overline{K}_i(1,f,d) e^\epsilon + \overline{K}_i(0,f,d)\right) \\ 
 & = \nu^{sr}(d,\tau) \ \overline{Z}(\epsilon)\ \frac{e^\epsilon - 1}{e^\epsilon + 1}.
\end{align*}
Note that we have $\beta = \beta_1 = \beta_0$ under equal priors assumption, and then it follows that
\begin{align*}
    & \sum_{f=0}^d \Pr(F_i \teq f | D_i \teq d ) \overline{K}_i(f,d) = \overline{Z}(\epsilon)  \frac{1}{(2\theta_0-1)(2\beta-1)}.
\end{align*}
Consequently, after some algebra, $\E_{\sigma^*}[\tilde{R}_i(\mathbf{X})|D_i \teq d]$ can be written as
\begin{equation*}
\E_{\sigma^*}[\tilde{R}_i(\mathbf{X})|D_i \teq d] \teq \frac{\overline{Z}(\epsilon)}{2\theta_0 \teg 1} \left( \frac{2 - 2 \beta}{2\beta - 1} + 2\nu^{\mathrm{nd}}(d,\tau) + 2 \lambda \nu^{\mathrm{sr}}(d,\tau) \right)
\end{equation*}
where $\lambda$ is defined in Theorem~\ref{Thm:Pmin} as $\lambda = (e^\epsilon \theta_0+ 1 \teg \theta_0)/(e^\epsilon\tsum 1)$.
Taking expectations of both sides over $\rho$ gives as:
\begin{align*}
\E_{\sigma^*}[\tilde{R}_i(\mathbf{X})] =  \frac{2\overline{Z}(\epsilon)}{2\theta_0 \teg 1} \left( \frac{1 - \beta}{2\beta - 1} + \E[\nu^{\mathrm{nd}}(D,\tau)] + \lambda \E[\nu^{\mathrm{sr}}(D,\tau)] \right).
\end{align*}
Note that we have $Z_1 = Z_0$ under equal priors assumptions as follows:
\begin{equation*}
    Z = Z_1 = Z_0 = \frac{2\overline{Z}(\epsilon)}{(2\beta-1)(2\theta_0-1)}.
\end{equation*}
Thus, we have that
\begin{equation*}
\E_{\sigma^*}[\tilde{R}_i(\mathbf{X})] = Z\left(1-\beta + \frac{\mu_1(\sigma^*)}{2\beta -1}\right).
\end{equation*}
As a result, we have that
\begin{equation*}
    \sum_{i=1}^N \E_{\sigma^*}[\tilde{R}_i(\mathbf{X})] = Z\left(1-\beta + \frac{\mu_1(\sigma^*)}{2\beta -1}\right)N.
\end{equation*}

\section{Proof of Proposition 3} \label{Appendix:EntND}
Recall that $\sigma^\mathrm{nd}$ is defined as the strategy profile at which the users determine their reported data as the majority bit of their group signals:
\begin{align*}
    &\Pr_{\sigma^{\mathrm{nd}}} (X_i \teq 1|S_i \teq s_i, F_i \teq f_i) = \begin{cases} 1 & \text{ if } f_i > \nicefrac{d_i}{2}, \\
    0 & \text{ if } f_i < \nicefrac{d_i}{2}, \\
    0.5 & \text{ else;}
    \end{cases} \\
    & \Pr_{\sigma^{\mathrm{nd}}} (X_i \teq 0|S_i \teq s_i, F_i\teq f_i) = 1 - \Pr_{\sigma^{\mathrm{nd}}} (X_i \teq 0|S_i \teq s_i, F_i\teq f_i).
\end{align*}
\textbf{(I)} First, we evaluate $\Pr_{\sigma^{\mathrm{nd}}}(M_{-i}=1|W=1)$ and $\Pr_{\sigma^{\mathrm{nd}}}(M_{-i}=1|W=1)$. Following the rationale presented in the Proof of Proposition~\ref{Prop:EqualPriorsKappa}, we can find $\mu_1(\sigma^{\mathrm{nd}})$ and $\mu_0(\sigma^{\mathrm{nd}})$ as follows:
\begin{align*}
    \mu_1\big(\sigma^{\mathrm{nd}}\big) = 1\teg \mu_0\big(\sigma^{\mathrm{nd}}\big) = \E \left[\Gamma \left(\frac{D\tsum 1}{2}, D; D, \theta_1 \right) + \frac{1}{2} \gamma \left(\frac{D}{2}; D, \theta_1 \right) \right].
\end{align*}
When users $i$ and $j$ friends without any common friend, their reported data is conditionally independent given $W \teq w$ at $\sigma^{\mathrm{nd}}$ since $X_i$ is independent from $S_i$ and $X_j$ is independent from $S_j$ given $W$:
\begin{align*}
    & \Pr_{\sigma^{\mathrm{nd}}}(X_i \teq X_j \teq w|W \teq w, S_i \teq s_i, S_j \teq s_j, \mathcal{E}_{ij} \teq 1, B_{ij} \teq 0) = \\
    & \Pr_{\!\sigma^{\mathrm{nd}}} (X_i \teq w|W \teq w,\! S_j \!\teq s_j,\! \mathcal{E}_{ij} \teq 1)\Pr_{\!\sigma^{\mathrm{nd}}} (X_j \teq w|W \teq w,\! S_i \!\teq s_i,\! \mathcal{E}_{ij} \teq 1).
\end{align*}
Thus, 
\begin{align*}
\varsigma_1\big(\sigma^{\mathrm{nd}}\big) & =  \varsigma_0\big(\sigma^{\mathrm{nd}}\big) = \Pr^2_{\sigma^{\mathrm{nd}}} (X_i \teq w|W \teq w, D_i > 0 ), \\
& = \E^2_{\tilde{\rho}} \left[\Gamma \left(\nicefrac{(D\tsum 1)}{2}, D; D, \theta_1 \right) + 0.5 \gamma \left(\nicefrac{D}{2}; D, \theta_1 \right) \right].
\end{align*}
Next, we evaluate $\tilde{\varsigma}_1$ and $\tilde{\varsigma}_0$. For convenience, define $\mathcal{J}_{k}$ for $k \in \{0,1\}$ as follows:
\begin{align*}
\mathcal{J}_k \vcentcolon = & \Pr_{\sigma^{\mathrm{nd}}}(X_i \teq 1|W \teq 1, \mathcal{E}_{i\ell} \teq 1, C_{i\ell} \teq k)\\
= & \E_{\tilde{\rho}} \left[ \Gamma\left(\frac{D\tsum 1}{2} \teg k,D\teg 1; D\teg 1, \theta_1 \right) \tsum \frac{1}{2} \gamma\left(\frac{D}{2} \teg k; D \teg 1, \theta_1 \right) \right].
\end{align*}
Furthermore, note that
\begin{equation*}
\mathcal{J}_{k} = \Pr_{\sigma^{\mathrm{nd}}}(X_i = 0|W = 0, \mathcal{E}_{i\ell} = 1, C_{i\ell} = 1-k).
\end{equation*}
Thus, we have $\tilde{\varsigma}_1 (\sigma^{\mathrm{nd}}) = \tilde{\varsigma}_0(\sigma^{\mathrm{nd}})$ and $\tilde{\varsigma}_1 (\sigma^{\mathrm{nd}})$ can be written as
\begin{align*}
    \tilde{\varsigma}_1 (\sigma^{\mathrm{nd}}) = \theta_0 \left( \alpha \mathcal{J}_0 \tsum (1\teg \alpha) \mathcal{J}_1 \right)^2 \tsum (1\teg \theta_0) \left( \alpha \mathcal{J}_1 \tsum (1\teg \alpha) \mathcal{J}_0 \right)^2.
\end{align*}
Consequently, we have that 
\begin{equation*}
    \kappa_1(\sigma^{\mathrm{nd}}) = \kappa_0(\sigma^{\mathrm{nd}}).
\end{equation*}
Appealing to Lemma~\ref{CLTLemma}, for sufficiently large $N$, we have that
\begin{equation*}
    \Pr_{\sigma^{\mathrm{nd}}}(M_{-i}=1|W=1) = \Pr_{\sigma^{\mathrm{nd}}}(M_{-i}=0|W=0) = \beta^{\mathrm{nd}}
\end{equation*}
where
\begin{equation*}
    \beta^{\mathrm{nd}} = \Phi\left(\sqrt{\frac{N - 1}{\kappa_1(\sigma^{\mathrm{nd}})}} \left(\mu_1\big(\sigma^\mathrm{nd}\big) - \frac{1}{2} \right) \right).
\end{equation*}

Next we consider a payment mechanism $\mathbf{R}^{\mathrm{nd}}$ that is constructed based on the payment mechanism $\tilde{\mathbf{R}}$ where the mechanism design parameters $Z_0$ and $Z_1$ are determined as the following:
\begin{align*}
    Z_0 & = \delta \frac{\Pr_W(1)\beta^{\mathrm{nd}} + \Pr_W(0)(1-\beta^{\mathrm{nd}})}{(2\beta^{\mathrm{nd}} - 1) (2 \theta_0 \teg 1) \Pr_W(1) \Pr_W(0))}, \\
    Z_1 & = \delta \frac{\Pr_W(1)(1-\beta^{\mathrm{nd}}) + \Pr_W(0)\beta^{\mathrm{nd}}}{(2\beta^{\mathrm{nd}} - 1) (2 \theta_0 \teg 1) \Pr_W(1) \Pr_W(0))},
\end{align*}
where $\delta>0$ is an arbitrarily small positive number. Note that, $Z_0$ and $Z_1$ are well defined since $\mu_1\big(\sigma^\mathrm{nd}\big) > 0.5$ and $\beta^{\mathrm{nd}}>0.5$.

\textbf{(II)} Next, we prove that $\sigma^{\mathrm{nd}}$ is a BNE in $\mathbf{R}^{\mathrm{nd}}$ payment mechanism. From the Proof of Theorem~\ref{MVbne}, it follows that the expected utility of user $i$ at strategy $\sigma'_i$ when other users use $\sigma^{\mathrm{nd}}_{-i}$ can be found as
\begin{align*}
    & \E_{(\sigma'_i,\sigma^{\mathrm{nd}}_{-i})}  \left [R^\mathrm{nd}_i(\mathbf{X}) - g\left(\zeta(\sigma'_i,C_i)\right) | D_i \teq d_i \right] \teq \sum_{c_i} \Pr(C_i = c_i | D_i = d_i) \cdot \\ 
    & \hspace{5mm} \left ( p(1,c_i) \overline{K}(1,c_i) + p(0,c_i)\overline{K}(0,c_i) + \overline{K}(c_i) - g (p(1,c_i),p(0,c_i)) \right )
\end{align*}
where $\overline{K}(s_i,c_i)$ and $\overline{K}(c_i)$ are found as
\begin{equation*}
\overline{K}(1,c_i) \teq \frac{\delta}{2\theta_0 \teg 1} \frac{\theta_0 \theta_1^{f_i} (1 \teg \theta_1)^{d_i \! - f_i} \teg (1 - \theta_0) (1 \teg \theta_1)^{f_i} \theta_1^{d_i \! - f_i}}{\Pr_W(1) \theta_1^{f_i} (1 \teg \theta_1)^{d_i \! - f_i} + \Pr_W(0) (1 \teg \theta_1)^{f_i} \theta_1^{d_i \! - f_i}},
\end{equation*}
\begin{equation*}
\overline{K}(0,c_i) \teq \frac{\delta}{2\theta_0 \teg 1} \frac{(1-\theta_0) \theta_1^{f_i} (1 \teg \theta_1)^{d_i \! - f_i} - \theta_0 (1 \teg \theta_1)^{f_i} \theta_1^{d_i \! - f_i}}{\Pr_W(1) \theta_1^{f_i} (1 \teg \theta_1)^{d_i \! - f_i} + \Pr_W(0) (1 \teg \theta_1)^{f_i} \theta_1^{d_i \! - f_i}}.
\end{equation*}
By Lemma~\ref{Lemma:SRND}, it suffices to consider SR and ND strategies for $\sigma'_i(t_i)$. From the Proof of Theorem~\ref{MVbne}, we have the following results:

\textbf{(a)} SR Strategies: The optimal privacy level $\xi*(c_i)$ can be found by solving the first order condition $J'(\varphi) = 0$:
\begin{equation*}
J'(\varphi) = \left (\overline{K}(1,c_i)- \overline{K}(0,c_i) \right) \frac{e^\varphi}{\left (e^{\varphi} + 1 \right)^2} - g'(\varphi),\quad \varphi \geq 0.
\end{equation*}
For any $\varphi>0$, we have $g(\xi^*)>0$. Hence, $\xi^*(c_i) = 0$, since since $\delta$ is an arbitrarily small positive constant. It implies that user $i$ tosses a fair coin when she employs SR strategies.

\textbf{(b)} ND Strategies: When user $i$ employs the optimal ND strategy, her reported data as follows:
\begin{align*}
\Pr_{\sigma'_i}(X_i=1|T_i = t_i) = (\mathrm{sgn}(\overline{K}(1,c_i) + \overline{K}(0,c_i)) + 1)/2.
\end{align*}
Note that, this strategy corresponds to $\sigma^{\mathrm{nd}}$.

From Theorem~\ref{MVbne}, we have that the best response strategy of user $i$ is employing the SR strategy if $f_i \teq \nicefrac{d_i}{2}$. For $f_i \neq \nicefrac{d_i}{2}$, we defined $h(c_i)$ in \eqref{Eq:hci} such that, the best response strategy is the ND strategy if $h(c_i) < 0$ and the SR strategy if $h(c_i)>0$. In the payment mechanism $\mathbf{R}^{\mathrm{nd}}$, $h(c_i)$ renders to the following:
\begin{equation*}
h(c_i) \teq \begin{cases}
0.5 \overline{K}(1,c_i) +
0.5 \overline{K}(0,c_i) \ \mathrm{if } f_i < d_i/2 \vspace{2mm} \\ 
- 0.5 \overline{K}(1,c_i)  - 0.5\overline{K}_{0,c_i}   \ \mathrm{if } f_i > d_i/2.
\end{cases}
\end{equation*}
It is clear that $\overline{K}(1,c_i) + \overline{K}(0,c_i) > 0$ if $f_i>\nicefrac{d_i}{2}$ and $\overline{K}(1,c_i) + \overline{K}(0,c_i) < 0$ if $f_i<\nicefrac{d_i}{2}$. Consequently, $\sigma^{\mathrm{nd}}$is a BNE in $\mathbf{R}^{\mathrm{nd}}$.

\textbf{(III)} Finally, we compute the expected total payment at $\sigma^{\mathrm{nd}}$. Appealing to  Theorem~\ref{Thm:ExpectedPayment}, we have that
\begin{equation*}
    \sum_{i=1}^N \E_{\sigma^{\mathrm{nd}}}[R^{\mathrm{nd}}_i(\mathbf{X})] = \frac{2\delta}{(2\beta^{\mathrm{nd}}-1)(2\theta_0-1)} \left(1\teg \beta^{\mathrm{nd}} + \frac{\mu_1(\sigma^*)}{2\beta^{\mathrm{nd}} -1}\right)N.
\end{equation*}
Recall that $\delta>0$ is an arbitrarily small positive number. Thus,
\begin{equation*}
    \sum_{i=1}^N \E_{\sigma^{\mathrm{nd}}}[R^{\mathrm{nd}}_i(\mathbf{X})] =\tilde{\delta} N
\end{equation*}
where $\tilde{\delta}>0$ is an arbitrarily small positive number.

\end{document}